\documentclass[12pt]{article}
\pdfoutput=1
\usepackage{mathtools,amssymb,mathrsfs,microtype,ytableau,tikz}%,pgfplots}
\usepackage[
 linktoc=all,
 colorlinks=true,
 linkcolor=blue,
 urlcolor=blue,
 citecolor=red,
 ]{hyperref}
\usepackage[backend=bibtex,style=numeric-comp,sorting=none,maxbibnames=99, minbibnames=99]{biblatex}
\makeatletter\@addtoreset{equation}{section}\makeatother

\DeclareMathOperator{\tr}{tr}

\newcommand{\SU}{\mathrm{SU}}
\newcommand{\Spin}{\mathrm{Spin}}
\newcommand{\Sp}{\mathrm{Sp}}
\newcommand{\SO}{\mathrm{SO}}
\newcommand{\U}{\mathrm{U}}
\renewcommand{\O}{\mathrm{O}}

\renewcommand{\title}[1]{\vbox{\center\LARGE{#1}}\vspace{5mm}}
\renewcommand{\author}[1]{\vbox{\center\large#1}\vspace{5mm}}
\newcommand{\address}[1]{\vbox{\center\em#1}}

\begin{document}
%\pgfplotsset{compat=1.16}
%\bibliographystyle{utphys}
%\bibliographystyle{plain}

\begin{titlepage}
\begin{center}
\vspace{5mm}
%\hfill {\tt HU-EP-09/40}\\
\hfill {\tt }\\
\vspace{8mm}

\title{\makebox[\textwidth]{\huge{Domain Walls in $4d~{\mathcal N}=1$ SYM}}}
\vspace{10mm}
Diego Delmastro,${}^{ab}$\footnote{\href{mailto:ddelmastro@perimeterinstitute.ca}
{\tt ddelmastro@perimeterinstitute.ca}}
Jaume Gomis${}^{a}$\footnote{\href{mailto:jgomis@perimeterinstitute.ca}
{\tt jgomis@perimeterinstitute.ca}}
\vskip 7mm
\address{
${}^a$Perimeter Institute for Theoretical Physics,\\
Waterloo, Ontario, N2L 2Y5, Canada}
\address{
${}^b$ Department of Physics, University of Waterloo,\\ Waterloo, ON N2L 3G1, Canada}
\end{center}

\vspace{5mm}
\abstract{
$4d$ ${\mathcal N}=1$ super Yang-Mills (SYM) with simply connected gauge group $G$ has $h$ gapped vacua arising from the spontaneously broken discrete $R$-symmetry, where $h$ is the dual Coxeter number of $G$. Therefore, the theory admits stable domain walls interpolating between any two vacua, but it is a nonperturbative problem to determine the low energy theory on the domain wall. We put forward an explicit answer to this question for all the domain walls for $G=\SU(N),\Sp(N), \Spin(N)$ and $G_2$, and for the minimal domain wall connecting neighboring vacua for arbitrary $G$. We propose that the domain wall theories support specific nontrivial topological quantum field  theories (TQFTs), which include the Chern-Simons theory proposed long ago by Acharya-Vafa for $\SU(N)$. We provide nontrivial evidence for our proposals by exactly matching renormalization group invariant partition functions twisted by global symmetries of SYM computed in the ultraviolet with those computed in our proposed infrared TQFTs. A crucial element in this matching is constructing the Hilbert space of spin TQFTs, that is, theories that depend on the spin structure of spacetime and admit fermionic states -- a subject we delve into in some detail.
}
\vfill\eject

\vspace{20pt}

{\hypersetup{linkcolor=black}
\tableofcontents
\thispagestyle{empty}
}

\end{titlepage}

\section{Domain Walls in $\boldsymbol{4d}$ $\boldsymbol{\mathcal N=1}$ SYM}\label{sec:intro}
\setcounter{footnote}{0} 

\begin{figure}[!h]
\centering
\begin{tikzpicture}
\begin{scope}[scale=.76]

\draw[thick,<->,>=stealth] (-1,4) -- (9,4);
\draw[thick,<->,>=stealth] (4,-1) -- (4,9);
\fill[white] ({4+3*cos(15*12)},{4+3*sin(15*12)}) circle (3pt);

\draw[thick] ([shift=(170:3cm)]4,4) arc (170:-170:3cm);
\foreach \i in {0,...,10}
{
%\draw[gray,thin] (4,4) -- ({4+3*cos(15*\i)},{4+3*sin(15*\i)}) arc (15*\i:15*(\i+1):3cm) -- cycle;
%\draw[gray,thin] (4,4) -- ({4+3*cos(15*(\i+13))},{4+3*sin(15*(\i+13))}) arc (15*(\i+13):15*(\i+14):3cm) -- cycle;
\fill[white,draw=black] ({4+3*cos(15*\i)},{4+3*sin(15*\i)}) circle (2pt);
\fill[white,draw=black] ({4+3*cos(15*(\i+13))},{4+3*sin(15*(\i+13))}) circle (2pt);
}

\fill[white,draw=black] ({4+3*cos(15*0)},{4+3*sin(15*0)}) circle (2pt);
\fill[white,draw=black] ({4+3*cos(15*11)},{4+3*sin(15*11)}) circle (2pt);

\foreach \i in {-1,0,1}
{
\filldraw ({4+3*cos(180+3*\i)},{4+3*sin(180+3*\i)}) circle (1pt);
\filldraw ({4+3.3*cos(60+2*\i)},{4+3.3*sin(60+2*\i)}) circle (.3pt);
\filldraw ({4+3.3*cos(-45+2*\i)},{4+3.3*sin(-45+2*\i)}) circle (.3pt);
}

\draw[thick,|->,>=stealth] ([shift=(30+46:3.8cm)]4,4) arc (30+46:164:3.8cm);
\draw[thick,<-|,>=stealth] ([shift=(30+44:3.8cm)]4,4) arc (30+44:-194:3.8cm);
\draw[thick,<-|,>=stealth] ([shift=(30+46:4.6cm)]4,4) arc (30+46:164:4.6cm);

\foreach \i in {1,2,3}
{
\node[scale=.55] at ({4+3.3*cos(15*\i)},{4+3.3*sin(15*\i)}) {$\i$};
}
%\fill[white] ({4+3.3*cos(15*0)},{4+3.3*sin(15*0)}) circle (3pt);

\node[scale=.55, anchor=south] at ({4+3.3*cos(15*0)},{4+3.3*sin(15*0)}) {$0$};
\node[scale=.55] at ({4.1+3.3*cos(-15)},{3.9+3.3*sin(-15)}) {$h-1$};
\node[scale=.55] at ({4.13+3.3*cos(-2*15)},{3.98+3.3*sin(-2*15)}) {$h-2$};
\node[scale=.55] at ({4+3.3*cos(15*5)},{4+3.3*sin(15*5)}) {$a$};
\node[scale=.55] at ({4+3.3*cos(15*11)},{4+3.3*sin(15*11)}) {$b$};
\node at ({4+4.2*cos(15*8)},{4+4.2*sin(15*8)}) {\footnotesize$\mathrm W_n$};
\node at ({4+5.1*cos(15*8+2)},{4+5.1*sin(15*8+2)}) {\footnotesize$\overline{\,\mathrm W_n}$};
\node at ({4.2+4.3*cos(-15*4)},{4+4.3*sin(-15*4)}) {\footnotesize$\mathrm W_{h-n}$};
\end{scope}
\end{tikzpicture}\caption{Vacua of $4d$ $\mathcal N=1$ SYM realized as the $h$ roots of unity in the $\langle\tr \lambda \lambda\rangle$-plane, where $\mathbb Z_{2h}$ acts by a $2\pi/h$ rotation. $\mathrm W_n$ denotes the domain wall interpolating between vacua separated by $n$ steps counterclockwise, and $\overline{\,\mathrm W_n}$ the domain wall connecting the same vacua but with its orientation reversed. Clearly, $\overline{\,\mathrm W_n}=\mathrm W_{h-n}$.}
\label{fig:wallspic}
\end{figure}
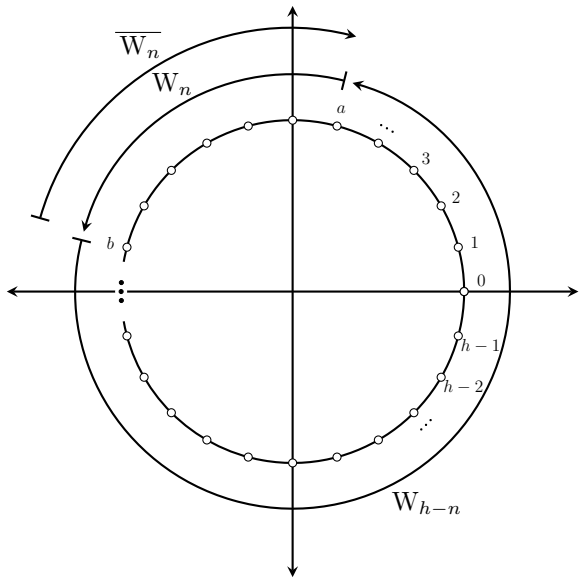

$4d$ $\mathcal N=1$ super Yang-Mills (SYM) -- Yang-Mills theory with a massless adjoint fermion -- is believed to share with QCD nonperturbative phenomena such as confinement, existence of a mass gap, and chiral symmetry breaking. $4d$ $\mathcal N=1$ SYM with simple and simply-connected gauge group $G$ has $h$ trivial vacua arising from the spontaneously broken $\mathbb Z_{2h}$ chiral $R$-symmetry down to $\mathbb Z_{2}$, where $h$ is the dual Coxeter number of $G$ (see table~\ref{tab:comarksone}). The vacua are distinguished by the value of the gluino condensate~\cite{SHIFMAN1988445,MOROZOV1988291,AFFLECK1985557}
\begin{equation}
\langle \tr\lambda \lambda \rangle=\Lambda^3 e^{2\pi i a/h}\,,\qquad a=0,1,\dots, h-1\,.
\end{equation}
A supersymmetric domain wall that interpolates between two arbitrary vacua $a$ and $b$ at $x_3\rightarrow \pm \infty$ can be defined. The $\mathbb Z_{2h}$ symmetry implies that the domain wall theory depends only on the difference between vacua, on $n\equiv a-b\mod~h$. We denote the resulting $3d$ low energy theory on the wall by $\mathrm W_n$ (see figure~\ref{fig:wallspic}).

While the $n$-wall tension is fixed by the supersymmetry algebra~\cite{Dvali:1996xe}, it is a nonperturbative problem to determine the low energy (i.e.~$E\ll\Lambda$) effective theory on the domain wall. A supersymmetric domain wall preserves $3d$ $\mathcal N=1$ supersymmetry and therefore a universal $3d$ ${\mathcal N}=1$ Goldstone multiplet describes the spontaneously broken translation and supersymmetry. The nontrivial dynamical question is whether anything else remains in the infrared, a topological quantum field theory (TQFT) or gapless modes and, if so, which one(s). 

In this paper we put forward a detailed answer to this question for all the $n$-domain walls for $G=\SU(N)$, $\Sp(N)$, $\Spin(N)$ and $G_2$, and for $n=1$ for arbitrary gauge group $G$.\footnote{Vacua of $4d$ ${\mathcal N}=1$ SYM when $G$ is not simply connected can be nontrivial~\cite{Aharony:2013hda,Tachikawa:2014mna}. The presence of a $4d$ TQFT means that there is no purely $3d$ wall theory.} The proposal for $G=\SU(N)$ was put forward long ago by Acharya-Vafa~\cite{Acharya:2001dz} motivated by brane constructions.\footnote{The $\Sp(N)$ case was mentioned in~\cite{GZS:unpublished,10.21468/SciPostPhys.6.4.044}. For a partial list of references on domain walls in $4d$ gauge theories see e.g.~\cite{Gaiotto:2013gwa,Dierigl:2014xta,Gaiotto:2017yup,Gaiotto:2017tne,Armoni:2009vv,Choi:2018tuh,Argurio:2018uup,Rocek:2019eve,Anber:2018xek,Wang:2019obe,Hsin:2018vcg,Bandos:2019qok}.} We provide nontrivial new evidence for the $\SU(N)$ proposal and for all the new proposals in this paper. The case $G=\Spin(N)$ is particularly subtle and rich. 

We conjecture that the infrared of the $n$-domain wall theory in $4d$ $\mathcal N=1$ SYM with gauge group $G$ is the infrared phase of $3d$ ${\mathcal N}=1$ SYM with gauge group $G$ and Chern-Simons level
\begin{equation}\label{leveljumpwall}
k=\tfrac12h-n\,.
\end{equation}
Since the $n$ and $h-n$ domain walls are related by time-reversal (see figure~\ref{fig:wallspic}), consistency of this proposal requires that the corresponding infrared phases must also be related by time-reversal, that is, by sending $k\rightarrow -k$ in the $3d$ theory. This requirement is indeed fulfilled by the identification between $n$ and $k$ in~\eqref{leveljumpwall}. In short, the proposal is the infrared duality
\begin{equation}\label{proposal}
\mathrm W_n~\text{in}~4d~{\mathcal N}=1~\text{SYM with}~G\ \longleftrightarrow\ 3d~{\mathcal N}=1~\text{SYM with}~G_{h/2-n}\,.
\end{equation}

Determining the infrared phase of $3d$ ${\mathcal N}=1$ SYM is also a nonperturbative problem. In~\cite{Gomis:2017ixy} it was proposed that this theory flows in the infrared to a nontrivial TQFT. The domain wall theories, we conjecture, are the ``quantum phases" put forward in~\cite{Gomis:2017ixy,Cordova:2017vab,Cordova:2018qvg}. This predicts the following domain wall theories:\footnote{The notation $G_k$ for Chern-Simons theories refers to Chern-Simons theory with gauge group $G$ at level $k\in \mathbb Z$.
The Chern-Simons theory $\U(n)_{k,k'}\equiv\frac{\SU(n)_k\times \U(1)_{nk'}}{\mathbb Z_n}$ has two levels, and the theory based on $\O(n)$ has three levels (see section~\ref{sec:orthogonal_chern_simons}).}
\begin{itemize}
\item $G=\SU(N)$. The $n$-domain wall theory is $\mathrm W_n=\U(n)_{N-n,N}$ Chern-Simons theory. This reproduces the proposal in~\cite{Acharya:2001dz}. 
\item $G=\Sp(N)$. The $n$-domain wall theory is $\mathrm W_n=\Sp(n)_{N+1-n}$ Chern-Simons theory. 
\item $G=\Spin(N)$. The $n$-domain wall theory is $\mathrm W_n=\O(n)^1_{N-2-n,N-n+1}$ Chern-Simons theory. We review the construction of this TQFT in section~\ref{sec:orthogonal_chern_simons}.
\item $G=G_2$. The theory has $h=4$ vacua and two independent walls: $n=1,2$. The $2$-domain wall theory is $\mathrm W_2=\SO(3)_3\times S^1$, with $\SO(3)_3$ Chern-Simons theory and $S^1$ the $3d$ sigma model on the circle. For the $n=1$ wall see below.
\end{itemize}
The expectation is that the $n$ and $(h-n)$-domain walls are related by time-reversal, that is $\mathrm W_{h-n}=\overline{\,\mathrm W_n}$ (see figure~\ref{fig:wallspic}). This is realized by virtue of the level-rank dualities of Chern-Simons theories~\cite{Hsin:2016blu,Aharony:2016jvv,Cordova:2017vab} and time-reversal flipping the sign of the Chern-Simons levels:
\begin{equation}
\begin{aligned}
\U(N-n)_{n,N}&\ \longleftrightarrow\ \U(n)_{-(N-n),-N}\\
\Sp(N+1-n)_{n}&\ \longleftrightarrow\ \Sp(n)_{-(N+1-n)}\\ 
\O(N-2-n)^1_{n,n+3}&\ \longleftrightarrow\ \O(n)^1_{-(N-2-n),-(N-n+1)}\,.
\end{aligned}
\end{equation}
The domain walls with $n=h/2$ are nontrivially time-reversal invariant. These TQFTs emerge in the infrared of $3d$ ${\mathcal N}=1$ SYM $G_0$, with vanishing Chern-Simons level, which is time-reversal invariant.

We also conjecture that:
\begin{itemize}
\item Arbitrary group $G$. The $n=1$ domain wall theory connecting neighboring vacua is $\mathrm W_1=G_{-1}$ Chern-Simons theory. This is consistent with the proposals put forward above due to the level-rank dualities $\U(1)_N\leftrightarrow \SU(N)_{-1}$, $\Sp(1)_N\leftrightarrow \Sp(N)_{-1}$ and $\O(1)^1_{N-3,N}=(\mathbb Z_2)_N\leftrightarrow \Spin(N)_{-1}$. 
\end{itemize}

We subject these proposals to a number of nontrivial quantitative tests. We exactly match renormalization-group invariant partition functions computed in the $3d$ ${\mathcal N}=1$ domain walls in the ultraviolet with the corresponding partition functions computed in the proposed infrared TQFTs. This lends nontrivial support for our domain wall proposals in $4d$ ${\mathcal N}=1$ SYM.

The most basic partition function in the $3d$ ${\mathcal N}=1$ $n$-domain wall theory is the Witten index~\cite{WITTEN1982253,Witten:1999ds}
\begin{equation}\label{eq:witten_index_def}
I_n=\tr_{\mathrm W_n}(-1)^F\,,
\end{equation}
where $\tr_{\mathrm W_n}$ denotes the trace over the torus Hilbert space of $\mathrm W_n$ with periodic boundary conditions, and $(-1)^F$ fermion parity. The partition function in the ultraviolet was first computed by Acharya-Vafa in~\cite{Acharya:2001dz}. 

We introduce and compute additional partition functions on the domain wall theory where the Witten index is twisted by a global symmetry of SYM. $4d$ $\mathcal N=1$ SYM with gauge group $G$ can have charge conjugation zero-form symmetry $\mathsf{C}$ and one-form symmetry $\Gamma$~\cite{Gaiotto:2014kfa}.\footnote{$\mathsf C$ is the outer automorphism group of the Dynkin diagram $\mathfrak g$ of $G$ while $\mathsf S$ is the outer automorphism group of the extended Dynkin diagram $\mathfrak g^{(1)}$ of the affine Lie algebra associated to $G$. The group $\Gamma$ is defined as the quotient $\mathsf S/\mathsf C$, i.e., the symmetries of $\mathfrak g^{(1)}$ that are \emph{not} symmetries of $\mathfrak g$.} $\Gamma$ is the center of $G$, since the fermion in $4d$ $\mathcal N=1$ SYM is in the adjoint representation of the gauge group. The symmetries $\mathsf{C}$ and $\Gamma$ do not commute when acting on Wilson lines, and combine into $\mathsf S=\Gamma\rtimes\mathsf C$ (see table~\ref{tab:comarksone}). $\mathsf{C}$ acts on local operators and Wilson lines, and $\Gamma$ on the Wilson lines of the theory. These symmetries are unbroken in each of the $h$ vacua of $4d$ $\mathcal N=1$ SYM. $\mathsf{S}$ is the unbroken symmetry at each vacuum, while $\mathbb Z_{2h}$ is spontaneously broken to $\mathbb Z_{2}$. This allows us to define the following twisted Witten indices on the $n$-domain wall theory\footnote{One could also twist by any element $\mathsf{c}\mathsf{g} \in \mathsf S$.}
\begin{equation}\label{eq:twistc_tr_def}
I^{\mathsf{c}}_n=\tr_{\mathrm W_n}(-1)^F\mathsf{c}\,,
\end{equation}
where $\mathsf c\in\mathsf C$, and 
\begin{equation}\label{eq:twistone_tr_def}
I^{\mathsf{g}}_n=\tr_{\mathrm W_n}(-1)^F\mathsf{g}\,,
\end{equation}
where $\mathsf{g}\in \Gamma$. We compute these partition functions in the ultraviolet in section~\ref{sec:UV} and in the infrared TQFTs in section~\ref{sec:IR_index}. 

Computing the domain wall untwisted and twisted Witten indices in the infrared requires understanding the Hilbert space of spin TQFTs (section~\ref{sec:spin}), and not merely counting the number of states of the Hilbert space of the TQFT on the torus, as has been often stated in the literature. We delve into the details of constructing the Hilbert space of spin TQFTs and determining the fermionic parity of the states in section~\ref{sec:spin}. The infrared (twisted) Witten  indices~\eqref{eq:witten_index_def},~\eqref{eq:twistc_tr_def},~\eqref{eq:twistone_tr_def} map to twisted partition functions in the infrared spin TQFT. Importantly, the dimension of the Hilbert space and the index differ in general, as we shall see. In particular, the index sometimes vanishes in theories of interest. While the index can vanish, the twisted indices are non-vanishing, and supersymmetry on the domain wall is unbroken.

\begin{table}[h!]
\begin{equation*}
\begin{array}{|c|c|c|c|c|c|c|c|c|c|c|}\hline
G&\SU(N)&\Sp(N)&\Spin(2N+1)&\Spin(4N)&\Spin(4N+2)&E_6&E_7&E_8&F_4&G_2\\\hline
h&N&N+1&2N-1&4N-2&4N&12&18&30&9&4\\\hline
\mathsf{C}&\mathbb Z_2&\cdot &\cdot & \mathbb Z_2&\mathbb Z_2&\mathbb Z_2&\cdot &\cdot &\cdot &\cdot \\\hline
\Gamma&\mathbb Z_N&\mathbb Z_2&\mathbb Z_2&\mathbb Z_2\times \mathbb Z_2&\mathbb Z_4&\mathbb Z_3& \mathbb Z_2&\cdot &\cdot &\cdot
\\\hline
\mathsf S& \mathbb D_N&\mathbb Z_2&\mathbb Z_2& \mathbb D_4& \mathbb D_4&\mathbb S_3 &\mathbb Z_2&\cdot &\cdot &\cdot\\\hline
\end{array}
\end{equation*}
\caption{Lie data for the simple Lie groups $G$. Here $h$ denotes the dual Coxeter number (defined as $\tr(t_\mathrm{adj}t'_\mathrm{adj})\equiv 2h (t,t')$, where $(\cdot,\cdot)$ denotes the Killing form on $\mathfrak g$, normalized so that the highest root has $(\theta,\theta)=2$). $\mathsf{C}$, $\Gamma$ are the zero-form and one-form symmetry groups of $4d$ ${\mathcal N}=1$ SYM with gauge group $G$, and $\mathsf S=\Gamma\rtimes\mathsf C$. $\mathbb D_N$ denotes the dihedral group with $2N$ elements, and $\mathbb S_N$ the symmetric group with $N!$ elements. For $\SU(2)$ the zero-form symmetry group is trivial, and for $\Spin(8)$ the zero-form symmetry group is enhanced to $\mathbb S_3$ and the total symmetry group to $\mathbb S_4$. The $\mathbb D_N$ symmetry of pure $\SU(N)$ YM was considered in~\cite{PhysRevD.100.085004}.}
\label{tab:comarksone}
\end{table}

We summarize here the results of our computations, performed both in the ultraviolet and infrared, and for which we find perfect agreement. We find it convenient to organize the results into master partition functions, which are defined as the generating functions for the twisted Witten indices. In other words, we sum the (twisted) partition functions over all $n$-walls:
\begin{equation}\label{eq:genZ_def}
Z^{\mathsf s}(q):=\sum_{n=0}^{h} I^{\mathsf s}_n q^n\,,%\,\qquad Z^{\mathsf{C}}(q)=\sum_{n=0}^{h} I^{\mathsf{C}}_n q^n\,, \qquad 
%Z^{\mathsf{g}}(q)=\sum_{n=0}^{h} I^{\mathsf{g}}_n q^n\,,
\end{equation}
where $q$ is a fugacity parameter, and where $\mathsf s\in\mathsf S$ is an element of the unbroken symmetry group. These partition functions have an elegant interpretation as twisted partition functions of a collection of free fermions in $0+1$ dimensions with energies determined by the Lie data of $G$ (see section~\ref{sec:UV}). Interestingly, the twisted partition function can be expressed as the untwisted partition function of an associated affine Lie algebra, whose extended Dynkin diagram is obtained by the ``folding procedure'' introduced in~\cite{Fuchs:1995zr}. 

The master partition functions take a rather simple form:
\begin{itemize}
\item $\SU(N)$:
\begin{align}
Z(q)&=(1-q)^N\\[+4pt]
Z^{\mathsf c}(q)&=\begin{cases}
(1-q)(1-q^2)^{(N-1)/2} \qquad\ N\text{ odd,}\\[+4pt]
(1-q)^2(1-q^2)^{(N-2)/2} \qquad N\text{ even,}
\end{cases}\\[+4pt]
Z^{\mathsf g}(q)&=\prod_{i=0}^{N-1}(1- \mathsf g^iq)\,,%\qquad\qquad \qquad {\mathsf g}^l\in \mathbb Z_N
\end{align}
where $\mathsf c$ denotes the non-trivial element of $\mathsf C=\mathbb Z_2$, and $\mathsf g$ is any element of $\Gamma=\mathbb Z_N$, thought of as an $N$-th root of unity.

\item $\Sp(N)$:
\begin{align}
Z(q)&=(1-q)^{N+1}\\[+4pt]
Z^{{\mathsf g}}(q)&=\begin{cases}
(1-q)(1-q^2)^{N/2} &\text{$N$ even,}\\[+4pt]
(1-q^2)^{(N+1)/2} &\text{$N$ odd,}%\qquad \qquad {\mathsf g} \in \mathbb Z_2
\end{cases}
\end{align}
where $\mathsf g$ denotes the non-trivial element of $\Gamma=\mathbb Z_2$.

\item $\Spin(N)$, $N$ odd:
\begin{align}
Z(q)&=(1-q)^3(1-q^2)^{(N-1)/2-2}\\[+4pt]
Z^{{\mathsf g}}(q)&= (1+q)(1-q)^2(1-q^2)^{(N-1)/2-2}\,,%\qquad \qquad\qquad {\mathsf g} \in \mathbb Z_2
\end{align}
where $\mathsf g$ denotes the non-trivial element of $\Gamma=\mathbb Z_2$.

\item $\Spin(N)$, $N$ even:
\begin{align}
Z(q)&=(1-q)^4(1-q^2)^{N/2-3}\\[+4pt]
Z^{\mathsf c}(q)&= (1-q)^2(1-q^2)^{N/2-2}\,,
\end{align}
where $\mathsf c$ denotes the non-trivial element of $\mathsf C=\mathbb Z_2$.
 
\begin{itemize}
\item $N=0\mod4$: $\Gamma=\mathbb Z_2\times \mathbb Z_2=\{\boldsymbol1,{\mathsf g}_1,{\mathsf g}_2,{\mathsf g}_1{\mathsf g}_2 \}$
 \begin{equation}
 \begin{aligned}
 Z^{{\mathsf g}_1}(q)&=(1-q^2)^{N/2-1} \\[+4pt]
 Z^{{\mathsf g}_2}(q)&=Z^{{\mathsf g}_1{\mathsf g}_2}(q)=(1-q^2)^3(1-q^4)^{N/4-2} \,.
\end{aligned}
\end{equation}

\item $N=2\mod4$: $\Gamma=\mathbb Z_4=\{\boldsymbol1,\mathsf g,\mathsf g^2,\mathsf g^3\}$
 \begin{equation}
\begin{aligned}
 Z^{{\mathsf g}}(q)&= Z^{{\mathsf g}^3}(q)=(1-q^4)^{(N/2-1)/2} \\[+4pt]
 Z^{{\mathsf g}^2}(q)&= (1-q^2)^{N/2-1}\,.%\qquad\qquad \qquad \qquad {\mathsf g}^l\in \mathbb Z_4
\end{aligned}
\end{equation}

\end{itemize}

\item $G_2$:
\begin{equation}\label{eq:g_2_UV_Z}
Z(q)=(1 - q)^2 (1 - q^2)\,.
\end{equation}
\end{itemize}
Expanding these formulas in a series in $q$ yields $I^{\mathsf s}_n$ (see section~\ref{sec:UV}). See also section~\ref{sec:min} for the $n=1$ domain wall twisted Witten indices for arbitrary simply-connected $G$.

The plan of the rest of the paper is as follows. In section~\ref{sec:UV} we review the calculation of the untwisted Witten index for general domain walls in $4d$ $\mathcal N=1$ SYM, develop the necessary tools to study the twisted indices, and present a detailed calculation thereof, for all the classical Lie groups. In section~\ref{sec:spin} we explain how the Hilbert space of a spin Chern-Simons theory is constructed and, in particular, how to determine the fermion parity $(-1)^F$ of the different states. In section~\ref{sec:IR_index} we use this refined understanding of spin Chern-Simons theories to compute the twisted partition functions of the $3d$ TQFTs that, conjecturally, describe the infrared dynamics of the domain walls, and show exact agreement. We end with some forward-looking comments in section~\ref{sec:fin}. We delegate to appendices some technical details that are needed in the computation of the twisted partition functions in section~\ref{sec:IR_index} and some additional material.

\section{Twisted Witten Indices}\label{sec:UV}

In this section we study the twisted Witten indices on the $3d$ ${\mathcal N}=1$ domain walls in the ultraviolet. This requires considering $4d$ ${\mathcal N}=1$ SYM on a two-torus and quantizing the space of zero energy states. This leads to a $2d$ ${\mathcal N}=(2,2)$ sigma model on the moduli space of flat $G$-connections on a two-torus, which is the weighted projective space ${\bf WCP}^{r}_{a^{\vee}_0,a^{\vee}_1,\dots,a^{\vee}_r}$, where $a^{\vee}_i$ is the comark for the $i$-th node in the extended Dynkin diagram $\mathfrak g^{(1)}$ of the affine Lie algebra associated to $G$ and $r=\operatorname{rank}(G)$~\cite{Looijenga1976,Friedman:1997yq}. Just as $4d$ ${\mathcal N}=1$ SYM, this $2d$ theory also has $h$ quantum vacua. A supersymmetric domain wall in $4d$ ${\mathcal N}=1$ SYM corresponds to a supersymmetric soliton in the $2d$ ${\mathcal N}=(2,2)$ sigma model~\cite{Acharya:2001dz}. 

Using the $2d$ ${\mathcal N}=(2,2)$ sigma model, Acharya and Vafa argued that the Witten index of the domain wall is encoded in the Hilbert space of $r+1$ free fermions in $0+1$ dimensions. Each fermion $\psi_i$ is associated to the $i$-th node of the extended Dynkin diagram $\mathfrak g^{(1)}$ of $G$ and the energy of each fermion is $a^{\vee}_i$. The fermion Hilbert space is graded by the energy of the states
\begin{equation}\label{eq:graded_H_F}
{\mathcal H}_\text{F}=\bigoplus_{n=0}^{h} {\mathcal H}^n_\text{F}\,,
\end{equation}
where the maximal energy is $h$ since $h=\sum_{i=0}^r a^\vee_i$. ${\mathcal H}^n_\text{F}$ denotes the subspace of energy $n$, that is, the configurations such that 
\begin{equation}\label{eq:energy_fermion}
\sum_{i=0}^r \lambda_i a^\vee_i=n\,,
\end{equation}
where $\lambda_i\in\{0,1\}$ is the occupation number of the $i$-th fermion. The Witten index for the $n$-domain wall~\eqref{eq:witten_index_def}, with the Goldstino multiplet contribution removed, is the trace over the fermion Hilbert space ${\mathcal H}^n_\text{F}$~\cite{Acharya:2001dz}
\begin{equation}
I_n=\tr_{\mathrm W_n}(-1)^F\equiv\tr_{{\mathcal H}^n_\text{F}}(-1)^F \,.
\end{equation}
The Witten index of all $n$-domain walls is encoded in the partition function of the fermions with periodic boundary conditions on a circle, corresponding to a sum over all states weighted by the energy:
\begin{equation} 
Z(q)=\tr_{{\mathcal H}_\text{F}}(-1)^Fq^H=\sum_{n=0}^{h}I_n q^n\,.
\end{equation}
This partition function is readily evaluated
\begin{equation}\label{eq:numberwalls}
Z(q)=\prod_{i=0}^{r}(1-q^{a^{\vee}_i})\,,
\end{equation}
which implies, in particular, that the Witten index for the $n$ and $h-n$ wall are the same ($I_n=(-1)^{r+1}I_{h-n}$) since the fermionic Hilbert space for the $n$ and $h-n$ walls are related by particle-hole symmetry. This beautifully reproduces the expectation that the $n$ domain wall and the $h-n$ domain wall (cf.~figure~\ref{fig:wallspic}) are related to each other by time-reversal!
 
A symmetry $\mathsf s\in \mathsf S$ of $4d$ ${\mathcal N}=1$ SYM acts in a simple way on the Wilson lines of the gauge theory.
 A Wilson line is labeled by a representation of $G$ with highest weight $\lambda\equiv\lambda_1\omega_1+\lambda_2\omega_2+\cdots+\lambda_r\omega_r$, where $\omega_i$ is the fundamental weight associated to the $i$-th node of the Dynkin diagram $\mathfrak g$. The Wilson line $W_i$ labeled by the fundamental weight $\omega_i$ transforms under $\mathsf c \in \mathsf C$ as
\begin{equation}
\mathsf c\colon W_i\mapsto \, W_{\mathsf c(i)}\,,
\end{equation}
where $\omega_{\mathsf c(i)}$ is the fundamental weight which is charge conjugate to $\omega_i$. An element 
$\mathsf g\in\Gamma$ acts by
\begin{equation}\label{eq:one_form_act_W}
\mathsf g\colon W_i\mapsto \alpha_{\mathsf g}(\omega_i) \, W_i\,,
\end{equation}
where $\alpha_{\mathsf g}(\omega_i) \in \Gamma^*$ is the charge of $\omega_i$ under the center $\Gamma$ of $G$. The action of a symmetry on the fundamental Wilson lines $W_i$ induces an action on the fermions $\psi_i$, which are labeled by a node in the extended Dynkin diagram $\mathfrak g^{(1)}$. We recall that $\mathsf C$ acts as an outer automorphism of $\mathfrak g$, $\mathsf S$ acts as an outer automorphism of $\mathfrak g^{(1)}$ and $\Gamma= \mathsf S/\mathsf C$.

We now proceed to compute the Witten index on the domain wall twisted by the symmetries of the system, $\mathsf S$. This group is identified with the group of symmetries of the extended Dynkin diagram $\mathfrak g^{(1)}$, i.e., a given $\mathsf s\in\mathsf S$ can be thought of as a permutation of the nodes $i\mapsto \mathsf s(i)$ that leaves the diagram $\mathfrak g^{(1)}$ invariant. The induced action on the effective $0+1$ system of fermions is
\begin{equation}
\mathsf s\colon\psi_i\mapsto\psi_{\mathsf s(i)}\,,
\end{equation}
where $\psi_i$ is the fermion associated to the $i$-th node of $\mathfrak g^{(1)}$. This means that the symmetry $\mathsf s$ lifts to a map $\mathcal H_\text{F}\to\mathcal H_\text{F}$ which, by definition, commutes with the Hamiltonian,
\begin{equation}
[H,\mathsf s]=0\,,
\end{equation}
inasmuch as $a_i^\vee\equiv a_{\mathsf s(i)}^\vee$. Thus, $\mathsf s$ restricts to a well-defined action on $\mathcal H^n_{\mathsf F}$, i.e., it preserves the grading~\eqref{eq:graded_H_F}. The twisted Witten index is
\begin{equation}\label{eq:def_twist_s_n}
I^{\mathsf s}_n=\tr_{\mathrm W_n}(-1)^F\mathsf s\equiv\tr_{{\mathcal H}^n_\text{F}}(-1)^F\mathsf s\,.
\end{equation}
Similarly, the twisted partition function computes the generating function of twisted indices:
\begin{equation}
Z^{\mathsf s}(q)=\tr_{{\mathcal H}_\text{F}}(-1)^F\mathsf s\,q^H =\sum_{n=0}^{h}I^{\mathsf s}_n q^n\,.
\end{equation}

An efficient way to compute this partition function is as follows. Take the $i$-th fermion, and consider its orbit under $\mathsf s$:
\begin{equation}\label{eq:actfermi}
\psi_i\mapsto \psi_{\mathsf s(i)}\mapsto \psi_{\mathsf s^2(i)}\mapsto\cdots\mapsto\psi_{\mathsf s^{N_i}(i)}\equiv\psi_i\,,
\end{equation}
where $N_i$ denotes the length of the orbit of the $i$-th node under the symmetry $\mathsf s$, i.e., the minimal integer such that $\mathsf s^{N_i}(i)\equiv i$. In the trace~\eqref{eq:def_twist_s_n}, the only configurations that contribute are those where the occupation number $\lambda_i$ in~\eqref{eq:energy_fermion} is constant along the orbit:
\begin{equation}
\lambda_i=\lambda_{\mathsf s(i)}=\lambda_{\mathsf s^2(i)}=\cdots=\lambda_{\mathsf s^{N_i-1}(i)}\,.
\end{equation}
This means that we may restrict the sum over $\mathcal H_\text{F}^n$ in~\eqref{eq:energy_fermion} to those configurations where this identity is satisfied. We enforce this by dropping all but one of these labels, and multiplying its energy by $N_i$, i.e., we replace~\eqref{eq:energy_fermion} by
\begin{equation}
\sum_{i=0}^{r'} \lambda_i a'_i{}^{\vee}=n\,,
\end{equation}
where the sum is over one representative for each orbit, $r'$ is the number of orbits of $\mathsf s$, and $a'_i{}^{\vee}=N_i a_i^{\vee}$ is the combined energy of all the elements of the orbit of $\lambda_i$. With this, the twisted Witten index~\eqref{eq:twistc_tr_def} on the domain wall can be computed as the untwisted partition function of $r'+1$ free fermions with energies $a'_i{}^{\vee}$:
\begin{equation} \label{eq:numberwallsC}
Z^\mathsf s(q)=\prod_{i=0}^{r'}(1-q^{a'_i{}^{\vee}})\,.
\end{equation}
Since $h=\sum_{i=0}^{r'} a'^\vee_i$ we see that $I^\mathsf s_n=(-1)^{r'+1}I^\mathsf s_{h-n}$, as required by time-reversal.

Diagrammatically, twisting by a symmetry folds the Dynkin diagram $\mathfrak g^{(1)}$ according to the action of $\mathsf s$ on the nodes~\cite{Fuchs:1995zr}. This yields a new affine Dynkin diagram, which has $r'+1<r+1$ nodes, and comarks $a'_i{}^{\vee}=N_ia_i^\vee$. %(Note that the normalization of the nodes after twisting depends on conventions. It is sometimes convenient, for example, to divide the new comarks by the length of the orbit, or the order of $\mathsf s$, cf.~ref??; we shall not do so here).
 The twisted Witten index is identical to the untwisted Witten index of the folded diagram. 

A quick remark is in order. Let $\lambda'_i$ be a node in the folded diagram, and let $N_i$ be the number of nodes in the original diagram that folded into $\lambda'_i$. The node $\lambda'_i$ is therefore a bound state of $N_i$ fermions, and thus has fermion parity $(-1)^F=(-1)^{N_i}$. Moreover, the symmetry $\mathsf s$ permutes these fermions, which generates an extra sign corresponding to the signature of the permutation. When the permutation is cyclic, which is the case relevant to this paper, the signature is just $N_i-1$. All in all, the contribution of $\lambda'_i$ to the twisted trace is $(-1)^F\mathsf s=(-1)^{(N_i+N_i-1)}=-1$. Therefore, in the folded diagram the node behaves as a regular fermion, just with more energy, and so~\eqref{eq:numberwallsC} is correct as written: the fermionic signs are all taken care of automatically by the folding.

The twisted Witten index can also be computed by diagonalizing the action of $\mathsf s$~\eqref{eq:actfermi} by a direct sum of unitary transformations, one for each orbit, which is a symmetry of the collection of fermions. In this basis, $\mathsf s$ acts with eigenvalue $\mathsf s_i$ on the $i$-th fermion, where $\mathsf s_i$ is an $N_i$-th root of unity. The twisted partition function can, therefore, also be expressed as
\begin{equation}\label{eq:numberwallsCa}
Z^\mathsf s(q)=\prod_{i=0}^{r}(1-\mathsf s_i \,q^{a^{\vee}_i})\,.
\end{equation}
This also makes the action of time-reversal symmetry on the domain walls manifest, cf.~$I^\mathsf s_n=(-1)^{r+1}\det(\mathsf s)I^\mathsf s_{h-n}$, where $\det(\mathsf s)=\mathsf s_0\mathsf s_1\cdots\mathsf s_r=(-1)^{r+r'}$ is the parity of the permutation induced by $\mathsf s$. %Needless to say, this alternative representation is only useful if we are able to determine the eigenvalues $\mathsf s_i$ efficiently. We shall see below that this is indeed the case. 
We now discuss zero-form symmetries and one-form symmetries in turn.

\paragraph{Zero-form symmetries.} $4d$ ${\mathcal N}=1$ SYM has charge conjugation symmetry $\mathsf C$ if and only if the (unextended) Dynkin diagram $\mathfrak g$ of the Lie algebra of $G$ has a symmetry. This corresponds to an outer automorphism of the Lie algebra of $G$ (see table~\ref{tab:comarksone}). Such symmetry is present for the $A_r,D_r,E_6$ algebras, where $\mathsf C$ acts as a transposition (order-two permutation) on the nodes of the Dynkin diagram (with low-rank exceptions $A_1,D_4$). In this case, folding the diagram by $\mathsf C$ gives rise to what is usually called the twisted affine Dynkin diagram $\mathfrak g^{(2)}$~\cite{fuchs1995affine,kac_1990}, which is constructed by identifying the nodes of $\mathfrak g$ that are permuted by $\mathsf C$ (and adding the extending node).

As $\mathsf C=\mathbb Z_2$, the eigenvalues in~\eqref{eq:numberwallsCa} are trivial to determine: if a node $i$ is fixed by $\mathsf c\in\mathsf C$, then its eigenvalue is $\mathsf c_i=+1$. On the other hand, if the pair of nodes $i,j$ are swapped, then the eigenvalues are $\pm1$, which can be assigned as $\mathsf c_i=+1$ and $\mathsf c_j=-1$ (or vice-versa).

\paragraph{One-form symmetries.} As discussed in section~\ref{sec:intro}, $4d$ $\mathcal N=1$ has a one-form symmetry group $\Gamma$ given by the center of the gauge group $G$. Here, the eigenvalues $\mathsf s_i$ in~\eqref{eq:numberwallsCa} have a very natural interpretation. An element $\mathsf g\in \Gamma$ acts as an outer automorphism of $\mathfrak g^{(1)}$, a permutation of the $r+1$ nodes. Diagonalizing this permutation results on an eigenvalue $\mathsf g_i$ on the fermion $\psi_i$ associated to the $i$-th node, which is the charge of the element of the center $\mathsf g\in \Gamma$ on the $i$-th fundamental weight of $\mathfrak g$~\cite{Schellekens:1990xy,OLIVE1983470,BERNARD1987628,Felder:1988sd,PhysRevD.41.2558}, that is 
 \begin{equation}
\mathsf g_i=\alpha_{\mathsf g}(\omega_i)\,.
\label{permueqcent}
\end{equation}
Note that this is precisely how $\mathsf g\in \Gamma$ acts on the ultraviolet Wilson loops $W_i$ (cf.~\eqref{eq:one_form_act_W}).

%The underlying reason for this is the following. The occupation numbers $\lambda_i$ not only define a representation of $\mathfrak g$, but they in fact define an integrable representation of the extended algebra $\mathfrak g^{(1)}$, at level $n$. Indeed, a representation $\lambda$ of $\mathfrak g$ lifts to a representation of $\mathfrak g^{(1)}$ if and only if $(\lambda,\theta)\le n$ which, when expanded in a basis of coroots, is precisely the condition~\eqref{eq:energy_fermion}. The group $\Gamma$ naturally acts on the Dynkin diagram of $\mathfrak g^{(1)}$. In the language of two-dimensional RCFTs, the map $i\mapsto \mathsf g(i)$ is what is usually referred to as the \emph{spectral flow} induced by $\mathsf g$. This permutation is generated, via fusion, by the simple current associated to the element $\mathsf g$. It is a well-known fact that the fusion rules are diagonalized by the $S$-matrix which, in our context, is the statement that the eigenvalue $\mathsf g_i$ is the monodromy phase of the element $\lambda_i$ with respect to the simple current associated to $\mathsf g$. This monodromy phase is indeed the charge of $\lambda_i$ under the center. The two perspectives~\eqref{eq:numberwallsC} and~\eqref{eq:numberwallsCa} corresponds, in this two-dimensional language, to inserting the symmetry line $\mathsf g$ along the $b$- and $a$-cycles of the two-torus, respectively. In the first case the symmetry acts via fusion, i.e., a permutation, and in the second case via monodromy, i.e., diagonally, through a phase.

We now proceed to compute the twisted Witten indices for $G=\SU(N)$, $\Sp(N)$, $\Spin(N)$, and $G_2$ respectively.

\subsection[$G=\SU(N)$]{$\boldsymbol{G=\SU(N)}$}

Consider the algebra $A_{N-1}=\mathfrak{su}_N$. The symmetries of this algebra are as follows:
\begin{itemize}
\item The group $\SU(N)$ has a $\mathbb Z_2$ zero-form symmetry, which corresponds to complex conjugation. It acts by interchanging the $i$-th node with the $(N-i)$-th node in $\mathfrak g$. The associated diagonal action can be chosen as follows: take $\mathsf c_0=+1$ for the extended node, and $\mathsf c_i=+1$ for the first half of the unextended nodes, and $\mathsf c_i=-1$ for the second half.

\item The group $\SU(N)$ has a $\mathbb Z_N$ one-form symmetry, whose associated charge is the $N$-ality (the number of boxes in the Young diagram modulo $N$). If $\mathsf g$ denotes a primitive root of unity, then a generic element $\mathsf g^t\in \mathbb Z_N$ acts on the extended diagram $\mathfrak g^{(1)}$ as a cyclic permutation by $t$ units, $\lambda_i\mapsto \lambda_{i+t\mod N}$. The center acts on a representation with highest weight $\lambda$ as follows:
\begin{equation}\label{eq:twist_SU}
\alpha_{\mathsf g}(\lambda)\equiv \prod_{i=0}^{N-1}\mathsf g^{i \lambda_i}\,,
\end{equation}
which means that the eigenvalues in~\eqref{eq:numberwallsCa} are $\mathsf g_i=\mathsf g^i$. %\footnote{This depends on a choice of generator. The center of $G$, and the group of symmetries of $\mathfrak g^{(1)}$, are isomorphic, although the isomorphism is not canonical.}
 
%and so $\alpha^i_{\mathsf g^k}=e^{2\pi i k i/N}$. The folded diagram is that of $\SU(\gcd(N,\mathsf z))$ (see~\eqref{eq:fold_SU_mu}), with comarks $N/\gcd(N,\mathsf z)$. Note that the total energy is conserved: $h\equiv 1\times N=N/\gcd(N,\mathsf z)\times \gcd(N,\mathsf z)$.

\end{itemize}

Let us begin by computing the untwisted partition function. The comarks are all $a_i^\vee=1$. Plugging this into equation~\eqref{eq:numberwalls} we obtain the untwisted partition function 
\begin{equation}
Z(q)=(1-q)^N\,, 
\end{equation}
and, expanding, the Witten index 
\begin{equation}\label{eq:I_SU_UV}
I_n=(-1)^n\binom{N}{n}\,.
\end{equation}
This result was also obtained, by an entirely different method, in~\cite{Ritz:2002fm}.

We now move on to the twisted indices. Charge conjugation acts on the extended Dynkin diagram $A_{N-1}^{(1)}$ as follows:
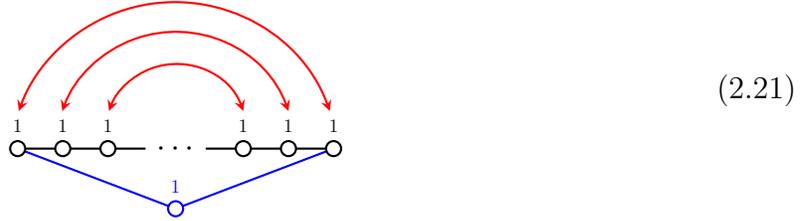
\begin{equation}
\begin{tikzpicture}[baseline=(current bounding box.center)]
%\node at (-1,3.1) {$A_{N-1}^{{\color{blue}(1)}}\colon$};
%\draw[thick,blue] (0,3) circle (.1cm) node[anchor=south,inner ysep=10pt,scale=.6] {$1$};
%\draw[thick,blue] (.1,3) -- (.5,3);
\draw[thick,blue] (.6,3) -- (2.7,2.2) -- (0+.6*8,3);
\draw[thick,blue,fill=white] (2.7,2.2) circle (.1cm) node[anchor=south,inner ysep=10pt,scale=.6] {$1$};

\foreach \i in {1,2,3,6,7}{
\draw[thick,fill=white] (0+.6*\i,3) circle (.1cm) node[anchor=south,inner ysep=10pt,scale=.6] {$1$};
\draw[thick] (.1+.6*\i,3) -- (.5+.6*\i,3);
}
\draw[thick,fill=white] (0+.6*8,3) circle (.1cm) node[anchor=south,inner ysep=10pt,scale=.6] {$1$};
\draw[thick] (.1+.6*5,3) -- (.5+.6*5,3);
\node at (2.73,3) {$\cdots$};

\draw[<->,thick,red,>=stealth] (4.75,3.5) arc[radius=2.2, start angle=20, end angle=160];
\draw[<->,thick,red,>=stealth] (4.2,3.5) arc[radius=1.6, start angle=20, end angle=160];
\draw[<->,thick,red,>=stealth] (3.6,3.5) arc[radius=.95, start angle=20, end angle=160];
\end{tikzpicture}
\end{equation}

\noindent where the blue node denotes the affine root, and the integers denote the comarks $a_i^\vee$. The automorphism folds the diagram in half.\footnote{For $N$ odd, the $(N\pm1)/2$-th nodes would naively fold into a loop, which does not yield a valid Dynkin diagram. The correct folding is given by the theory of twisted Ka\v{c}-Moody algebras~\cite{fuchs1995affine,kac_1990}. We henceforth fold
the diagrams following~\cite{fuchs1995affine,kac_1990}. The only information we need from the diagram are the comarks. \label{ftn:su_n_fold}} The result is
\begin{equation}\label{eq:fold_C_SU}
\begin{tikzpicture}[baseline=(current bounding box.center)]
\node at (-1.5,2.1) {$A_{2m-1}^{{\color{blue}(1)}}\mapsto A_{2m-1}^{{\color{blue}(2)}}\colon$};

%\foreach \i in {1,2}{
%\draw[thick] ({0+.6*(\i+1)},2) circle (.1cm) node[anchor=south,inner ysep=10pt,scale=.6] {$2$};
%\draw[thick] ({.1+.6*(\i+1)},2) -- ({.5+.6*(\i+1)},2);
%}
%\draw[thick] (3.5,2) circle (.1cm) node[anchor=south,inner ysep=10pt,scale=.6] {$2$};
%\draw[thick,blue] (4.2,2) circle (.1cm) node[anchor=south,inner ysep=10pt,scale=.6] {$1$};
%
%\node at (2.8,2) {$\cdots$};
%\draw[thick] (3.3,2) -- (3.4,2);
%
%\draw[thick] (0+.6,2) circle (.1cm) node[anchor=south,inner ysep=10pt,scale=.6] {$1$};
%\draw[thick] (1,2.2) -- (0.8,2) -- (1,1.8);
%\draw[thick] (1.2,1.9) -- (.6,1.9);
%\draw[thick] (1.2,2.1) -- (.6,2.1);
%
%\draw[thick,blue] (3.5,1.9) -- (4.2,1.9);
%\draw[thick,blue] (3.5,2.1) -- (4.2,2.1);
%\draw[thick,blue] (3.75,2.2) -- (3.95,2) -- (3.75,1.8);

\begin{scope}[xscale=-1,shift={(-5,0)}]
\begin{scope}[shift={(-1,0)}]
\foreach \i in {1,2}{
\draw[thick] ({0+.6*(\i+1)},2) circle (.1cm) node[anchor=south,inner ysep=10pt,scale=.6] {$2$};
\draw[thick] ({.1+.6*(\i+1)},2) -- ({.5+.6*(\i+1)},2);
}
\node at (2.8,2) {$\cdots$};
\draw[thick] (0+.6,2) circle (.1cm) node[anchor=south,inner ysep=10pt,scale=.6] {$1$};
\draw[thick] (0.8,2.2) -- (1,2) -- (0.8,1.8);
\draw[thick] (1.2,1.9) -- (.6,1.9);
\draw[thick] (1.2,2.1) -- (.6,2.1);
\end{scope}

\draw[thick] (3.5,2) circle (.1cm) node[anchor=south,inner ysep=10pt,scale=.6] {$2$};
\draw[thick,blue] (4.2,2) circle (.1cm) node[anchor=south,inner ysep=10pt,scale=.6] {$1$};

\draw[thick,fill=white] (3.5-.6,2) circle (.1cm) node[anchor=south,inner ysep=10pt,scale=.6] {$2$};
\draw[thick] (3.6-.6,2) -- (4.2-.8,2);
\draw[thick] (3.6-1.2,2) -- (4.2-1.4,2);

\draw[thick,blue] (3.5,1.9) -- (4.2,1.9);
\draw[thick,blue] (3.5,2.1) -- (4.2,2.1);
\draw[thick,blue] (3.95,2.2) -- (3.75,2) -- (3.95,1.8);

\end{scope}

\begin{scope}[shift={(0,-1.5)}]
\node at (-1.5,2.1) {$A_{2m\hphantom{-1}}^{{\color{blue}(1)}}\mapsto A_{2m\hphantom{-1}}^{{\color{blue}(2)}}\colon$};
\end{scope}

\begin{scope}[xscale=-1,shift={(-5,-1.5)}]
\begin{scope}[shift={(-1,0)}]
\foreach \i in {1,2}{
\draw[thick] ({0+.6*(\i+1)},2) circle (.1cm) node[anchor=south,inner ysep=10pt,scale=.6] {$2$};
\draw[thick] ({.1+.6*(\i+1)},2) -- ({.5+.6*(\i+1)},2);
}
\node at (2.8,2) {$\cdots$};
\draw[thick] (0+.6,2) circle (.1cm) node[anchor=south,inner ysep=10pt,scale=.6] {$2$};
\draw[thick] (1,2.2) -- (0.8,2) -- (1,1.8);
\draw[thick] (1.2,1.9) -- (.6,1.9);
\draw[thick] (1.2,2.1) -- (.6,2.1);
\end{scope}

\draw[thick] (3.5,2) circle (.1cm) node[anchor=south,inner ysep=10pt,scale=.6] {$2$};
\draw[thick,blue] (4.2,2) circle (.1cm) node[anchor=south,inner ysep=10pt,scale=.6] {$1$};

\draw[thick,fill=white] (3.5-.6,2) circle (.1cm) node[anchor=south,inner ysep=10pt,scale=.6] {$2$};
\draw[thick] (3.6-.6,2) -- (4.2-.8,2);
\draw[thick] (3.6-1.2,2) -- (4.2-1.4,2);

\draw[thick,blue] (3.5,1.9) -- (4.2,1.9);
\draw[thick,blue] (3.5,2.1) -- (4.2,2.1);
\draw[thick,blue] (3.95,2.2) -- (3.75,2) -- (3.95,1.8);

\end{scope}

\end{tikzpicture}
\end{equation}
where the folded diagrams both have $m+1=\lfloor N/2\rfloor +1$ nodes. From this we conclude that the folded diagram has $r'+1=(N+1)/2$ and $r'=(N+2)/2$ nodes for $N$ odd and $N$ even, respectively. In the first case, one node has comark equal to $1$, and the rest equal to $2$; while in the second case, there are two nodes with comark $1$, and the rest equal to $2$. Using~\eqref{eq:numberwallsC}, the $\mathsf c$-twisted partition function is\footnote{In $\SU(2)$ the action of $\mathsf c$ is a gauge transformation and $\mathsf c$ is not a symmetry; indeed, $Z^{\mathsf c}(q)=Z(q)$.}
\begin{equation}\label{eq:Z_UV_C}
\begin{aligned}
Z^{\mathsf c}(q)&=\begin{cases}
(1-q)(1-q^2)^{(N-1)/2} \qquad\ N\text{ odd,}\\[+8pt] 
(1-q)^2(1-q^2)^{(N-2)/2} \qquad N\text{ even,}
\end{cases} 
\end{aligned}
\end{equation}
and, expanding, the $\mathsf c$-twisted Witten indices are
\begin{equation}\label{eq:I_UV_C}
\begin{aligned}
I_n^{\mathsf c}&=\begin{cases}
\displaystyle(-1)^{n/2} \binom{(N-1)/2}{n/2}&\text{$N$ odd, $n$ even,}\\[+15pt] 
\displaystyle(-1)^{(n+1)/2} \binom{(N-1)/2}{(n-1)/2}&\text{$N$ odd, $n$ odd,}\\[+15pt] 
\displaystyle(-1)^{n/2}\bigg[ \binom{(N-2)/2}{n/2}-\binom{(N-2)/2}{n/2-1}\bigg]&\text{$N$ even, $n$ even,}\\[+15pt] 
\displaystyle2 (-1)^{(n+1)/2} \binom{(N-2)/2}{(n-1)/2}&\text{$N$ even, $n$ odd.}
\end{cases}
\end{aligned}
\end{equation}
One can also compute the partition function in the diagonal basis, where $\mathsf c_i=+1$ for the first half of the nodes, and $\mathsf c_i=-1$ for the second half. Plugging this into~\eqref{eq:numberwallsCa} yields the same expression for the twisted partition function.

Let us now consider the partition function twisted by the $\Gamma=\mathbb Z_N$ one-form symmetry. If $\mathsf g$ denotes a primitive $N$-th root of unity, then a generic element $\mathsf g^t\in\mathbb Z_N$ acts on the extended diagram as follows:
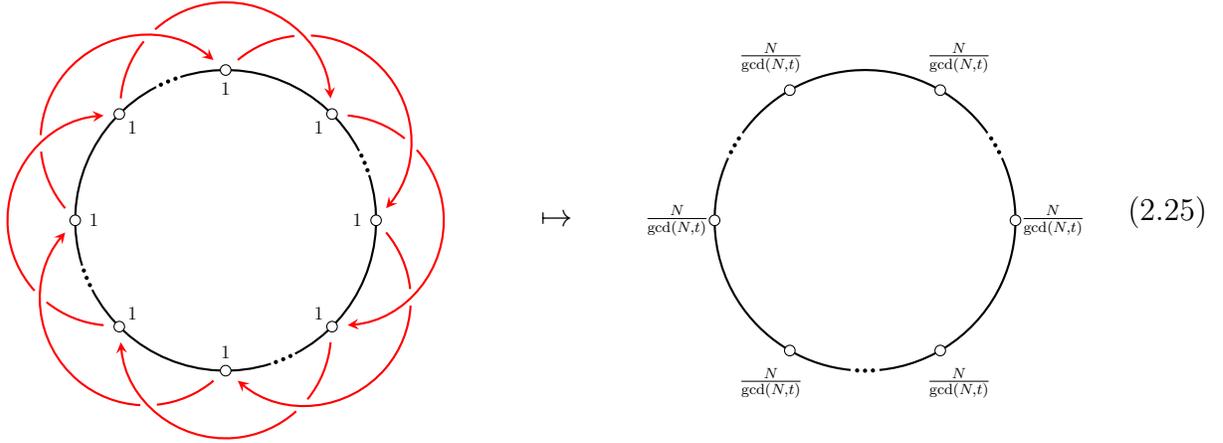
\begin{equation}\label{eq:fold_SU_mu}
\begin{tikzpicture}[baseline=(current bounding box.center)]

\foreach \i in {0,...,3}
{\draw[thick] ([shift=({(\i-.5)*90+5-.5*45}:2cm)]4,4) arc ((\i-.5)*90+5-.5*45:-5+45+\i*90-22.5:2cm);
\foreach \j in {-3,0,3}
\fill ({4+2*cos(90*(\i+.25)+\j)},{4+2*sin(90*(\i+.25)+\j)}) circle (.8pt);}
\foreach \i in {0,...,7}
{\fill[white,draw=black] ({4+2*cos(45*\i)},{4+2*sin(45*\i)}) circle (2pt);
\node[scale=.6] at ({4+1.75*cos(45*\i)},{4+1.75*sin(45*\i)}) {$1$};%\draw[thick,<-,>=stealth,red] ([shift=(-90+5+\i*45:1.4cm)]{{4+1.5*cos(45*\i)},{4+1.5*sin(45*\i)}}) arc (-90+5+\i*45:90-5+\i*45:1.4cm);
\draw[thick,<-,>=stealth,red] ([shift=(-90+5+\i*45:1.4cm)]{{4+1.5*cos(45*\i)},{4+1.5*sin(45*\i)}}) arc (-90+5+\i*45:90-5+\i*45-43:1.4cm);
\draw[thick,red] ([shift=(-90+5+\i*45-10+125+21:1.4cm)]{{4+1.5*cos(45*\i)},{4+1.5*sin(45*\i)}}) arc (-90+5+\i*45-10+125+21:90-5+\i*45:1.4cm);
}

\node at (8.4,4) {$\mapsto$};

\begin{scope}[shift={(8.5,0)}]

\foreach \i in {0,...,2}
{\draw[thick] ([shift=({(\i-.5)*120-2-.5*45}:2cm)]4,4) arc ((\i-.5)*120-2-.5*45:2+45+\i*120-22.5:2cm);
\foreach \j in {-3,0,3}
\fill ({4+2*cos(120*(\i+.25)+\j)},{4+2*sin(120*(\i+.25)+\j)}) circle (.8pt);}

\foreach \i in {0,...,5}
{\fill[white,draw=black] ({4+2*cos(60*\i)},{4+2*sin(60*\i)}) circle (2pt);
\node[scale=.7] at ({4+2.5*cos(60*\i)},{4+2.5*sin(60*\i)}) {$\frac{N}{\gcd(N,t)}$};}

\end{scope}
\end{tikzpicture}
\end{equation}
 where the folded diagram has $\gcd(N,t)$ nodes, each with energy $N/\gcd(N,t)$. In other words, $\mathsf g^t$ folds the diagram into the affine diagram of $\SU(\gcd(N,t))$, with comarks $N/\gcd(N,t)$. This immediately yields the twisted partition function as~\eqref{eq:numberwallsC}
\begin{equation}\label{eq:one_form_twist_SU}
Z^{\mathsf g^t}(q)=(1-q^{N/\gcd(N,t)})^{\gcd(N,t)}\,.
\end{equation}
The twisted index reads
\begin{equation}\label{eq:q_gcd_binom_UV}
I_n^{\mathsf g^t}= \begin{cases}
\displaystyle(-1)^{n\gcd(N,t)/N}\binom{\gcd(N,t)}{n\gcd(N,t)/N} & N|n\gcd(N,t),\\[+15pt] 
\displaystyle0&\text{otherwise.}
\end{cases}
\end{equation}
Naturally, for $t=0$ this reduces to the untwisted result.

Alternatively, we may compute the same partition function in the diagonal basis. Using equations~\eqref{eq:numberwallsCa} and~\eqref{eq:twist_SU}, the twisted partition function is given by
\begin{equation}
Z^{\mathsf g}(q)=\prod_{i=0}^{N-1}(1- \mathsf g^iq)\equiv (q;\mathsf g)_N\,,
\end{equation}
the so-called $q$-Pochhammer symbol, essentially defined by this product. One may prove that this is in fact identical to~\eqref{eq:one_form_twist_SU}. Expanding the product, the twisted index becomes
\begin{equation}\label{eq:q_binom_UV}
I_n^{\mathsf g}= (-1)^n \mathsf g^{\frac12n(n-1)} \binom{N}{n}_{\!\!\mathsf g},\qquad \binom{N}{n}_{\!\!\mathsf g}:=\frac{(\mathsf g;\mathsf g)_N}{(\mathsf g ;\mathsf g)_n(\mathsf g ;\mathsf g)_{N-n}}\,,
\end{equation}
where the term in parentheses denotes the so-called $q$-binomial coefficient. This is again identical to~\eqref{eq:q_gcd_binom_UV}.

\subsection[$G=\Sp(N)$]{$\boldsymbol{G=\Sp(N)}$}

Consider the algebra $C_N=\mathfrak{sp}_N$. The symmetries of this algebra are as follows:
\begin{itemize}
\item The group $\Sp(N)$ has no zero-form symmetry.

\item The group $\Sp(N)$ has a $\mathbb Z_2$ one-form symmetry, whose charged representations are the pseudo-real ones. The non-trivial element $\mathsf g\in\mathbb Z_2$ acts on the extended diagram by reversing the nodes $\lambda_i\mapsto\lambda_{N-i}$. The center acts on a representation $\lambda$ as follows:
\begin{equation}\label{eq:twist_C_series}
\alpha_{\mathsf g}(\lambda)\equiv (-1)^{\sum_{i=0}^{\lfloor (N-1)/2\rfloor}\lambda_{2i+1}}\,,
\end{equation}
which means that the eigenvalues in~\eqref{eq:numberwallsCa} are $\mathsf g_i=(-1)^i$.% (or, more succinctly, $\mathsf z_i=(-1)^i$). %Note that this is just the mod $2$ reduction of the action of $\mathbb Z_{2N}$ in $\Sp(N)=\SU(2N)\cap \Sp(N;\mathbb C)$.
%The folded diagram associated to $\mathsf z=1$ is that of $\Sp(\lfloor N/2\rfloor)$, but with its comarks doubled, except for the last one if $N$ is even, which stays at $1$ (corresponding to the middle node, which is fixed; see~\eqref{eq_fig:fold_C}). Note that the total energy is conserved: $h\equiv (N+1)\times 1=(\lfloor N/2\rfloor+1-x)\times2+x$, where $x=1$ if $N$ is even, and $x=0$ if odd.

\end{itemize}

Let us begin by computing the untwisted partition function. The comarks for $\Sp(N)$ are all equal to one, i.e. $a_i^\vee=1$ for $i=0,1,\dots,N$. Plugging this into equation~\eqref{eq:numberwalls} we obtain the untwisted partition function 
\begin{equation}
Z(q)=(1-q)^{N+1}\,, 
\end{equation}
and, expanding, the Witten index 
\begin{equation}\label{eq:I_Sp_UV}
I_n=(-1)^n\binom{N+1}{n}\,.
\end{equation}

$4d$ $\Sp(N)$ ${\mathcal N}=1$ SYM has no charge conjugation symmetry. We can consider instead the index twisted by the $\Gamma=\mathbb Z_2$ one-form center symmetry, which acts on the extended Dynkin diagram as follows:
\begin{equation}\label{eq_fig:fold_C}
\begin{tikzpicture}[baseline=(current bounding box.center)]
%\draw[thick,blue] (0,3) circle (.1cm) node[anchor=south,inner ysep=10pt,scale=.6] {$1$};
%\draw[thick,blue] (.1,3) -- (.5,3);
\draw[thick,blue] (0+.6*1,3) circle (.1cm) node[anchor=south,inner ysep=10pt,scale=.6] {$1$};
%\draw[thick,blue] (.1+.6*1,3) -- (.5+.6*1,3);
\draw[thick,blue] (.6*1,2.9) -- (.6+.6*1,2.9);
\draw[thick,blue] (.6*1,3.1) -- (.6+.6*1,3.1);
\foreach \i in {2,3,6}{
\draw[thick] (0+.6*\i,3) circle (.1cm) node[anchor=south,inner ysep=10pt,scale=.6] {$1$};
\draw[thick] (.1+.6*\i,3) -- (.5+.6*\i,3);
}
\draw[thick] (0+.6*7,3) circle (.1cm) node[anchor=south,inner ysep=10pt,scale=.6] {$1$};
\draw[thick] (0+.6*8,3) circle (.1cm) node[anchor=south,inner ysep=10pt,scale=.6] {$1$};
\draw[thick] (.6*7,3.1) -- (.6+.6*7,3.1);
\draw[thick] (.6*7,2.9) -- (.6+.6*7,2.9);
\draw[thick] (.1+.6*5,3) -- (.5+.6*5,3);
\node at (2.73,3) {$\cdots$};

\draw[<->,thick,red,>=stealth] (4.75,3.5) arc[radius=2.2, start angle=20, end angle=160];
\draw[<->,thick,red,>=stealth] (4.2,3.5) arc[radius=1.6, start angle=20, end angle=160];
\draw[<->,thick,red,>=stealth] (3.6,3.5) arc[radius=.95, start angle=20, end angle=160];

\draw[thick] (3.8+.6,2.2+1) -- (4+.6,2+1) -- (3.8+.6,1.8+1);
\draw[thick,blue] (3.8+.6-3.4,2.2+1) -- (4+.6-.4-3.4,2+1) -- (3.8+.6-3.4,1.8+1);

\end{tikzpicture}
\end{equation}

\noindent where the blue node denotes the affine root, and the integers denote the comarks $a_i^\vee$. The automorphism folds the diagram in half (see footnote~\ref{ftn:su_n_fold}).
%\footnote{As in the case of $\SU(N)$, the folding is ambiguous for odd $N$, inasmuch as the middlemost nodes fold into a loop (see footnote~\ref{ftn:su_n_fold}). For our purposes, the value of the comarks is the only important information, and the details of the folding are irrelevant.}. 
The result is
\begin{equation}
\begin{tikzpicture}[baseline=(current bounding box.center)]
\node at (-1.5,2.1) {$C_{2m\hphantom{-1}}^{{\color{blue}(1)}}\mapsto A_{2m}^{{\color{blue}(2)}}\ \colon$};

\begin{scope}[xscale=-1,shift={(-4.8,0)}]
\begin{scope}[shift={(-1,0)}]
\foreach \i in {1,2}{
\draw[thick] ({0+.6*(\i+1)},2) circle (.1cm) node[anchor=south,inner ysep=10pt,scale=.6] {$2$};
\draw[thick] ({.1+.6*(\i+1)},2) -- ({.5+.6*(\i+1)},2);
}
\node at (2.8,2) {$\cdots$};
\draw[thick] (0+.6,2) circle (.1cm) node[anchor=south,inner ysep=10pt,scale=.6] {$1$};
\draw[thick] (0.8,2.2) -- (1,2) -- (0.8,1.8);
\draw[thick] (1.2,1.9) -- (.6,1.9);
\draw[thick] (1.2,2.1) -- (.6,2.1);
\end{scope}

\draw[thick] (3.5,2) circle (.1cm) node[anchor=south,inner ysep=10pt,scale=.6] {$2$};
\draw[thick,blue] (4.2,2) circle (.1cm) node[anchor=south,inner ysep=10pt,scale=.6] {$2$};

\draw[thick,fill=white] (3.5-.6,2) circle (.1cm) node[anchor=south,inner ysep=10pt,scale=.6] {$2$};
\draw[thick] (3.6-.6,2) -- (4.2-.8,2);
\draw[thick] (3.6-1.2,2) -- (4.2-1.4,2);

\draw[thick,blue] (3.5,1.9) -- (4.2,1.9);
\draw[thick,blue] (3.5,2.1) -- (4.2,2.1);
\draw[thick,blue] (3.75,2.2) -- (3.95,2) -- (3.75,1.8);
\end{scope}

\begin{scope}[shift={(0,-1.5)}]
\node at (-1.5,2.1) {$C_{2m-1}^{{\color{blue}(1)}}\mapsto C_{m-1}^{{\color{blue}(1)}}\colon$};
\begin{scope}[xscale=-1,shift={(-4.8,0)}]
\begin{scope}[shift={(-1,0)}]
\foreach \i in {1,2}{
\draw[thick] ({0+.6*(\i+1)},2) circle (.1cm) node[anchor=south,inner ysep=10pt,scale=.6] {$2$};
\draw[thick] ({.1+.6*(\i+1)},2) -- ({.5+.6*(\i+1)},2);
}
\node at (2.8,2) {$\cdots$};
\draw[thick] (0+.6,2) circle (.1cm) node[anchor=south,inner ysep=10pt,scale=.6] {$2$};
\draw[thick] (1,2.2) -- (0.8,2) -- (1,1.8);
\draw[thick] (1.2,1.9) -- (.6,1.9);
\draw[thick] (1.2,2.1) -- (.6,2.1);
\end{scope}

\draw[thick] (3.5,2) circle (.1cm) node[anchor=south,inner ysep=10pt,scale=.6] {$2$};
\draw[thick,blue] (4.2,2) circle (.1cm) node[anchor=south,inner ysep=10pt,scale=.6] {$2$};

\draw[thick,fill=white] (3.5-.6,2) circle (.1cm) node[anchor=south,inner ysep=10pt,scale=.6] {$2$};
\draw[thick] (3.6-.6,2) -- (4.2-.8,2);
\draw[thick] (3.6-1.2,2) -- (4.2-1.4,2);

\draw[thick,blue] (3.5,1.9) -- (4.2,1.9);
\draw[thick,blue] (3.5,2.1) -- (4.2,2.1);
\draw[thick,blue] (3.75,2.2) -- (3.95,2) -- (3.75,1.8);
\end{scope}

\end{scope}

\end{tikzpicture}
\end{equation}
where the folded diagrams have $m+1=\lfloor N/2\rfloor+1$ nodes.

From this we learn that the folded diagram has $r'+1=(N+2)/2$ and $r'+1=(N+1)/2$ nodes, for $N$ even and $N$ odd, respectively. In the first case, one of these nodes has energy equal to $1$, and the rest equal to $2$; while in the second case, they are all of energy $2$. Plugging this into~\eqref{eq:numberwallsC} the one-form twisted partition function is
\begin{equation}
 Z^{{\mathsf g}}(q)=\begin{cases}
 (1-q)(1-q^2)^{N/2}&\text{$N$ even,}\\[+4pt]
(1-q^2)^{(N+1)/2}&\text{$N$ odd,} 
\end{cases}
\end{equation}
and, expanding, the twisted Witten index 
\begin{equation}\label{eq:I_Sp_tw_UV}
I^{\mathsf g}_n=\begin{cases}
\,\displaystyle(-1)^{(n+1)/2} \binom{N/2}{(n-1)/2} &\text{$N$ even, $n$ odd,}\\[+15pt] 
\,\displaystyle(-1)^{n/2} \binom{N/2}{n/2} &\text{$N$ even, $n$ even,}\\[+15pt] 
\,\displaystyle0 &\text{$N$ odd, $n$ odd,}\\[+15pt] 
\,\displaystyle(-1)^{n/2} \binom{(N+1)/2}{n/2} &\text{$N$ odd, $n$ even.}
\end{cases}
\end{equation}

One can also compute the partition function in the diagonal basis, where $\mathsf g_i=+1$ for the even nodes, and $\mathsf g_i=-1$ for the odd ones. Plugging this into~\eqref{eq:numberwallsCa} yields the same expression for the twisted partition function.

\subsection[$G=\Spin(2N+1)$]{$\boldsymbol{G=\Spin(2N+1)}$}

Consider the algebra $B_N=\mathfrak{so}_{2N+1}$. The symmetries of this algebra are as follows:
\begin{itemize}
\item The group $\Spin(2N+1)$ has no zero-form symmetry.

\item The group $\Spin(2N+1)$ has a $\mathbb Z_2$ one-form symmetry, whose charged representations are the spinors. The non-trivial element $\mathsf g\in\mathbb Z_2$ acts on the extended diagram by permuting the zeroth and first nodes, $\lambda_0\leftrightarrow\lambda_1$. The center acts on a representation $\lambda$ as follows:
\begin{equation}\label{eq:twist_B_series}
\alpha_{\mathsf g}(\lambda)\equiv (-1)^{\lambda_N}
\end{equation}
which means that the eigenvalues in~\eqref{eq:numberwallsCa} are $\mathsf g_i=(-1)^{\delta_{i,N}}$. %The folded diagram associated to $\mathsf z=1$ is that of $\Sp(N-1)$, but with comarks with value $2$ instead of $1$, except for the last one, which stays at $1$ (see~\eqref{eq:fold_B_mu}). Note that the total energy is conserved: $h\equiv 1\times3+(N-2)\times 2=1\times1+(N-1)\times2$.

\end{itemize}

Let us begin by computing the untwisted partition function. The comarks for $\Spin(2N+1)$ are $a_i^\vee=1$ for $i=0,1,N$, and $a_i^\vee=2$ for $i=2,3,\dots,N-1$. Plugging this into equation \eqref{eq:numberwalls} we obtain the untwisted partition function 
\begin{equation}
Z(q)=(1-q)^3(1-q^2)^{N-2}\,,
\end{equation}
and, expanding, the Witten index 
\begin{equation}\label{eq:I_UV_Spin_odd}
%\begin{aligned}
%I_{2n}&=(-1)^n\bigg[\binom{N-2}{n}-3\binom{N-2}{n-1}\bigg]\\[+8pt]
%I_{2n+1}&=(-1)^n\bigg[\binom{N-2}{n-1}-3\binom{N-2}{n}\bigg]\,.
%\end{aligned}
I_n=\begin{cases}
\displaystyle(-1)^{n/2}\bigg[\binom{N-2}{n/2}-3\binom{N-2}{n/2-1}\bigg]&\text{$n$ even,}\\[+15pt]
\displaystyle(-1)^{(n-1)/2}\bigg[\binom{N-2}{(n-1)/2-1}-3\binom{N-2}{(n-1)/2}\bigg]&\text{$n$ odd.}
\end{cases}
\end{equation}
Note that the index vanishes for $N=1\mod4$ and $n=(N-1)/2$ and by time-reversal for $n'=h-n=(3N-1)/2$. This clearly illustrates the crucial difference between the dimension of the Hilbert space and the index.

$4d$ $\Spin(2N+1)$ ${\mathcal N}=1$ SYM has no charge conjugation symmetry. We can consider instead the index twisted by the $\Gamma=\mathbb Z_2$ one-form center symmetry. The non-trivial element $\mathsf g\in\mathbb Z_2$ acts on the extended Dynkin diagram as follows:
\begin{equation}\label{eq:fold_B_mu}
\begin{tikzpicture}[baseline=1.9cm]

\node at (-1.5,2.1) {$B_N^{{\color{blue}(1)}}\colon$};
\begin{scope}[yscale=1,xscale=-1,shift={(-4.2,0)}]
\draw[thick] (0+.6*0,2) circle (.1cm) node[anchor=south,inner ysep=10pt,scale=.6] {$1$};
\draw[thick] (0,2.1) -- (.6*1,2.1);
\draw[thick] (0,1.9) -- (.6*1,1.9);
\draw[thick] (.2,2.2) -- (.4,2) -- (.2,1.8);
\draw[thick,blue] (3.67,2.05) -- (4.12,2.25);
\draw[thick] (3.67,1.95) -- (4.12,1.75);

\draw[thick] (3.6,2) circle (.1cm) node[anchor=south,inner ysep=10pt,scale=.6] {$2$};
\draw[thick,blue] (4.2,2.3) circle (.1cm) node[anchor=east,inner xsep=10pt,scale=.6] {$1$};
\draw[thick] (4.2,1.7) circle (.1cm) node[anchor=east,inner xsep=10pt,scale=.6] {$1$};

\foreach \i in {1,2,3}{
\draw[thick] (0+.6*\i,2) circle (.1cm) node[anchor=south,inner ysep=10pt,scale=.6] {$2$};
\draw[thick] (.1+.6*\i,2) -- (.5+.6*\i,2);
}
\draw[thick] (3.2,2) -- (3.5,2);
\node at (2.73,2) {$\cdots$};

\draw[<->,thick,red,>=stealth] (4.75,2.35) arc[radius=.35, start angle=70, end angle=-70];
\end{scope}

\node at (5.5,2) {$\mapsto$};

\begin{scope}[shift={(7.5,0)}]
\node at (-.4,2.1) {$ A_{2N}^{{\color{blue}(2)}}\colon$};
\foreach \i in {1,2}{
\draw[thick] ({0+.6*(\i+1)},2) circle (.1cm) node[anchor=south,inner ysep=10pt,scale=.6] {$2$};
\draw[thick] ({.1+.6*(\i+1)},2) -- ({.5+.6*(\i+1)},2);
}
\draw[thick] (3.5,2) circle (.1cm) node[anchor=south,inner ysep=10pt,scale=.6] {$2$};
\draw[thick] (4.2,2) circle (.1cm) node[anchor=south,inner ysep=10pt,scale=.6] {$1$};

\node at (2.8,2) {$\cdots$};
\draw[thick] (3.3,2) -- (3.4,2);

\draw[thick,blue] (0+.6,2) circle (.1cm) node[anchor=south,inner ysep=10pt,scale=.6] {$2$};
\draw[thick,blue] (1,2.2) -- (0.8,2) -- (1,1.8);
\draw[thick,blue] (1.2,1.9) -- (.6,1.9);
\draw[thick,blue] (1.2,2.1) -- (.6,2.1);

\draw[thick] (3.5,1.9) -- (4.2,1.9);
\draw[thick] (3.5,2.1) -- (4.2,2.1);
\draw[thick] (3.95,2.2) -- (3.75,2) -- (3.95,1.8);

\end{scope}

\end{tikzpicture}
\end{equation}

\noindent where the blue node denotes the affine root, and the integers denote the comarks $a_i^\vee$.

From this we learn that the folded diagram has $r'+1=N$ nodes, one of which has energy equal to $1$, and the rest all energy equal to $2$. Plugging this into~\eqref{eq:numberwallsC} the one-form twisted partition function is
\begin{equation}
Z^{\mathsf g}(q)= (1-q)(1-q^2)^{N-1}\,,
\end{equation}
and, expanding, the Witten index 
\begin{equation}\label{eq:I_UV_Spin_odd_g}
%\begin{aligned}
%I^{\mathsf g}_{2n}&=(-1)^n \binom{N-1}{n} \\[+8pt]
%I^{\mathsf g}_{2n+1}&=(-1)^{n+1} \binom{N-1}{n}\,.
%\end{aligned}
I^{\mathsf g}_n=\begin{cases}
\displaystyle(-1)^{n/2} \binom{N-1}{n/2}&\text{$n$ even,}\\[+15pt]
\displaystyle(-1)^{(n+1)/2} \binom{N-1}{(n-1)/2}&\text{$n$ odd.}
\end{cases}
\end{equation}
One can also compute the partition function in the diagonal basis, where $\mathsf g_i=+1$ for all the nodes except for the last one, which has $\mathsf g_N=-1$. Plugging this into~\eqref{eq:numberwallsCa} yields the same expression for the twisted partition function.

\subsection[$G=\Spin(2N)$]{$\boldsymbol{G=\Spin(2N)}$}

Consider the algebra $D_N=\mathfrak{so}_{2N}$. The symmetries of this algebra are as follows:
\begin{itemize}
\item The group $\Spin(2N)$ has a $\mathbb Z_2$ zero-form symmetry. The corresponding charge is the chirality of the representation. This symmetry acts by permuting the last two nodes in the unextended Dynkin diagram. The associated diagonal action can be chosen as follows: take $\mathsf c_i=+1$ for all but the last two nodes, and $\mathsf c_{N-1}=+1$ and $\mathsf c_N=-1$.

\item The group $\Spin(2N)$ has a $\mathbb Z_2\times \mathbb Z_2$ one-form symmetry if $N$ is even, and $\mathbb Z_4$ if odd. They act on the extended Dynkin diagram as follows: one of the $\mathbb Z_2$'s for $N$ even, and the $\mathbb Z_2$ subgroup of $\mathbb Z_4$ for $N$ odd, acts as the permutation $\lambda_0\leftrightarrow\lambda_1$ and $\lambda_{N-1}\leftrightarrow \lambda_N$, while fixing the rest of Dynkin labels in the extended diagram. The other $\mathbb Z_2$ factor reverses the order of the extended Dynkin labels, while $\mathbb Z_4$ acts as $\lambda_0\mapsto\lambda_N\mapsto\lambda_1\mapsto\lambda_{N-1}\mapsto\lambda_0$, and it reverses the order of the rest of Dynkin labels. 

For $N$ even, let $(\mathsf g_1,\mathsf g_2)\in\mathbb Z_2\times \mathbb Z_2$; and, for $N$ odd, let $\mathsf g\in\mathbb Z_4$; all thought of as roots of unity. The center acts on a representation $\lambda$ as follows:
\begin{equation}
\begin{aligned}
\alpha_{\mathsf g_1,\mathsf g_2}(\lambda)&\equiv \mathsf{g}_1^{\lambda_{N-1}+\lambda_N}
\mathsf g_2^{(N/2-1)\lambda_{N-1}+(N/2)\lambda_N+\sum_{i=0}^{N/2-2}\lambda_{2i+1}}\\
\alpha_{\mathsf g}(\lambda)&\equiv \mathsf g^{-(N-2)\lambda_{N-1}-N\lambda_N+2\sum_{i=0}^{(N-3)/2}\lambda_{2i+1}}\,.
\end{aligned}
\end{equation}
Therefore, the eigenvalues in~\eqref{eq:numberwallsCa} are
\begin{equation}\label{eq:twist_D_series}
\begin{aligned}
(\mathsf g_1,\mathsf g_2)_{2i}&=1,\hspace{45pt} i\in[0,N/2-1)\\
(\mathsf g_1,\mathsf g_2)_{2i+1}&=\mathsf g_2,\hspace{40pt} i\in[0,N/2-1)\\
(\mathsf g_1,\mathsf g_2)_{N-1}&=\mathsf g_1\mathsf g_2^{N/2-1}\\
(\mathsf g_1,\mathsf g_2)_N&=\mathsf g_1\mathsf g_2^{N/2}\\
\mathsf g_{2i+1}&=\mathsf g^2,\hspace{42pt} i\in[0,(N-1)/2)\\
\mathsf g_{2i}&=1,\hspace{47pt} i\in[1,(N-1)/2)\\
\mathsf g_{N-1}&=\mathsf g^{N-2}\\
\mathsf g_N&=\mathsf g^N\,.
\end{aligned}
\end{equation}
\end{itemize}

Let us begin by computing the untwisted partition function. The comarks of $\Spin(2N)$ are $a_i^\vee=1$ for $i=0,1,N-1,N$, and $a_i^\vee=2$ for $i=2,3,\dots,N-2$. Plugging this into equation~\eqref{eq:numberwalls} we obtain the untwisted partition function 
\begin{equation}
Z(q)=(1-q)^4(1-q^2)^{N-3}\,, 
\end{equation}
and, expanding, the Witten index 
\begin{equation}\label{eq:I_Spin_even_UV}
%\begin{aligned}
%I_{2n}&=(-1)^n\bigg[\binom{N-3}{n}-6\binom{N-3}{n-1}+\binom{N-3}{n-2}\bigg]\\[+8pt]
%I_{2n+1}&=4(-1)^n\bigg[\binom{N-3}{n-1}-\binom{N-3}{n}\bigg]\,.
%\end{aligned}
I_n=\begin{cases}
\displaystyle(-1)^{n/2}\bigg[\binom{N-3}{n/2}-6\binom{N-3}{n/2-1}+\binom{N-3}{n/2-2}\bigg]&\text{$n$ even,}\\[+15pt]
\displaystyle4(-1)^{(n-1)/2}\bigg[\binom{N-3}{(n-1)/2-1}-\binom{N-3}{(n-1)/2}\bigg]&\text{$n$ odd.}
\end{cases}
\end{equation} 
Note that the index vanishes when $N$ is even and $n$ corresponds to the time-reversal symmetric wall $n=h/2=N-1$. It also vanishes for the exceptional pairs $(N,n)$ such that $2+4n+2n^2-3N-4nN+N^2=0$.

Let us now consider the index twisted by charge conjugation. Its action on the extended Dynkin diagram, and the resulting folded diagram, are as follows:
\begin{equation}\label{eq:fold_C_D}
\begin{tikzpicture}[baseline=1.9cm]

%\node at (-1,2.1) {$D_N^{{\color{blue}(1)}}\colon$};
%
%\draw[thick,blue] (0,2.3) circle (.1cm) node[anchor=east,inner xsep=10pt,scale=.6] {$1$};
%\draw[thick] (0,1.7) circle (.1cm) node[anchor=east,inner xsep=10pt,scale=.6] {$1$};
%
%\draw[thick] (3.67,2.05) -- (4.12,2.25);
%\draw[thick] (3.67,1.95) -- (4.12,1.75);
%
%\begin{scope}[yscale=1,xscale=-1,shift={(-4.2,0)}]
%\draw[thick] (3.67,2.05) -- (4.12,2.25);
%\draw[thick] (3.67,1.95) -- (4.12,1.75);
%\end{scope}
%
%%\draw[thick] (-3.2+3.67,2.05) -- (-4.2+4.12,2.25);
%%\draw[thick] (-3.2+3.67,1.95) -- (-4.2+4.12,1.75);
%
%\draw[thick] (3.6,2) circle (.1cm) node[anchor=south,inner ysep=10pt,scale=.6] {$2$};
%\draw[thick] (4.2,2.3) circle (.1cm) node[anchor=west,inner xsep=10pt,scale=.6] {$1$};
%\draw[thick] (4.2,1.7) circle (.1cm) node[anchor=west,inner xsep=10pt,scale=.6] {$1$};
%
%\foreach \i in {1,2,3}{
%\draw[thick] (0+.6*\i,2) circle (.1cm) node[anchor=south,inner ysep=10pt,scale=.6] {$2$};
%\draw[thick] (.1+.6*\i,2) -- (.5+.6*\i,2);
%}
%\draw[thick] (3.2,2) -- (3.5,2);
%\node at (2.73,2) {$\cdots$};

\begin{scope}[shift={(-1,0)}]
\node at (-.3,2.1) {$D_N^{{\color{blue}(1)}}\colon$};
\draw[thick] (3.5+1.3,2) -- (4.2+1.4,1.75);
\draw[thick] (3.5+1.3,2) -- (4.2+1.4,2.25);
\draw[thick,fill=white] (4.2+1.4,2.25) circle (.1cm) node[anchor=west,inner xsep=10pt,scale=.6] {$1$};
\draw[thick,fill=white] (4.2+1.4,2-.25) circle (.1cm) node[anchor=west,inner xsep=10pt,scale=.6] {$1$};
\begin{scope}[xscale=-1,shift={(-5,0)}]
\begin{scope}[shift={(-1,0)}]
\foreach \i in {1,2}{
\draw[thick,fill=white] ({0+.6*(\i+1)},2) circle (.1cm) node[anchor=south,inner ysep=10pt,scale=.6] {$2$};
\draw[thick,fill=white] ({.1+.6*(\i+1)},2) -- ({.5+.6*(\i+1)},2);
}
\node at (2.8,2) {$\cdots$};
%\draw[thick] (3.3,2) -- (3.4,2);

%\draw[thick] (0+.6,2) circle (.1cm) node[anchor=south,inner ysep=10pt,scale=.6] {$2$};
%\draw[thick] (0.8,2.2) -- (1,2) -- (0.8,1.8);
%\draw[thick] (1.2,1.9) -- (.6,1.9);
%\draw[thick] (1.2,2.1) -- (.6,2.1);
\end{scope}

\draw[thick,fill=white] (3.5-.6,2) circle (.1cm) node[anchor=south,inner ysep=10pt,scale=.6] {$2$};
\draw[thick] (3.6-.6,2) -- (4.2-.6,2);
\draw[thick] (3.6-1.2,2) -- (4.2-1.4,2);

\draw[thick] (3.5,2) -- (4.2,1.75);
\draw[thick,blue] (3.5,2) -- (4.2,2.25);
\draw[thick,fill=white,draw=blue] (4.2,2.25) circle (.1cm) node[anchor=east,inner xsep=10pt,scale=.6] {$\color{blue}1$};
\draw[thick,fill=white] (4.2,2-.25) circle (.1cm) node[anchor=east,inner xsep=10pt,scale=.6] {$1$};
\draw[thick,fill=white] (3.5,2) circle (.1cm) node[anchor=south,inner ysep=10pt,scale=.6] {$2$};

\end{scope}
\end{scope}

\draw[<->,thick,red,>=stealth] (4.75+.25,2.35) arc[radius=.35, start angle=70, end angle=-70];

\node at (5.9,2) {$\mapsto$};

\begin{scope}[shift={(7.5,0)}]
\node at (-.3,2.1) {$D_N^{{\color{blue}(2)}}\colon$};

\begin{scope}[xscale=-1,shift={(-5,0)}]
\begin{scope}[shift={(-1,0)}]
\foreach \i in {1,2}{
\draw[thick] ({0+.6*(\i+1)},2) circle (.1cm) node[anchor=south,inner ysep=10pt,scale=.6] {$2$};
\draw[thick] ({.1+.6*(\i+1)},2) -- ({.5+.6*(\i+1)},2);
}
\node at (2.8,2) {$\cdots$};
%\draw[thick] (3.3,2) -- (3.4,2);

\draw[thick] (0+.6,2) circle (.1cm) node[anchor=south,inner ysep=10pt,scale=.6] {$2$};
\draw[thick] (1,2.2) -- (0.8,2) -- (1,1.8);
\draw[thick] (1.2,1.9) -- (.6,1.9);
\draw[thick] (1.2,2.1) -- (.6,2.1);
\end{scope}

\draw[thick,fill=white] (3.5-.6,2) circle (.1cm) node[anchor=south,inner ysep=10pt,scale=.6] {$2$};
\draw[thick] (3.6-.6,2) -- (4.2-.6,2);
\draw[thick] (3.6-1.2,2) -- (4.2-1.4,2);

\draw[thick] (3.5,2) -- (4.2,1.75);
\draw[thick,blue] (3.5,2) -- (4.2,2.25);
\draw[thick,fill=white,draw=blue] (4.2,2.25) circle (.1cm) node[anchor=east,inner xsep=10pt,scale=.6] {$\color{blue}1$};
\draw[thick,fill=white] (4.2,2-.25) circle (.1cm) node[anchor=east,inner xsep=10pt,scale=.6] {$1$};
\draw[thick,fill=white] (3.5,2) circle (.1cm) node[anchor=south,inner ysep=10pt,scale=.6] {$2$};

\end{scope}
\end{scope}

\end{tikzpicture}
\end{equation}
where the folded diagram has $N$ nodes. From this we learn that the folded diagram has $r'+1=N$ nodes, two of which have energy equal to $1$, and the rest all energy equal to $2$. Plugging this into~\eqref{eq:numberwallsC} the zero-form twisted partition function is
\begin{equation}
Z^{\mathsf c}(q)=(1-q)^2(1-q^2)^{N-2}
\end{equation}
and, expanding, the twisted Witten index 
\begin{equation}\label{eq:I_Spin_even_c_UV}
%\begin{aligned}
%I^{\mathsf c}_{2n}&=(-1)^n\bigg[\binom{N-2}{n}-\binom{N-2}{n-1}\bigg]\\[+8pt]
%I^{\mathsf c}_{2n+1}&=2(-1)^{n+1}\binom{N-2}{n}\,.
%\end{aligned}
I^{\mathsf c}_n=\begin{cases}
\displaystyle(-1)^{n/2}\bigg[\binom{N-2}{n/2}-\binom{N-2}{n/2-1}\bigg]&\text{$n$ even,}\\[+15pt]
\displaystyle2(-1)^{(n+1)/2}\binom{N-2}{(n-1)/2}&\text{$n$ odd.}
\end{cases}
\end{equation}
One can also compute the partition function in the diagonal basis, where $\mathsf c_i=+1$ for all the nodes except for the last one, which has $\mathsf c_N=-1$. Plugging this into~\eqref{eq:numberwallsCa} yields the same expression for the twisted partition function.

Let us now consider the one-form-twisted partition functions. The symmetry depends on whether $N$ is even or odd, which we consider in turn.

\paragraph{$\boldsymbol N$ even.} Here the symmetry is $\mathbb Z_2\times\mathbb Z_2$. $\mathsf g_1$ and $\mathsf g_2$ act as follows:
\begin{equation}
\begin{tikzpicture}[baseline=1.9cm]

\draw[thick,blue] (0,2.3) circle (.1cm) node[anchor=east,inner xsep=10pt,scale=.6] {$1$};
\draw[thick] (0,1.7) circle (.1cm) node[anchor=east,inner xsep=10pt,scale=.6] {$1$};

\draw[thick] (3.67,2.05) -- (4.12,2.25);
\draw[thick] (3.67,1.95) -- (4.12,1.75);

\begin{scope}[yscale=1,xscale=-1,shift={(-4.2,0)}]
\draw[thick,blue] (3.67,2.05) -- (4.12,2.25);
\draw[thick] (3.67,1.95) -- (4.12,1.75);
\end{scope}

%\draw[thick] (-3.2+3.67,2.05) -- (-4.2+4.12,2.25);
%\draw[thick] (-3.2+3.67,1.95) -- (-4.2+4.12,1.75);

\draw[thick] (4.2,2.3) circle (.1cm) node[anchor=west,inner xsep=10pt,scale=.6] {$1$};
\draw[thick] (4.2,1.7) circle (.1cm) node[anchor=west,inner xsep=10pt,scale=.6] {$1$};

\foreach \i in {1,2}{
\draw[thick] (0+.6*\i,2) circle (.1cm) node[anchor=south,inner ysep=10pt,scale=.6] {$2$};
\draw[thick] (.1+.6*\i,2) -- (.5+.6*\i,2);
}
\draw[thick] (3.2,2) -- (3.5,2);
\node at (2.15,2) {$\cdots$};
\foreach \i in {4,5}{
\draw[thick] (0+.6*\i+.6,2) circle (.1cm) node[anchor=south,inner ysep=10pt,scale=.6] {$2$};
\draw[thick] (.5+.6*\i,2) -- (.1+.6*\i,2);
}

\draw[<->,thick,red,>=stealth] (4.75,2.35) arc[radius=.35, start angle=70, end angle=-70];

\begin{scope}[yscale=1,xscale=-1,shift={(-4.2,0)}]
\draw[<->,thick,red,>=stealth] (4.75,2.35) arc[radius=.35, start angle=70, end angle=-70];
\end{scope}

\begin{scope}[shift={(8,0)}]
\draw[thick,blue] (0,2.3) circle (.1cm) node[anchor=east,inner xsep=10pt,scale=.6] {$1$};
\draw[thick] (0,1.7) circle (.1cm) node[anchor=east,inner xsep=10pt,scale=.6] {$1$};

\draw[thick] (3.67,2.05) -- (4.12,2.25);
\draw[thick] (3.67,1.95) -- (4.12,1.75);

\begin{scope}[yscale=1,xscale=-1,shift={(-4.2,0)}]
\draw[thick,blue] (3.67,2.05) -- (4.12,2.25);
\draw[thick] (3.67,1.95) -- (4.12,1.75);
\end{scope}

%\draw[thick] (-3.2+3.67,2.05) -- (-4.2+4.12,2.25);
%\draw[thick] (-3.2+3.67,1.95) -- (-4.2+4.12,1.75);

\draw[thick] (4.2,2.3) circle (.1cm) node[anchor=west,inner xsep=10pt,scale=.6] {$1$};
\draw[thick] (4.2,1.7) circle (.1cm) node[anchor=west,inner xsep=10pt,scale=.6] {$1$};

\foreach \i in {1,2}{
\draw[thick] (0+.6*\i,2) circle (.1cm) node[anchor=south,inner ysep=10pt,scale=.6] {$2$};
\draw[thick] (.1+.6*\i,2) -- (.5+.6*\i,2);
}
\draw[thick] (3.2,2) -- (3.5,2);
\node at (2.15,2) {$\cdots$};
\foreach \i in {4,5}{
\draw[thick] (0+.6*\i+.6,2) circle (.1cm) node[anchor=south,inner ysep=10pt,scale=.6] {$2$};
\draw[thick] (.5+.6*\i,2) -- (.1+.6*\i,2);
}

\begin{scope}[yscale=-1,xscale=1,shift={(0,-4)}]
\draw[<->,thick,red,>=stealth] (0,1.4) to[in=210,out=-30] (4.2,1.4);
\end{scope}

\draw[<->,thick,red,>=stealth] (1.2,1.8) to[in=210,out=-30] (3,1.8);
\draw[<->,thick,red,>=stealth] (.6,1.6) to[in=210,out=-30] (3.6,1.6);
\draw[<->,thick,red,>=stealth] (0,1.4) to[in=210,out=-30] (4.2,1.4);
\end{scope}

\end{tikzpicture}
\end{equation}
The folded diagrams are
\begin{equation}
\hspace{-10pt}
\begin{tikzpicture}[baseline=1.9cm]

\node at (-1.5,2.05) {$C_{N-2}^{{\color{blue}(1)}}\ \colon$};
\node at (7,2.05) {$B_{N/2}^{{\color{blue}(1)}}\ \colon$};

\draw[thick] (3.6,2.1) -- (4.2,2.1);
\draw[thick] (3.6,1.9) -- (4.2,1.9);
\draw[thick] (4.2,2) circle (.1cm) node[anchor=south,inner ysep=10pt,scale=.6] {$2$};

\draw[thick,blue] (0,2) circle (.1cm) node[anchor=south,inner ysep=10pt,scale=.6] {$2$};
\draw[thick,blue] (3.6-3.6,2.1) -- (4.2-3.6,2.1);
\draw[thick,blue] (3.6-3.6,1.9) -- (4.2-3.6,1.9);
\draw[thick,blue] (4-3.6,2.2) -- (3.8-3.6,2) -- (4-3.6,1.8);

\foreach \i in {1,2}{
\draw[thick] (0+.6*\i,2) circle (.1cm) node[anchor=south,inner ysep=10pt,scale=.6] {$2$};
\draw[thick] (.1+.6*\i,2) -- (.5+.6*\i,2);
}
%\draw[thick] (3.2,2) -- (3.5,2);
\node at (2.15,2) {$\cdots$};
\foreach \i in {4,5}{
\draw[thick] (0+.6*\i+.6,2) circle (.1cm) node[anchor=south,inner ysep=10pt,scale=.6] {$2$};
\draw[thick] (.5+.6*\i,2) -- (.1+.6*\i,2);
}

\draw[thick] (3.8,2.2) -- (4,2) -- (3.8,1.8);

\begin{scope}[shift={(8.6,0)}]
\draw[thick,blue] (0,2.3) circle (.1cm) node[anchor=east,inner xsep=10pt,scale=.6] {$2$};
\draw[thick] (0,1.7) circle (.1cm) node[anchor=east,inner xsep=10pt,scale=.6] {$2$};

\begin{scope}[yscale=1,xscale=-1,shift={(-4.2,0)}]
\draw[thick,blue] (3.67,2.05) -- (4.12,2.25);
\draw[thick] (3.67,1.95) -- (4.12,1.75);
\end{scope}

%\draw[thick] (-3.2+3.67,2.05) -- (-4.2+4.12,2.25);
%\draw[thick] (-3.2+3.67,1.95) -- (-4.2+4.12,1.75);

\foreach \i in {1,2}{
\draw[thick] (0+.6*\i,2) circle (.1cm) node[anchor=south,inner ysep=10pt,scale=.6] {$4$};
\draw[thick] (.1+.6*\i,2) -- (.5+.6*\i,2);
}

\node at (2.15,2) {$\cdots$};
\draw[thick] (0+.6*4+.6,2) circle (.1cm) node[anchor=south,inner ysep=10pt,scale=.6] {$4$};
\draw[thick] (.5+.6*4,2) -- (.1+.6*4,2);
\draw[thick] (0+.6*5+.6,2) circle (.1cm) node[anchor=south,inner ysep=10pt,scale=.6] {$2$};

\draw[thick] (4-.6,2.2) -- (3.8-.6,2) -- (4-.6,1.8);
\draw[thick] (3.6,2.1) -- (4.2-1.2,2.1);
\draw[thick] (3.6,1.9) -- (4.2-1.2,1.9);
\end{scope}

\end{tikzpicture}
\end{equation}
which have $N-1$ and $N/2+1$ nodes, respectively. The folding by $\mathsf g_1\mathsf g_2$ is in fact identical to that of $\mathsf g_2$, i.e., the second diagram.

The twisted partition functions read
\begin{equation}
\begin{aligned}
Z^{\mathsf g_1}(q)&=(1-q^2)^{N-1}\\
Z^{\mathsf g_2}(q)&=(1-q^2)^3(1-q^4)^{N/2-2}\,,
\end{aligned}
\end{equation}
and, expanding, the twisted Witten indices
\begin{equation}\label{eq:I_Spin_even_g_UV_even}
\begin{aligned}
%I^{\mathsf g_1}_{2n}&=(-1)^n \binom{N-1}{n}\\
%I^{\mathsf g_1}_{2n+1}&=0\\
%I^{\mathsf g_2}_{4n}&=(-1)^n\bigg[\binom{N/2-2}{n}-3\binom{N/2-2}{n-1}\bigg]\\
%I^{\mathsf g_2}_{4n+2}&=(-1)^n\bigg[\binom{N/2-2}{n-1}-3\binom{N/2-2}{n}\bigg]\\
%I^{\mathsf g_2}_{2n+1}&=0
I^{\mathsf g_1}_n&=\begin{cases}
\displaystyle(-1)^{n/2} \binom{N-1}{n/2}&\text{$n$ even,}\\[+15pt]
\displaystyle0&\text{$n$ odd.}
\end{cases}\\[+5pt]
I^{\mathsf g_2}_n&=\begin{cases}
\displaystyle(-1)^{n/4}\bigg[\binom{N/2-2}{n/4}-3\binom{N/2-2}{n/4-1}\bigg]&n\equiv0\mod4,\\[+15pt]
\displaystyle(-1)^{(n-2)/4}\bigg[\binom{N/2-2}{(n-2)/4-1}-3\binom{N/2-2}{(n-2)/4}\bigg]&n\equiv2\mod4,\\[+15pt]
\displaystyle0&\text{$n$ odd,}
\end{cases}
\end{aligned}
\end{equation}
while $I^{\mathsf g_1\mathsf g_2}_n= I^{\mathsf g_2}_n$.

\paragraph{$\boldsymbol N$ odd.} Here the one-form symmetry is $\mathbb Z_4$, whose action on the extended Dynkin diagram, and the corresponding folded diagram, are as follows:
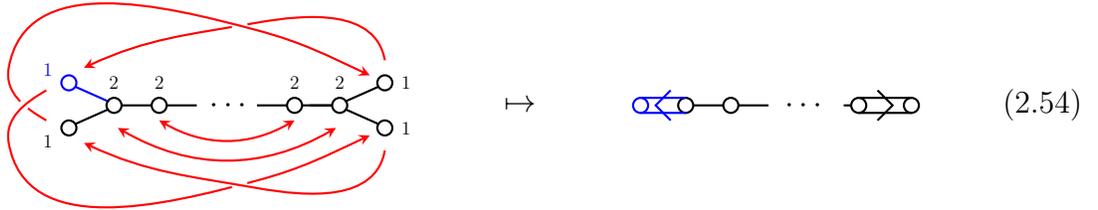
\begin{equation}
\begin{tikzpicture}[baseline=1.9cm]

\draw[thick,blue] (0,2.3) circle (.1cm) node[anchor=south east,inner xsep=10pt,scale=.6] {$1$};
\draw[thick] (0,1.7) circle (.1cm) node[anchor=north east,inner xsep=10pt,scale=.6] {$1$};

\draw[thick] (3.67,2.05) -- (4.12,2.25);
\draw[thick] (3.67,1.95) -- (4.12,1.75);

\begin{scope}[yscale=1,xscale=-1,shift={(-4.2,0)}]
\draw[thick,blue] (3.67,2.05) -- (4.12,2.25);
\draw[thick] (3.67,1.95) -- (4.12,1.75);
\end{scope}

%\draw[thick] (-3.2+3.67,2.05) -- (-4.2+4.12,2.25);
%\draw[thick] (-3.2+3.67,1.95) -- (-4.2+4.12,1.75);

\draw[thick] (4.2,2.3) circle (.1cm) node[anchor=west,inner xsep=10pt,scale=.6] {$1$};
\draw[thick] (4.2,1.7) circle (.1cm) node[anchor=west,inner xsep=10pt,scale=.6] {$1$};

\foreach \i in {1,2}{
\draw[thick] (0+.6*\i,2) circle (.1cm) node[anchor=south,inner ysep=10pt,scale=.6] {$2$};
\draw[thick] (.1+.6*\i,2) -- (.5+.6*\i,2);
}
\draw[thick] (3.2,2) -- (3.5,2);
\node at (2.15,2) {$\cdots$};
\foreach \i in {4,5}{
\draw[thick] (0+.6*\i+.6,2) circle (.1cm) node[anchor=south,inner ysep=10pt,scale=.6] {$2$};
\draw[thick] (.5+.6*\i,2) -- (.1+.6*\i,2);
}

\draw[thick,red,->,>=stealth] (4.2,2.6) to[in=10,out=90+10] (2.5,3.1) to[in=25,out=180+10] (.2,2.5);
\fill[white] (2.27,3.06) circle (3pt);

\draw[thick,red,->,>=stealth] (-.3,1.8) to[in=-90-10,out=90+60] (-.8,2.5) to[in=180-15,out=90-10] (2.5,3) to[in=180-25,out=-15] (4,2.4);
\fill[white] (-.6,2) circle (2pt);

\begin{scope}[yscale=-1,xscale=1,shift={(0,-4)}]
\draw[thick,red,->,>=stealth] (-.3,1.8) to[in=-90-10,out=90+60] (-.8,2.5) to[in=180-15,out=90-10] (2.5,3) to[in=180-25,out=-15] (4,2.4);
\fill[white] (2.27,3.06) circle (3pt);
\draw[thick,red,->,>=stealth] (4.2,2.6) to[in=10,out=90+10] (2.5,3.1) to[in=25,out=180+10] (.2,2.5);\
\end{scope}

\draw[<->,thick,red,>=stealth] (1.2,1.8) to[in=210,out=-30] (3,1.8);
\draw[<->,thick,red,>=stealth] (.65,1.7) to[in=210,out=-30] (3.55,1.7);

\node at (6,2) {$\mapsto$};

\begin{scope}[shift={(7,0)}]
\foreach \i in {1,2}{
\draw[thick] ({0+.6*(\i+1)},2) circle (.1cm);% node[anchor=south,inner ysep=10pt,scale=.6] {$4$};
\draw[thick] ({.1+.6*(\i+1)},2) -- ({.5+.6*(\i+1)},2);
}
\draw[thick] (3.5,2) circle (.1cm);% node[anchor=south,inner ysep=10pt,scale=.6] {$4$};
\draw[thick] (4.2,2) circle (.1cm);% node[anchor=south,inner ysep=10pt,scale=.6] {$4$};

\node at (2.8,2) {$\cdots$};
\draw[thick] (3.3,2) -- (3.4,2);

\draw[thick,blue] (0+.6,2) circle (.1cm);% node[anchor=south,inner ysep=10pt,scale=.6] {$4$};
\draw[thick,blue] (1,2.2) -- (.8,2) -- (1,1.8);
\draw[thick,blue] (1.2,1.9) -- (.6,1.9);
\draw[thick,blue] (1.2,2.1) -- (.6,2.1);

\draw[thick] (3.5,1.9) -- (4.2,1.9);
\draw[thick] (3.5,2.1) -- (4.2,2.1);
\draw[thick] (3.75,2.2) -- (3.95,2) -- (3.75,1.8);
\end{scope}

\end{tikzpicture}
\end{equation}
where the comarks are all $4$ if we fold by a generator of $\mathbb Z_4$, and all $2$ if we fold by a generator squared. The number of nodes is $(N-1)/2$ in the first case, and $N-1$ in the second case. %Note that the total energy is conserved: $4\times1+2\times(N-3)\equiv 2\times (N-1)$.
The folded diagram corresponds to $C^{{\color{blue}(1)}}_{(N-1)/2}$ and $C^{{\color{blue}(1)}}_{N-1}$, respectively.

If we let $\mathsf g$ denote a generator of $\mathbb Z_4$, the twisted partition function is
\begin{equation} 
\begin{aligned}
Z^{{\mathsf g}}(q)&= Z^{{\mathsf g}^3}(q)=(1-q^4)^{(N-1)/2}\,, \\[+4pt]
Z^{{\mathsf g}^2}(q)&= (1-q^2)^{N-1} \,,
\end{aligned}
\end{equation}
and, expanding, the twisted Witten indices
\begin{equation}\label{eq:I_Spin_even_g_UV_odd}
\begin{aligned}
%I^{{\mathsf g}}_{4n}&=I^{{\mathsf g}^3}_{4n}=(-1)^{n} \binom{(N-1)/2}{n}\,, \\[+4pt]
%I^{{\mathsf g}}_{4n+1}&=I^{{\mathsf g}^3}_{4n+1}=I^{{\mathsf g}}_{4n+2}=I^{{\mathsf g}^3}_{4n+2}=0\,, \\[+4pt]
%I^{{\mathsf g}^2}_{2n}&=(-1)^{n} \binom{N-1}{n}\,, \\[+4pt]
%I^{{\mathsf g}^2}_{2n+1}&=0\,.
I^{{\mathsf g}}_n&=I^{{\mathsf g^3}}_n=\begin{cases}
\displaystyle(-1)^{n/4} \binom{(N-1)/2}{n/4}&n\equiv0\mod4,\\[+15pt]
\displaystyle0&\text{otherwise,}
\end{cases}\\[+5pt]
I^{{\mathsf g}^2}_n&=\begin{cases}
\displaystyle(-1)^{n/2} \binom{N-1}{n/2}&\text{$n$ even,}\\[+15pt]
\displaystyle0&\text{$n$ odd.}
\end{cases}
\end{aligned}
\end{equation}

As usual, one may also compute these partition functions in the diagonal basis. Using the phases~\eqref{eq:twist_D_series} in~\eqref{eq:numberwallsCa} yields the same expressions for the twisted partition functions, as expected.

\subsection[$G=G_2$]{$\boldsymbol{G=G_2}$}

$G_2$ has no zero-form or one-form symmetry. The comarks for $G_2$ are $a_0^\vee=a_2^\vee=1$ and $a_1^\vee=2$. Plugging this into equation~\eqref{eq:numberwalls} we obtain the untwisted partition function 
\begin{equation}
Z(q)=(1-q)^2(1-q^2)\,, 
\end{equation}
and, expanding, the Witten indices 
\begin{equation}\label{eq:G2}
\begin{aligned}
I_1&=-2\\ 
I_2&=0\\ 
I_3&=2\,.
\end{aligned}
\end{equation} 
Note that $I_3=-I_1$, as expected from the action of time-reversal on domain walls.

\subsection{Minimal Wall for Arbitrary Gauge Group}
\label{sec:min}

The domain wall theory for $n=1$ admits a uniform description for all simply-connected groups, including the exceptional ones. Indeed, the only fermion configurations with total energy equal to $1$, that is the solutions to~\eqref{eq:energy_fermion} 
\begin{equation}\label{eq:minimal_wall_G}
\sum_{i=0}^r \lambda_i a^\vee_i=1\,,
\end{equation}
are clearly of the form $\lambda_i=1$ for one $i$ such that $a_i^\vee=1$, and $\lambda_j=0$ for all $j\neq i$. In other words, in each configuration there is only one excited fermion, which moreover necessarily has energy $a_i^\vee=1$. All these configurations have the same fermion number, namely $(-1)^F=-1$, which means that the index is
\begin{equation}\label{eq:minimal_wall_G_index}
I_1\equiv -m_1
\end{equation}
where $m_1$ denotes the number of nodes in the extended Dynkin diagram $\mathfrak g^{(1)}$ with comark equal to $1$. The values of $m_1$ are given in the following table:
\begin{equation}
\begin{array}{|c|c|c|c|c|c|c|c|c|c|}\hline
G & \SU(N) &\Sp(N)&\Spin(2N+1)&\Spin(2N)& E_6&E_7&E_8&F_4&G_2\\\hline
m_1&N&N+1&3&4&3&2&1&2&2\\\hline
\end{array}
\end{equation}
Note that, for simply-laced $G$, $m_1$ is the order of $\Gamma$.

The index twisted by a symmetry $\mathsf s\in \mathsf S$ is
\begin{equation}
I^{\mathsf s}_1\equiv -m^{\mathsf s}_1\,,
\end{equation}
where $m^{\mathsf s}_1$ denotes the number of nodes in the extended Dynkin diagram $\mathfrak g^{(1)}$ with comark equal to $1$ that are fixed by $\mathsf s$. 
$m^{\mathsf s}_1$ has already been computed for the classical groups $\SU(N), \Sp(N),\Spin(2N+1)$ and $\Spin(2N)$. For the exceptional groups, only $E_6$ and $E_7$ have non-trivial symmetry group $\mathsf S$ (see table~\ref{tab:comarksone}). In $E_6$, the zero-form charge-conjugation symmetry leaves invariant the extended node, which has comark $1$, and permutes the other two nodes with comark $1$. In $E_6$ and $E_7$, the one-form center symmetry permutes all the nodes with comark $1$. Therefore, letting $\mathsf c$ denote the non-trivial element of $\mathsf C$, and $\mathsf g$ any non-trivial element of $\Gamma$, the indices are
\begin{equation}
\begin{array}{rll}
E_6:& I_1^{\mathsf c}=-1,\quad I_1^{\mathsf g}=0&\quad\mathsf g\in \Gamma=\mathbb Z_3\\
E_7:& I_1^{\mathsf g}=0&\quad\mathsf g\in \Gamma=\mathbb Z_2\,.
\end{array}
\label{twistEE}
\end{equation}

The (twisted) indices for the exceptional groups $E_6,E_7,E_8,F_4$ and arbitrary $n$ have been included in appendix~\ref{sec:exceptional_indices} for completeness.

\medskip

This concludes our discussion of the twisted Witten indices in the ultraviolet. A rather nontrivial consistency test of our proposal is that the Witten indices on the domain walls we just computed are reproduced by the corresponding partition functions of our conjectured TQFTs in the infrared. Computing the image of the (twisted) Witten indices in the infrared TQFT is nontrivial and requires understanding in detail the Hilbert space of spin TQFTs and the action of $(-1)^F$ on it, a subject to which we now turn.

\section{Hilbert Space of Spin TQFTs}\label{sec:spin}

The domain wall theory preserves $3d$ $\mathcal N=1$ supersymmetry and observables depend on the choice of a spin structure. Therefore, the TQFT that emerges in the deep infrared of the domain wall must also depend on a choice of a spin structure, that is, it must be a spin TQFT~\cite{Dijkgraaf:1989pz}.

The data of a TQFT in $3d$ includes the set of anyons $\mathcal A$ (or Wilson lines) and the braiding matrix $B\colon\mathcal A\times\mathcal A\to U(1)$ encoding their braiding. A spin TQFT is a TQFT that has an abelian\footnote{An abelian line is one that yields a single line in its fusion with any line in ${\mathcal A}$.} line $\psi$ that braids trivially with all lines in $\mathcal A$ and has half-integral spin.\footnote{A TQFT that has a line with half-integral spin which braids nontrivially with at least one line in the theory is not spin. There is no unambiguous way to assign a sign to the fermion as we move it around a circle since the phase it acquires depends on which lines link with the circle, unlike when the fermion is transparent.} Transparency of $\psi$ implies that it fuses with itself into the vacuum, that is $\psi\times\psi=\boldsymbol1$. Since $\psi$ is transparent and has half-integral spin, the observables of a spin TQFT depend on the choice of a spin structure. 

A spin TQFT can be constructed from a parent bosonic TQFT which has an abelian, non-transparent fermion $\psi$ with $\psi\times \psi=\boldsymbol1$, that is a bosonic TQFT that has a $\mathbb Z_2^\psi$ one-form symmetry generated by a fermion~\cite{Gu:2013gma,beliakova2014,Gaiotto:2015zta,Bhardwaj:2016clt,Aasen:2017ubm}. The bosonic parent theory defines a spin TQFT upon gauging its $\mathbb Z_2^{\psi}$ one-form symmetry generated by $\psi$ 
\begin{equation}\label{spinquotient}
\text{spin TQFT}=\frac{\text{bosonic TQFT}}{\mathbb Z_2^{\psi}}\,.
\end{equation}
This procedure is an extension of the notion of bosonic ``anyon condensation"~\cite{MOORE1989422,PhysRevB.79.045316,PhysRevB.90.195130}.\footnote{More precisely, the parent bosonic TQFT must be attached to a suitable $4d$ SPT phase so that the combined system is non-anomalous, and the symmetry can be gauged.} 
Upon gauging, the fermion $\psi$ in the parent bosonic theory becomes the transparent fermion $\psi$ in the spin TQFT. The gauged one-form symmetry $\mathbb Z_2^{\psi}$ of the parent bosonic theory gives rise to an emergent zero-form symmetry $\mathbb Z_2$ in the spin TQFT that is generated by the fermion parity operator $(-1)^F$, and which acts on the ``twisted sector". We will discuss the action of $(-1)^F$ on the Hilbert space of spin TQFTs shortly.

The lines of the parent bosonic theory ${\mathcal A}$ can be arranged as the disjoint union of two sets ${\mathcal A}={\mathcal A}_\text{NS} \cup {\mathcal A}_\text{R}$ according to their braiding with $\psi$. Lines in ${\mathcal A}_\text{NS}$, by definition, braid trivially with $\psi$ while lines in ${\mathcal A}_\text{R}$ have braiding $-1$ with $\psi$. This partitions the lines of the bosonic TQFT according to their $\mathbb Z_2^{\psi}$ quantum number. The lines in each set can be organized into orbits of $\mathbb Z_2^{\psi}$, generated by fusion with $\psi$. The orbits can be either two- or one-dimensional. The lines in one-dimensional orbits are referred to as ``Majorana lines'' in that they can freely absorb the fermion $\psi$:
\begin{equation}\label{eq:majorana}
\psi\times m=m\,.
\end{equation}
The Majorana lines, if any, are necessarily in ${\mathcal A}_\text{R}$.\footnote{To prove this we compute the braiding of $\psi$ with a Majorana line $m$ and show that it necessarily has braiding $-1$ with $\psi$
\begin{equation}
B(\psi,m)=e^{2\pi i (h_\psi+h_m-h_{\psi\times m})}=e^{2\pi h_\psi}=-1\,,\quad\Longrightarrow\quad m\in {\mathcal A}_\text{R}\,,
\end{equation}
where $h$ denotes spin of lines and in the second equality we have used the defining relation for a Majorana line~\eqref{eq:majorana}. See also~\cite{Aasen:2017ubm}.} The lines of the bosonic parent theory thus split as
\begin{equation}
\begin{aligned}
{\mathcal A}_\text{NS}&=\{\{a,a\times \psi\}\ |\ B(\psi,a)=+1\}\\
{\mathcal A}_\text{R}&=\{\{x,x\times \psi\},\ \{m\}\ |\ B(\psi,x)=B(\psi,m)=-1\}\,.
\end{aligned}
\label{orbits}
\end{equation}
 The first set, referred to as the Neveu-Schwarz (NS) lines, is what is usually regarded as the set of Wilson line operators in the spin TQFT. The second set, the Ramond (R) lines, change the spin structure background. This decomposition will be useful shortly in the construction of the Hilbert space of the spin TQFT.

The Hilbert space of the spin TQFT on the spatial torus depends on the choice of spin structure. There are two equivalence classes of spin structures on the torus (or, more generally, on any Riemann surface): even and odd spin structures. Consider the even and odd spin structure Hilbert spaces ${\mathcal H}_\text{NS-NS}$ and ${\mathcal H}_\text{R-R}$. ${\mathcal H}_\text{NS-NS}$ correspond to choosing antiperiodic boundary conditions on the two circles while ${\mathcal H}_\text{R-R}$ corresponds to periodic boundary conditions. The other two even spin-structure Hilbert spaces ${\mathcal H}_\text{NS-R}$ and ${\mathcal H}_\text{R-NS}$ can be obtained from ${\mathcal H}_\text{NS-NS}$ by the action of the mapping class group. This group is a non-trivial extension of the modular group $\mathrm{SL}_2(\mathbb Z)$ by the $\mathbb Z_2$ fermion parity symmetry. It is known as the \emph{metaplectic group} $\mathrm{Mp}_1(\mathbb Z)$. It does not preserve the individual spin structures but it does preserve their equivalence class. The Hilbert spaces of spin TQFTs realize a unitary representation of this group. 
 
The states in the Hilbert space ${\mathcal H}_B$ of a bosonic TQFT are constructed from the path integral on a solid torus by inserting lines $M\in {\mathcal A}$ along the non-contractible cycle~\cite{Witten:1988hf}. This defines conformal blocks on the torus. We represent this pictorially by
\begin{equation}\label{eq:states_bosonic}
|M\rangle=\ \tikz[baseline=-3]{
\draw[thick] (-.5,-.4) -- (-.5,.4);
\draw[thick] (.5,.4) -- (.65,0) -- (.5,-.4);
\draw[thick] (0,0) circle (.25 cm);
\node[anchor=south] at (0,-.8) {\footnotesize$M$};
} \in {\mathcal H}_B\,.
\end{equation}
The Hilbert space of the spin TQFT can be constructed from its definition as a quotient of the bosonic parent TQFT~\eqref{spinquotient}.\footnote{More details of the explicit construction of the Hilbert space of spin TQFTs will appear elsewhere~\cite{delmastro2021global}.} The states in ${\mathcal H}_\text{NS-NS}$ are labeled by $a\in {\mathcal A}_\text{NS}$, and are represented as 
\begin{equation}
|a\rangle_\text{spin}=\tikz[baseline=-3]{
\node[anchor=south] at (-1.2,-.65) {$\frac{1}{\sqrt2}\ \bigg($};
\draw[thick] (-.5,-.4) -- (-.5,.4);
\draw[thick] (.5,.4) -- (.65,0) -- (.5,-.4);
\draw[thick] (0,0) circle (.25 cm);
\node[anchor=south] at (0,-.8) {\footnotesize$a$};
\node[anchor=south] at (1.3,-.3) {$+$};
\node[anchor=south] at (3.5,-.65) {$\vphantom{\frac{1}{\sqrt2}}\bigg)$};
\begin{scope}[shift={(2.5,0)}]
\draw[thick] (-.5,-.4) -- (-.5,.4);
\draw[thick] (.5,.4) -- (.65,0) -- (.5,-.4);
\draw[thick] (0,0) circle (.25 cm);
\node[anchor=south] at (0,-.8) {\footnotesize$a\times\psi$};
\end{scope}
}\in {\mathcal H}_\text{NS-NS}%,\qquad a=1\ldots,N_a
\,.
\end{equation}

The states in ${\mathcal H}_\text{R-R}$ are constructed from conformal blocks of the bosonic parent TQFT on the torus and, in the presence of Majorana lines~\eqref{eq:majorana}, from the once-punctured torus conformal blocks of the bosonic parent TQFT. By virtue of $m$ being a Majorana line obeying the fusion rule $\psi\times m=m$, the one-point conformal block on the torus with $m$ along the cycle and $\psi$ at the puncture is nontrivial, as it is allowed by the fusion rules. The states in ${\mathcal H}_\text{R-R}$ are labeled by $x,m\in {\mathcal A}_\text{R}$, and are represented as\footnote{Unlike in bosonic anyon condensation, where a fixed line in the parent theory yields multiples states in the quotient theory, a Majorana line is in an irreducible representation of $\operatorname{Cliff}(1|1)$ and yields a unique state in the quotient (spin) TQFT.}
\begin{equation}
\begin{aligned}
|\,x\,\rangle_\text{spin}&=\tikz[baseline=-3]{
\node[anchor=south] at (-1.2,-.65) {$\frac{1}{\sqrt2}\ \bigg($};
\draw[thick] (-.5,-.4) -- (-.5,.4);
\draw[thick] (.5,.4) -- (.65,0) -- (.5,-.4);
\draw[thick] (0,0) circle (.25 cm);
\node[anchor=south] at (0,-.8) {\footnotesize$x$};
\node[anchor=south] at (1.3,-.3) {$-$};
\node[anchor=south] at (3.5,-.65) {$\vphantom{\frac{1}{\sqrt2}}\bigg)$};
\begin{scope}[shift={(2.5,0)}]
\draw[thick] (-.5,-.4) -- (-.5,.4);
\draw[thick] (.5,.4) -- (.65,0) -- (.5,-.4);
\draw[thick] (0,0) circle (.25 cm);
\node[anchor=south] at (0,-.8) {\footnotesize$x\times\psi$};
\end{scope}
}\in {\mathcal H}_\text{R-R}\,,% \qquad &\alpha=1,\ldots,N_\alpha\,,
\\
%\end{equation}
 %\begin{equation}
|m\rangle_\text{spin}&=\ \tikz[baseline=-3]{
\draw[thick] (-.5,-.4) -- (-.5,.4);
\draw[thick] (.5,.4) -- (.65,0) -- (.5,-.4);
\draw[thick] (0,.25) -- (0,.45);
\draw[thick] (0,0) circle (.25 cm);
\draw[thick] (0,.25) circle (1pt);
\node[anchor=south] at (0,-.8) {\footnotesize$m$};
\node[anchor=south] at (.2,.2) {\footnotesize$\psi$};
}
\in {\mathcal H}_\text{R-R}%\,\qquad &m=1,\ldots,N_m
\,.
\end{aligned}
\label{Ramondstates}
\end{equation}
Modular transformations preserve the odd spin structure, i.e., they map $\mathcal H_\text{R-R}$ into itself. The negative sign in~\eqref{Ramondstates} guarantees that under under modular transformations the states in ${\mathcal H}_\text{R-R}$ are mapped into themselves.

Note that the pair of lines $a$ and $a\times\psi$ in the bosonic parent descend to a pair of lines in the spin TQFT, because these are distinct anyons, being distinguishable by their spin. On the other hand, the pair of states $|a\rangle$ and $|a\times\psi\rangle$ descend to a \emph{single} state in the spin TQFT. Thus, while in a bosonic TQFT the number of states is the same as the number of lines, in a spin TQFT there are twice as many lines as there are states. 

Our next task is to compute the action of fermion parity, i.e.~$(-1)^F$, on the Hilbert space of the spin TQFT. The $\mathbb Z_2$ symmetry generated by $(-1)^F$ is the emergent zero-form symmetry that appears upon quotienting the parent bosonic theory by $\mathbb Z_2^\psi$ in~\eqref{spinquotient}. The charged states are therefore those constructed from the once-punctured torus in the bosonic theory 
\begin{equation}
\begin{aligned}
(-1)^F|m\rangle_\text{spin}&=-|m\rangle_\text{spin}\,,\\
(-1)^F|\,a\,\rangle_\text{spin}&=+|\,a\,\rangle_\text{spin}\,,\\
(-1)^F|\,x\,\rangle_\text{spin}&=+|\,x\,\rangle_\text{spin}\,.
\end{aligned}
\label{actionfermion}
\end{equation}
 $(-1)^F$ acts nontrivially on the $\psi$ puncture in the once-punctured torus.

Depending on the choice of spin structure on the ``time" circle we can define the following $2^3=8$ partition functions for spin TQFTs:\footnote{$-$ represents antiperiodic boundary condition while $+$ period boundary conditions.}
\begin{equation}
\begin{aligned}
\tr_{-,-}(\mathcal O)&\equiv \tr_{ {\mathcal H}_\text{NS-NS}}(\mathcal O)\,,\label{firstone}\\
\tr_{-,+}(\mathcal O)&\equiv \tr_{ {\mathcal H}_\text{NS-R}}(\mathcal O)\,,\\
\tr_{+,-}(\mathcal O)&\equiv \tr_{ {\mathcal H}_\text{R-NS}}(\mathcal O)\,,\\
\tr_{+,+}(\mathcal O)&\equiv \tr_{ {\mathcal H}_\text{R-R}}(\mathcal O)\,,
\end{aligned}
\end{equation}
and
\begin{equation}
\begin{aligned}
\tr_{-,-}((-1)^F\mathcal O)&\equiv \tr_{ {\mathcal H}_\text{NS-NS}}((-1)^F\mathcal O)\,,\\
\tr_{-,+}((-1)^F\mathcal O)&\equiv \tr_{ {\mathcal H}_\text{NS-R}}((-1)^F\mathcal O)\,,\\
\tr_{+,-}((-1)^F\mathcal O)&\equiv \tr_{ {\mathcal H}_\text{R-NS}}((-1)^F\mathcal O)\,,\\
\tr_{+,+}((-1)^F\mathcal O)&\equiv \tr_{ {\mathcal H}_\text{R-R}}((-1)^F\mathcal O)\label{eq:oddinsert}\,,
\end{aligned}
\end{equation}
where $\mathcal O$ is an operator in the theory. We will be interested in the case when $\mathcal O$ is a symmetry of the TQFT. We note that $(-1)^F$ is only non-trivial in the R-R sector, because this is the only Hilbert space that may contain Majorana states. This is the most subtle and rich sector, and the one of interest as far as the twisted Witten indices is concerned.
%All Hilbert spaces have the same dimension, and so $\tr_{\sigma,\sigma'}(\boldsymbol1)=\dim(\mathcal H_F)$ for all $(\sigma,\sigma')$. Similarly, note that $(-1)^F$ is trivial in all spaces with even spin structure (because there are no punctures, so everything is bosonic), and so $ \tr_{\sigma,\sigma'}(-1)^F=\dim(\mathcal H_F)$ for all but $(\sigma,\sigma')=(+,+)$. Thus, there are really only two independent partition functions.

\subsection{Partition Function of Spin TQFTs}\label{sec:Z_spin_TQFT}

The twisted Witten index of the domain wall theory is computed by considering the odd spin structure on the spatial torus and periodic boundary condition on the time circle. This implies that the twisted Witten indices in the ultraviolet must be reproduced by appropriate odd spin structure partition functions of our conjectured infrared spin TQFTs. In other words, a nontrivial check that our proposed infrared spin TQFTs describe the $n$-domain wall theories is proving that
\begin{equation}\label{eq:oddinsertb}
 I^{\mathsf s}_n\equiv \tr_{ {\mathcal H}_\text{R-R}}(-1)^F\mathsf s
\end{equation}
for symmetries $\mathsf s\in\mathsf S$. This requires, in particular, identifying the image of the symmetries $\mathsf s\in\mathsf S$ in the infrared TQFT.

Let us begin by considering the untwisted partition function. Given the construction of the Hilbert space ${\mathcal H}_\text{R-R}$ in~\eqref{Ramondstates} and the action of $(-1)^F$ in~\eqref{actionfermion} we can compute the desired partition function as follows
\begin{equation}
\tr_{{\mathcal H}_\text{R-R}}(-1)^F=N_{x}-N_{m}\,.
\label{indexdiff}
\end{equation}
This requires determining in the bosonic parent theory the number $N_x$ of two-dimensional orbits and the number $N_m$ of one-dimensional orbits (the number of Majorana lines) in ${\mathcal A}_\text{R}$ (see~\eqref{orbits}) under fusion with $\psi$.

Let us illustrate this in a simple example. The simplest spin TQFT is $\SO(N)_1$ Chern-Simons theory, which is is a trivial, invertible spin TQFT with lines $\{\boldsymbol 1,\psi\}$. The bosonic parent theory is $\Spin(N)_1$ Chern-Simons theory:
\begin{itemize}
\item For $N$ odd, $\Spin(N)_1$ is the Ising category, which has three lines $\{\boldsymbol 1,\sigma,\psi\}$: the vacuum $\boldsymbol 1$, the spin operator $\sigma$, and the energy operator $\psi$. These primaries have spins $h=0,\frac{N}{16},\frac12$, and fusion rules $\sigma^2=\boldsymbol1+\psi$, $\psi^2=\boldsymbol 1$, $\psi\times\sigma=\sigma$.

\item For $N$ even, $\Spin(N)_1$ has four lines, with spins $h=0,\frac{N}{16},\frac{N}{16},\frac12$, and which we denote by $\{\boldsymbol1,\mathsf e,\mathsf m,\psi\}$, which correspond to the trivial representation, the two fundamental spinor representations, and the vector representation, respectively. The theory for $N\equiv0\mod4$ has $\mathbb Z_2^2$ fusion rules, with $\mathsf e^2=\mathsf m^2=\psi^2=\boldsymbol 1$ and $\mathsf e\times\mathsf m=\psi$. For $N\equiv2\mod4$ the fusion ring is $\mathbb Z_4$, with $\mathsf e^2=\mathsf m^2=\psi$, and $\psi^2=\mathsf e\times\mathsf m=\boldsymbol 1$.
\end{itemize}
The theory $\SO(N)_1$ is obtained by condensing $\psi$, that is $\SO(N)_1=\Spin(N)_1/\mathbb Z_2^{\psi}$.
Using the fusion rules and the spins we see that $\boldsymbol1, \psi$ are neutral under $\mathbb Z_2^{\psi}$ and $\mathsf e,\mathsf m$ and $\sigma$ are charged. In other words, the Neveu-Schwarz sector is 
\begin{equation}
\mathcal A_\text{NS}=\{\boldsymbol1,\psi\}\,,
\end{equation}
while the Ramond sector is 
\begin{equation}
\mathcal A_\text{R}=\begin{cases}
\{\sigma\} & \text{$N$ odd,}\\
\{\mathsf e,\mathsf m\} & \text{$N$ even.}
\end{cases}
\end{equation}
This implies that $N_a=1$ in the NS sector. On the other hand, in the R sector, $(N_x,N_m)=(0,1)$ for odd $N$ and $(N_x,N_m)=(1,0)$ for even $N$. Thus there is a unique state in each spin structure, and all states are bosonic except in ${\mathcal H}_\text{R-R}$, where $(-1)^F=(-1)^N$, since there is a Majorana line for odd $N$. The $(-1)^F$ odd state is created by the well-known once-punctured torus conformal block in the Ising category with the insertion of $\sigma$, and $\psi$ at the puncture.

Let us now consider the spin TQFT $\SO(3)_3$ Chern-Simons theory, which is the simplest non-trivial spin TQFT. The bosonic parent theory is $\SU(2)_6$ since 
 \begin{equation}
\SO(3)_3=\frac{\SU(2)_6}{\mathbb Z_2^\psi}\,,
\end{equation}
where the abelian line $\psi$ is the line in $\SU(2)_6$ with $j=3$ and spin $h=3/2$. The lines in ${\mathcal A}=\{j=0,\tfrac12,1,\dots,3\}$ which have braiding $-1$ with $\psi$ are those with half-integral isospin: $\mathcal A_\text{R}=\{j=\frac12,\frac32,\frac52\}$. Under fusion with $\psi$ we have the following $\mathbb Z_2^\psi$ orbits
\begin{equation}
\begin{aligned}
3\times \tfrac12&=\tfrac52\\
3\times \tfrac32&=\tfrac32\,.
\end{aligned}
\end{equation}
Therefore, in the R-R sector of $\SO(3)_3$ there is a length-2 orbit with $(j=\tfrac12,\tfrac52)$ and a Majorana line with $j= \tfrac32$. Thus, $N_x=N_m=1$. There are $N_x+N_m=2$ states, but one of them is a boson and the other is a fermion, which means that the partition function with periodic boundary conditions vanishes
 \begin{equation}
\tr_{ {\mathcal H}_\text{R-R}}(-1)^F=N_x-N_m\equiv0\,.
\end{equation}
The vanishing of this trace will be important when discussing the $2$-domain wall theory in $4d$ ${\mathcal N}=1$ SYM with gauge group $G_2$ (cf. section~\ref{sec:G2_TQFT}). This example clearly illustrates the importance of looking at the appropriate partition function and not merely at the dimension of the Hilbert space.

%Finally, let us consider the case where the spin TQFT factorizes into a bosonic theory times a trivial spin TQFT
%\begin{equation}
%G_F=\tilde G\times \SO(N)_1\,,
%\end{equation}
%where $\tilde G$ is a purely bosonic TQFT, and $\SO(N)_1=\{\boldsymbol 1,\psi\}$. As the Hilbert space is a tensor product, we find
%\begin{equation}\label{eq:index_factor_bos}
%\tr_{G_F}(-1)^F=\dim({\mathcal H}_{\tilde G})\times \tr_{\SO(N)_1}(-1)^F\equiv (-1)^N\dim({\mathcal H}_{\tilde G})\,,
%\end{equation}
%where we have used the fact that $(-1)^F$ is trivial in the bosonic theory $\tilde G$, while $(-1)^F=(-1)^N$ in the R-R sector of $\SO(N)_1$. This means that the partition function of a spin TQFT that factorizes as above is -- up to a sign -- just the dimension of the Hilbert space. 
%This explains the observation in Acharya-Vafa (and repeated many times in the literature) that the domain wall Witten index for $4d$ $\mathcal N=1$ $\SU(N)$ matches the number of states of $\U(n)$ Chern-Simons theory, even if the RR Hilbert space of the latter was not properly understood. 

We now discuss a different way to compute the partition function that does not rely on computing $N_x$ and $N_m$ directly. The basic idea is to gauge the emergent zero-form $\mathbb Z_2$ symmetry generated by $(-1)^F$ in the spin TQFT to obtain the bosonic parent theory back~\cite{Gaiotto:2015zta,Bhardwaj:2016clt,Kapustin:2017jrc,Thorngren:2018bhj,Hsin:2019gvb}
\begin{equation}\label{spinquotientga}
\frac{\text{spin TQFT}}{\mathbb Z_2}={\text{bosonic TQFT}}\,.
\end{equation}
Gauging this $\mathbb Z_2$ amounts to summing the spin TQFT over all spin structures of the three-manifold $M$. Taking 
$M$ to the three-torus, and summing over the $2^3=8$ spin structures, corresponding to either periodic or antiperiodic boundary conditions around each of the three circles, we find that 
\begin{equation}
\frac12\sum_{\pm,\pm}(\tr_{\pm,\pm}(\boldsymbol1)+\tr_{\pm,\pm}(-1)^F)=\tr_{\mathcal H_B}(\boldsymbol1)\,.
\end{equation}
 $\tr_{\pm,\pm}$ (see~\eqref{firstone}--\eqref{eq:oddinsert}) denotes the trace over the Hilbert space on the spatial torus with boundary conditions $\pm,\pm$, and $\tr_{\mathcal H_B}$ the trace over the torus Hilbert space of the bosonic parent theory.

Using the fact that the dimension of the torus Hilbert space is the same in all spin structures and that $(-1)^F$ acts nontrivially only in ${\mathcal H}_\text{R-R}$ (see~\eqref{actionfermion}), we find the formula 
\begin{equation}\label{eq:index_from_dims}
\tr_{ {\mathcal H}_\text{R-R}}(-1)^F=2\dim(\mathcal H_B)-7\dim(\mathcal H_F)\,,
\end{equation}
where $\dim(\mathcal H_B)$ is the dimension of the torus Hilbert space of the bosonic parent TQFT and $\dim(\mathcal H_F)$ the dimension of the torus Hilbert space in any one spin structure of the spin TQFT. This formula offers a significant advantage in that it requires computing the total number of states $\dim(\mathcal H_F)=N_x+N_m$, and not separately $N_x$ and $N_m$, as in formula~\eqref{indexdiff}. Even simpler, one may compute $\dim(\mathcal H_F)=N_a$ in the NS-NS sector directly, where all orbits are of length-2: the number of states is just half the number of lines of the spin TQFT.\footnote{In the $\SO(3)_3$ example $\dim(\mathcal H_B)=7$ and $\dim(\mathcal H_F)=2$, and using~\eqref{eq:index_from_dims} the partition function indeed vanishes.} 
 
As a consistency check, consider the case where $G_F$ is the product of a bosonic theory $\tilde G$ times a trivial/invertible spin TQFT 
\begin{equation}
G_F=\tilde G\times \SO(N)_1\,,
\end{equation}
whose bosonic parent is $G_B=\tilde G\times\Spin(N)_1$. As $\SO(N)_1$ is a trivial spin TQFT, we get $\dim(\mathcal H_F)=\dim(\mathcal H_{\tilde G})$. 
%We now compute $\dim(\mathcal H_B)$. As discussed the bosonic parent TQFT of this spin TQFT is $G_B=\tilde G\times \Spin(N)_1$. The $\mathbb Z_2^\psi$ gauging affects the second factor only, and consists of condensing the fermion in $\Spin(N)_1$, which indeed yields $\Spin(N)_1/\mathbb Z_2^\psi=\SO(N)_1$, a trivial spin TQFT. 
Similarly, using that $\Spin(N)_1$ has a four-dimensional Hilbert space if $N$ is even, and a three dimensional Hilbert space if $N$ is odd, we get $\dim(\mathcal H_B)=\frac12(7+(-1)^N)\dim(\mathcal H_{\tilde G})$. Plugging this into~\eqref{eq:index_from_dims}, we get
\begin{equation}\label{eq:index_from_dimsimple}
\tr_{ {\mathcal H}_\text{R-R}^{G_F}}(-1)^F=(-1)^N\dim(\mathcal H_{\tilde G})\,,
\end{equation}
which is precisely what one would expect, given the tensor product structure of $G_F$ and the fact that the trace over $\SO(N)_1$ is $(-1)^N$. Put differently, in the Hilbert space of $G_F=\tilde G\times \SO(N)_1$ we have $N_a=N_x$ and $N_m=0$ for $N$ even, and $N_a=N_m$ and $N_x=0$ for $N$ odd. That is, in the R-R sector, either no states are Majorana or all are, 
depending on the parity of $N$. This implies that 
\begin{equation}\label{eq:index_from_dimsimplea}
\tr_{ {\mathcal H}_\text{R-R}^{G_F}}(-1)^F=\begin{cases} +N_a&\text{ $N$ even}\\
 -N_a&\text{ $N$ odd}\,,
\end{cases}
\end{equation}
which indeed equals~\eqref{eq:index_from_dimsimple}. 

There are spin TQFTs which factorize in a nontrivial fashion into the product of a bosonic TQFT and a trivial spin TQFT by virtue of a level-rank duality, as for example $\U(1)_k\leftrightarrow \SU(k)_{-1}\times \{\boldsymbol 1,\psi\}$. In these theories $\tr_{ {\mathcal H}_\text{R-R}}(-1)^F$ also just measures the dimension of the Hilbert space (up to possibly a sign).

%precisely as in~\eqref{eq:index_factor_bos}.

%The bosonic parent of this theory can be taken as, for example, $G_B=\tilde G\times\mathbb Z_2$, where $\mathbb Z_2$ denotes the toric code. The $\mathbb Z_2^\psi$ gauging affects the second factor only, and consists of condensing the electric line $\psi=\mathsf e$, which indeed yields $\mathbb Z_2/\mathbb Z_2^\psi=\{\boldsymbol1,\psi\}$, a trivial spin TQFT. (There are other choices beyond the toric code, but this one has the advantage of having vanishing central charge, so we do not change the fermion parity of the quotient). Noting that the toric code has a four-dimensional Hilbert space, we get $\dim(\mathcal H_B)=4\dim(\mathcal H_{\tilde G})$.

%Finally, using~\eqref{eq:index_from_dims}, we get
%\begin{equation}\label{eq:index_from_dimsimple}
%\tr_{ {\mathcal H}_\text{R-R}}(-1)^F= \dim(\mathcal H_F)=\dim(\mathcal H_{\tilde G})\,.
%\end{equation}
%which is precisely what one would expect for a theory that has no Majorana lines.

%As the toric code is abelian, there are no Majorana lines and all states have bosonic parity. This means that $\dim(\mathcal H_B)=4\dim(\mathcal H_{\tilde G})=\dim(\mathcal H_F)$. The partition function of a factorizable spin TQFT is, therefore, the dimension of the Hilbert space 

%as expected for a theory with trivial $(-1)^F$. When the spin TQFT does not factorize (as in many of the cases we will consider), the fermion parity operator will in general be non-trivial, and the computation of the partition function is more subtle.
 
More generally, stacking a TQFT, spin or bosonic, with a trivial spin TQFT defines 
\begin{equation}
\text{TQFT}\times \SO(N)_1\,.
\end{equation}
This theory has the same number of states as the original TQFT but $\SO(N)_1$ can change the global sign of the action of $(-1)^F$ on all states in ${\mathcal H}_\text{R-R}$. The partition functions are the same up to possibly a sign 
\begin{equation}\label{eq:inv_phase_sign}
\tr_{ {\mathcal H}^{\text{TQFT}\times \SO(N)_1}_\text{R-R}}(-1)^F=(-1)^N \tr_{ {\mathcal H}^{\text{TQFT}}_\text{R-R}}(-1)^F\,.
\end{equation}
Indeed, the single state of $\SO(N)_1$ in ${\mathcal H}_\text{R-R}$ is a Majorana state and thus has odd fermion parity for $N$ odd only. Therefore, when comparing the spin TQFT partition function with the Witten index of the domain wall, we will match their absolute values, as if those match, the signs can be also be matched by stacking a suitable trivial spin TQFT, which can be thought of as a purely gravitational counterterm~\cite{10.1093/ptep/ptw083}.\footnote{Staking $\SO(N)_1$ for odd $N$ to a $3d$ theory has the same effect as stacking to a $2d$ theory the trivial spin TQFT known as the Arf-invariant, which changes the sign of the partition with odd spin structure.}

The generalization to twisted indices is straightforward. Given a symmetry $\mathsf s \in \mathsf S_{\text{TQFT}}$ of the TQFT, which acts $\mathsf s\colon\mathcal H_\text{R-R}\to\mathcal H_\text{R-R}$, the partition function
\begin{equation} 
\tr_{\mathcal H_\text{R-R}}((-1)^F\mathsf s)
\end{equation}
 counts the number of bosons fixed by $\mathsf s$, minus the number of fermions fixed by $\mathsf s$. That being said, there are some subtleties that must be kept in mind. A state fixed by $\mathsf s$ does not necessarily contribute with $\mathsf s=+1$ to the trace -- it might contribute with $\mathsf s=-1$ instead, the reason being that the symmetry $\mathsf s$ might be realized projectively in the Hilbert space.

The most common example where this may happen is charge-conjugation $\mathsf c$. We can illustrate this in $\U(1)_1$ Chern-Simons theory, the simplest theory where this phenomenon occurs. This is an invertible spin TQFT, which means that it has a unique state on any spin structure. This state is clearly fixed by $\mathsf c$ but, interestingly, it has $\mathsf c=-1$ in the odd-spin-structure Hilbert space $\mathcal H_\text{R-R}$. We can show this as follows. The bosonic parent theory is $\U(1)_4$ Chern-Simons theory, which has four states, labeled by $q=0,1,2,3$. The $\U(1)_1$ theory is obtained by condensing the fermion $\psi$, which has $q=2$. The Ramond lines are easily checked to be $q=1,3$, and they are paired by fusion with $\psi$ into a single two-dimensional orbit, since $1\times 2=3$. Thus, the unique state in the R-R sector is~(cf.~\eqref{Ramondstates})
\begin{equation}\label{eq:c_eq_-1}
|1\rangle_\text{spin}=\frac{1}{\sqrt2}(|1\rangle-|3\rangle)\,.
\end{equation}
This indeed satisfies $\mathsf c|1\rangle_\text{spin}=-|1\rangle_\text{spin}$, inasmuch as $\mathsf c\colon q\mapsto-q\mod 4$ in the bosonic parent, which exchanges $|1\rangle$ and $|3\rangle$.

\section{Domain Wall TQFT Partition Functions}\label{sec:IR_index}

In this section we calculate partition functions twisted by a symmetry $\mathsf s\in \mathsf S_{\text{TQFT}}$ \begin{equation}\label{eq:oddinsertaaaaa}
 \tr_{ {\mathcal H}_\text{R-R}}(-1)^F\mathsf s \,
\end{equation}
of the Chern-Simons TQFTs we proposed emerge in the infrared of the domain wall theories (see section~\ref{sec:intro}). Our calculations beautifully reproduce the results obtained in the ultraviolet of the domain wall theories in section~\ref{sec:UV} 
\begin{equation}\label{eq:oddinserta}
 \tr_{ {\mathcal H}_\text{R-R}}(-1)^F\mathsf s=I^{\mathsf s}_n\,,
\end{equation}
where $I^{\mathsf s}_n$ is the twisted Witten index on the $n$-domain wall $\mathrm W_n$ (cf.~\eqref{eq:def_twist_s_n}). 
We identify each symmetry $\mathsf s\in \mathsf S$ in $4d$ ${\mathcal N}=1$ SYM with a symmetry $\mathsf s\in \mathsf S_{\text{TQFT}}$ in the infrared TQFT.

In $4d$ ${\mathcal N}=1$ SYM with gauge group $\Sp(N)$ our proposed domain wall theory corresponds to a Chern-Simons theory based on a  group that is simple, connected, and simply-connected, whereas for SYM with $\SU(N)$, $\Spin(N)$ and $G_2$ gauge groups, the proposed infrared Chern-Simons theories are based on a group that is neither. We discuss both cases in turn.

Chern-Simons theory $G_k$, with $G$ simple, connected and simply-connected is always a bosonic TQFT. These theories are made spin by tensoring with the trivial spin TQFT $\SO(N)_1$. It follows from our discussion in section~\ref{sec:spin} that 
\begin{equation}
\tr_{ {\mathcal H}^{G_k\times \SO(N)_1}_\text{R-R}}(-1)^F=(-1)^N \tr_{{G_k} }(\boldsymbol1) \,,
\label{tracer}
\end{equation}
since all states have the same fermion parity -- all bosonic, or all fermionic, depending on the parity of $N$. Therefore the partition function of $G_k\times \SO(N)_1$ in~\eqref{tracer} is, up to possibly a sign, the dimension of the Hilbert space of $G_k$ Chern-Simons theory on the two-torus.

The states in the torus Hilbert space of $G_k$ Chern-Simons theory are conformal blocks on the torus, which are labeled by the integrable representations of the corresponding affine lie algebra $\mathfrak g^{(1)}$ at level $k$~\cite{Witten:1988hf,ELITZUR1989108}. By definition, the representations of $G$ that are integrable are those whose highest weight $\lambda$ satisfies $(\lambda,\theta)\le k$, with $\theta$ the highest root of $G$. Expanding the latter in a basis of simple coroots, and introducing an extended label $\lambda_0:=k-(\lambda,\theta)$, integrability can be expressed as
\begin{equation}\label{eq:boson_int_reps}
\sum_{i=0}^r \lambda_i a_i^\vee=k,\qquad \lambda_i\in\mathbb Z_{\ge0}\,.
\end{equation}
The dimension of the Hilbert space $\tr_{G_k}(\boldsymbol1)$ is equal to the number of solutions to this equation.

Much like the discussion in section~\ref{sec:UV}, where the Witten index on the domain wall was computed through an auxiliary system of free fermions, $\tr_{G_k}(\boldsymbol1)$ has a nice combinatorial interpretation in terms of a system of free bosons in $0+1$ dimensions. Indeed, the number of integrable representations $\tr_{G_k}(\boldsymbol1)$ is the number of ways of creating a state of energy $k$ from $r+1$ free bosons, each with energy $a_i^\vee$. Each boson is associated with a node in the extended Dynkin diagram $\mathfrak g^{(1)}$, and $\lambda_i\in\{0,1,2,\dots\}$ in~\eqref{eq:boson_int_reps} corresponds to the occupation number of the $i$-th boson. Introducing a fugacity parameter $q$ defines a generating function, which is the partition function of the bosons on the circle
\begin{equation}
Z(G,q)\equiv\sum_{k\ge0}\tr_{G_k}(\boldsymbol1)\,q^k\,.
\end{equation}
The partition function is thus
\begin{equation}\label{eq:Z_bos_twista}
\begin{aligned}
Z(G,q)=\prod_{i=0}^r(1- q^{a^\vee_i})^{-1}\,.
\end{aligned}
\end{equation}
The Chern-Simons trace $\tr_{G_k}(\boldsymbol1)$ is the coefficient of $q^k$ in~\eqref{eq:Z_bos_twista}.

In a similar fashion, we define the trace twisted by a symmetry $\mathsf s\in \mathsf S_{\text{TQFT}}$ of $G_k$ Chern-Simons theory:
\begin{equation}
\tr_{{G_k}}(\mathsf s)\,.
\end{equation}
When $\mathsf s=\mathsf c$ is   a zero-form symmetry, this corresponds to inserting a surface operator, i.e., the symmetry defect is supported on the whole spatial torus. On the other hand, if $\mathsf s=\mathsf g$ denotes a one-form symmetry, the symmetry defect is a line operator, and one must specify a homology cycle on the torus on which it is is supported. The states of $G_k$ Chern-Simons are created by wrapping on a   cycle Wilson lines labeled by integrable representations $\lambda$; if $\mathsf g$ is supported on the same cycle, it acts on the states via fusion:
\begin{equation}\label{eq:CS_one_b}
\begin{tikzpicture}[baseline=0]

\draw[thick] (0,0) ellipse (2cm and 1cm);
\draw[thick,domain=180:360,smooth,variable=\x] plot ({.9*cos(\x)},{.2+.4*sin(\x)});
\draw[thick,domain=23:157,smooth,variable=\x] plot ({.9*cos(\x)},{-.1+.4*sin(\x)});
\draw[thick,blue] (0,.1) ellipse (1.4cm and .6cm);

\node at (-1.2,.1) {\footnotesize$\color{blue}{\lambda}$};

\draw[thick,red] (0,.05) ellipse (1.7cm and .8cm);
\node at (-1.85,.05) {\footnotesize$\color{red}{\mathsf g}$};

%\begin{scope}[rotate=-60,shift={(.75,.2)}]
%\draw[thick,red,domain=10:150,smooth,variable=\x] plot ({.45*cos(\x)},{.3*sin(\x)});
%\draw[thick,red,dashed,domain=180:320,smooth,variable=\x] plot ({.05+.45*cos(\x)},{.1+.3*sin(\x)});
%\end{scope}
%\node at (1.1,-.65) {\footnotesize$\color{red}{\alpha}$};

\node at (3.3,0) {$=$};

\begin{scope}[shift={(7.6,0)}]
\draw[thick] (0,0) ellipse (2cm and 1cm);
\draw[thick,domain=180:360,smooth,variable=\x] plot ({.9*cos(\x)},{.2+.4*sin(\x)});
\draw[thick,domain=23:157,smooth,variable=\x] plot ({.9*cos(\x)},{-.1+.4*sin(\x)});
\draw[thick,blue] (0,.1) ellipse (1.4cm and .6cm);2
\node at (0,-.75) {\footnotesize$\color{blue}{\mathsf g\times\lambda}$};
\end{scope}

\end{tikzpicture}
\end{equation}
Conversely, if $\mathsf g$ is supported on the dual cycle, it acts on the states via braiding:
\begin{equation}\label{eq:CS_one_a}
\begin{tikzpicture}[baseline=0]

\draw[thick] (0,0) ellipse (2cm and 1cm);
\draw[thick,domain=180:360,smooth,variable=\x] plot ({.9*cos(\x)},{.2+.4*sin(\x)});
\draw[thick,domain=23:157,smooth,variable=\x] plot ({.9*cos(\x)},{-.1+.4*sin(\x)});
\draw[thick,blue] (0,.1) ellipse (1.4cm and .6cm);
\filldraw[white] (.8,-.39) circle (1.2pt);

\node at (-1.2,.1) {\footnotesize$\color{blue}{\lambda}$};

\begin{scope}[rotate=-60,shift={(.75,.2)}]
\draw[thick,red,domain=10:150,smooth,variable=\x] plot ({.45*cos(\x)},{.3*sin(\x)});
\draw[thick,red,dashed,domain=180:320,smooth,variable=\x] plot ({.05+.45*cos(\x)},{.1+.3*sin(\x)});
\end{scope}
\node at (1.1,-.65) {\footnotesize$\color{red}{\mathsf g}$};

\node at (4,0) {$=\quad \alpha_{{\color{red}\mathsf g}}({\color{blue}\lambda})$};

\begin{scope}[shift={(7.5,0)}]
\draw[thick] (0,0) ellipse (2cm and 1cm);
\draw[thick,domain=180:360,smooth,variable=\x] plot ({.9*cos(\x)},{.2+.4*sin(\x)});
\draw[thick,domain=23:157,smooth,variable=\x] plot ({.9*cos(\x)},{-.1+.4*sin(\x)});
\draw[thick,blue] (0,.1) ellipse (1.4cm and .6cm);2
\node at (0,-.75) {\footnotesize$\color{blue}{\lambda}$};
\end{scope}

\end{tikzpicture}
\end{equation}
where $\alpha_{\mathsf g}(\lambda)$ is the charge of $\lambda$ under the center of $G$ (cf.~\eqref{eq:one_form_act_W}). More generally, one can wrap a pair of symmetry defects on both cycles, but one can always conjugate such configuration via a modular transformation to either of the two options above. This operation, being a similarity transformation, does not affect the value of the trace. In other words, the value of $\tr_{G_k}(\mathsf g)$ is independent of which cycle we define $\mathsf g$ on. 

When $\mathsf s$ is a symmetry of the classical action of $G_k$ Chern-Simons theory, it is induced by an outer automorphism of the extended Dynkin diagram $\mathfrak g^{(1)}$, and it acts as a permutation of the nodes thereof. In that case, $\mathsf s$ induces an action on the system of bosons, which permutes them in the same way it permutes the nodes of the Dynkin diagram. As in the system of free fermions, the trace above can be obtained from the partition function of these bosons 
\begin{equation}
Z^{\mathsf s}(G,q)\equiv \sum_{k\ge0}\tr_{{G_k}}(\mathsf s)q^k \,,
\end{equation}
where $Z^{\mathsf s}(G,q)$ denotes the bosonic partition function twisted by the permutation $\mathsf s$. One can evaluate this partition function by the same methods as in section~\ref{sec:UV}, i.e., by diagram folding or directly in the diagonal basis
\begin{equation}\label{eq:Z_bos_twist}
\begin{aligned}
Z^{\mathsf s}(G,q) 
%&=\prod_{i=0}^r\sum_{\lambda_i\ge0}q^{a^\vee_i\lambda_i}\mathsf s_i\\
=\prod_{i=0}^r(1-\mathsf s_iq^{a^\vee_i})^{-1}\,,
\end{aligned}
\end{equation}
where $\mathsf s_i$ are the eigenvalues of the permutation. Note that, for $\mathsf s=\mathsf g$ a one-form symmetry, diagram folding naturally corresponds to $\mathsf g$ acting as a permutation, i.e.~\eqref{eq:CS_one_b}, while the diagonal action corresponds to $\mathsf g$ acting via braiding, i.e.~\eqref{eq:CS_one_a}. Indeed, it is a well-known fact that an $S$ modular transformation -- which     interchanges the two cycles -- diagonalizes the fusion rules.  

It should be noted that Chern-Simons theories can have ``quantum symmetries". These are symmetries of the entire TQFT data that are not symmetries of the Lagrangian. Many explicit examples of these symmetries have been found in~\cite{Delmastro:2019vnj}. These symmetries permute the Wilson lines of the theory, in a way that does not necessarily correspond to a permutation of their Dynkin labels. As such, the free boson representation cannot be used to evaluate the twisted trace, but it must be computed from the action of the symmetry on the Hilbert space of the TQFT. That being said, we find that the symmetries $\mathsf S$ in the ultraviolet domain wall map to classical symmetries of the infrared Chern-Simons theories, and we can compute the twisted index using~\eqref{eq:Z_bos_twist}.
 
Some of our proposed domain wall TQFTs are Chern-Simons theory with a group $G$ that is not connected and/or simply-connected, in which case the theory $G_k$ (with $k$ a set of integers that defines the Chern-Simons action) may depend on the spin structure of the underlying manifold. There are four distinct Hilbert spaces corresponding to the four spin structures on the spatial torus (see section~\ref{sec:spin}), but our interest here is in the Hilbert space ${\mathcal H}_\text{R-R}$. Now $G_k$ can have fermionic states, which correspond to once-punctured conformal blocks of the parent bosonic theory, and $(-1)^F$ is in general a non-trivial operator. Our goal is to compute 
\begin{equation}
\tr_{{\mathcal H}^{G_k}_\text{R-R}}((-1)^F\mathsf s)\equiv \tr_{G_k}((-1)^F\mathsf s)\,,
\label{twisttracenn}
\end{equation}
where we use the latter to simplify notation.

We shall next compute the twisted traces for all the Chern-Simons theories of interest. We begin by considering the simply-connected group $\Sp(n)$, and then we move on to the more subtle and interesting cases $\U(n),\O(n)$. We finally make a few remarks concerning the exceptional groups. The remaining simply-connected groups $\SU(n),\Spin(n)$ as well as $\SO(n)$ are studied in appendix~\ref{sec:app_su_spin}.

\subsection[$G=\Sp(n)$]{$\boldsymbol{G=\Sp(n)}$}

The $n$-domain wall theory for $4d$ ${\mathcal N}=1$ SYM with $G=\Sp(N)$ is proposed to be $\Sp(n)_{N+1-n}$ Chern-Simons theory. Let us proceed to study the partition functions of $\Sp(n)_k$. 

Consider the algebra $C_n=\mathfrak{sp}_n$. The comarks are all $a_i^\vee=1$. Plugging this into~\eqref{eq:Z_bos_twist} we obtain the generating function as
\begin{equation}\label{eq:Z_Sp_IR}
Z(\Sp(n),q)=(1-q)^{-(n+1)}\,,
\end{equation}
and, by expanding, the untwisted trace
\begin{equation}\label{eq:I_Sp_IR}
\tr_{\Sp(n)_k}(\boldsymbol 1)=\binom{n+k}{k}\,.
\end{equation}
This is the number of integrable representations of $\Sp(n)_k$, that is, the dimension of the torus Hilbert space of this Chern-Simons theory.

Let us also compute the partition function twisted by the one-form symmetry $\Gamma=\mathbb Z_2$. This symmetry reverses the order of the extended labels, and the charged representations are the pseudo-real ones. Denoting by $\mathsf g$ the non-trivial element of $\mathbb Z_2$, and using~\eqref{eq:Z_bos_twist} and~\eqref{eq:twist_C_series}, we get the twisted partition function
\begin{equation}
Z^{\mathsf g}(\Sp(n),q) =(1-q)^{-\lfloor n/2\rfloor-1}(1+q)^{-\lceil n/2\rceil}\,, 
\end{equation}
and by expanding 
\begin{equation}\label{eq:I_Sp_tw_IR}
\begin{aligned}
%Z(\Sp(N)_k)&=\sum_{j=0}^k (-1)^{j \mathsf g} \binom{\lceil N/2\rceil+j -1}{j} \binom{\lfloor N/2\rfloor+k-j}{k-j}\,.
\tr_{\Sp(n)_k}(\mathsf g)&=\begin{cases}
\,\displaystyle0&\text{$n$ odd, $k$ odd,}\\[+15pt] 
\,\displaystyle\binom{(n+k-1)/2}{k/2} &\text{$n$ odd, $k$ even,}\\[+15pt] 
\,\displaystyle\binom{(n+k-1)/2}{(k-1)/2} &\text{$n$ even, $k$ odd,}\\[+15pt] 
\,\displaystyle\binom{(n+k)/2}{k/2} &\text{$n$ even, $k$ even.}
\end{cases}
\end{aligned}
\end{equation}
Note that $Z^{\mathsf g}$ is nothing but the untwisted partition function associated to the Dynkin diagram given by folding the original diagram by the one-form symmetry~\eqref{eq_fig:fold_C}.

We are now ready to test our proposal. Recall that the conjectured infrared theory corresponding to the $n$-domain wall of $\Sp(N)$ SYM was $\mathrm W_n=\Sp(n)_k$, with $k=N+1-n$. Using this value of the level in~\eqref{eq:I_Sp_IR} and~\eqref{eq:I_Sp_tw_IR} indeed reproduces the (twisted) Witten indices computed in the ultraviolet, cf.~\eqref{eq:I_Sp_UV} and~\eqref{eq:I_Sp_tw_UV}.

\subsection[$G=\U(n)$]{$\boldsymbol{G=\U(n)}$}\label{sec:unitary_chern_simons}

The $n$-domain wall theory for $4d$ ${\mathcal N}=1$ SYM with $G=\SU(N)$ is proposed to be $\U(n)_{N-n,N}$ Chern-Simons theory. Let us proceed to study the partition functions of $\U(n)_{k,n+k}$. 

The Chern-Simons gauge group is not simply connected. The theory is defined as
\begin{equation}\label{eq:spin_TQFT_ZN}
\U(n)_{k,n+k}:=\frac{\SU(n)_k\times \U(1)_{n(n+k)}}{\mathbb Z_n}\,,
\end{equation}
where $\mathbb Z_n$ is the one-form symmetry generated by the line $\psi=[0,k,0,\dots,0]\otimes (n+k)$. Here and in what follows, $[\lambda_0,\lambda_1,\dots,\lambda_{n-1}]$ denotes the Dynkin labels of an $\SU(n)_k$ representation, and $(q)\in \mathbb Z$ the charge of a $\U(1)_{n(n+k)}$ representation. The spin of the generator is easily computed to be $h_\psi=\frac{k(n-1)}{2n}+\frac{n+k}{2n}=\frac{k+1}{2}$. This theory is spin if and only if $k$ is even.

We now proceed to compute the relevant traces. As the theory can be a spin TQFT, the theory may contain fermionic states, and $(-1)^F$ will in general be a non-trivial operator, which we need to understand to compute $\tr((-1)^F\mathsf s)$. In other words, we have to identify which of the states of this Chern-Simons theory are bosons, and which are fermions.

Note that, unlike the general discussion of section~\ref{sec:spin}, this theory is more conveniently presented as a $\mathbb Z_n$ quotient rather than a $\mathbb Z_2$ quotient, so let us slightly generalize the discussion in that section to such quotients. In section~\ref{sec:spin} we argued that the bosonic states after a $\mathbb Z_2$ fermionic quotient are the length-2 orbits, while the fermions are the fixed-points. We now claim that the general statement for $\mathbb Z_n$ fermionic quotients is that the bosonic states are the orbits of even length, while the fermions are the orbits of odd length.

To prove this claim, consider a general spin TQFT that can be written as
\begin{equation}
G_F=\frac{G_B}{\mathbb Z_n}\,,
\end{equation}
where $G_B$ is some bosonic TQFT, and where $\mathbb Z_n$ is a one-form symmetry generated by a fermion $\psi$. Since $h_{\psi^p}=ph_\psi$, $\psi^p$ is a fermion if $p$ is odd, and a boson if $p$ is even. This means that $n$ is necessarily even, because $\psi^n=\boldsymbol 1$ is a boson.

The braiding phase with respect to $\psi$ is always an $n$-th root of unity, it is the charge with respect to the $\mathbb Z_n$ symmetry. This fact allows us to partition the lines of $G_B$ into $n$ equivalence classes according to their $n$-ality $j$, i.e., the value of braiding $B(\alpha,\psi)=e^{2\pi i j/n}$, $j=0,1,\dots,n-1$. The lines with $j=0$ are the NS lines (so that $B(\alpha,\psi)=+1$), and those with $j=n/2$ are the R lines (so that $B(\alpha,\psi)=-1$). The rest of lines are projected out by the $\mathbb Z_n$ quotient (unless we turn on a suitable background for the dual $\mathbb Z_n$ zero-form). 
%The NS lines are those with $B(\alpha,f)^{N/2}=+1$, and the R lines are those with $B(\alpha,f)^{N/2}=-1$.
Furthermore, in each sector the lines are organized into $\mathbb Z_n$ orbits,
\begin{equation}
\{\alpha,\psi\alpha,\psi^2\alpha,\dots,\psi^{|\alpha|-1}\alpha\},
\end{equation}
where $|\alpha|\in[1,n]$ denotes the length of the orbit -- the minimal integer such that $\psi^{|\alpha|}\times\alpha=\alpha$. An orbit is Majorana if and only if its length is odd, for then and only then it may absorb a fermion. Indeed, the conformal block with puncture $\psi^{|\alpha|}$ is non-vanishing only if $\psi^{|\alpha|}\times \alpha=\alpha$. In conclusion, the fermionic states in the R-R sector of the quotient theory $G_F$ correspond to the orbits of $G_B$ R-lines with an odd number of elements, as claimed.

We are now in position to study the theory $\U(n)_{k,n+k}$. The discussion above has taught us how to identify fermionic states in the Hilbert space of the theory. Rather anticlimactically, we shall now argue that this theory has, in fact, no fermionic states at all! This means that the trace $\tr(-1)^F$ actually just counts the number of states of the theory, much like in a bosonic theory. This explains why the counting of states in \cite{Acharya:2001dz} matched the domain wall index -- because all states are bosonic. This, importantly, is not always the case for other spin TQFTs, such as $\O(n)$ (see below, section~\ref{sec:orthogonal_chern_simons}).

Let us prove that the theory has no fermionic states. $\U(n)_{k,n+k}$ is level-rank dual to $\U(k)_{-n,-(n+k)}$ as a spin TQFT. Therefore, if either $k$ or $n$ is odd, the theory factorizes as a bosonic theory times an invertible spin TQFT, and so the theory clearly has no Majorana states. The only non-trivial case is, therefore, that of $n,k$ both even, which we assume in what follows.

The theory in the numerator of the quotient description of $\U(n)_{k,n+k}$ in~\eqref{eq:spin_TQFT_ZN} is bosonic (recall that $\U(1)_K$ is spin for $K$ odd and bosonic for $K$ even; here $K=n(n+k)$, which is even). The states of $\U(n)_{k,n+k}$ are $\mathbb Z_n$ orbits of $\SU(n)_k\times \U(1)_{n(n+k)}$ representations. If we manage to prove that there are no orbits of odd length, we succeed in proving that the theory has no Majorana states. In fact, we show that, more generally, all orbits have length-$n$, i.e., all orbits are long. This implies that the states correspond to conformal blocks with no punctures, i.e., all states are bosonic, $(-1)^F\equiv+1$.

Write $\alpha=(R,q)$, where $R$ is an $\SU(n)_k$ representation, and $q\in[0,n(n+k))$ labels a $\U(1)_{n(n+k)}$ representation. The abelian part of the condition $\psi^{|\alpha|}\times\alpha=\alpha$ reads
\begin{equation}
q+|\alpha|(n+k)=q\mod n(n+k)\,,
\end{equation}
which can be written as $|\alpha|=0\mod n$, i.e., $|\alpha|=n$, as claimed. This proves that all orbits are long, which indeed implies the absence of Majorana lines.

Let us now use this information to compute the different $\U(n)_{k,n+k}$ partition functions. The untwisted trace is the number of conformal blocks (in any of the $2^3$ spin structures). Counting this is a straightforward exercise in combinatorics: we have a factor of $n(n+k)$ due to $\U(1)_{n(n+k)}$, times a factor of $\binom{n+k-1}{k}$ due to $\SU(n)_k$ (cf.~\eqref{eq:Z_SU_TQFT}), and a factor of $1/n^2$ due to the quotient $\mathbb Z_n$ (one factor of $n$ is due to the projecting out of lines, and the other one because the neutral lines are organized into length-$n$ orbits). All in all, the number of states -- the untwisted trace -- is
\begin{equation}\label{eq:Z_U_IR}
\tr_{\U(n)_{k,n+k}}(\boldsymbol 1)=\frac{n(n+k)}{n^2}\binom{n+k-1}{k}\equiv\binom{n+k}{k}\,.
\end{equation}
This standard argument was already used in~\cite{Acharya:2001dz}. An important aspect of this computation, much overlooked in the literature, is that this equals $\tr(-1)^F$ only because all the states have trivial fermion parity, which is nontrivially true in this theory. This shall not be the case in the orthogonal group $\O(n)$, where $\tr(-1)^F$ does not just count the total number of states, but rather the bosons minus the fermions, both sets being typically nonempty. 

Recall that the conjectured infrared TQFT corresponding to the $n$-domain wall of $\SU(N)$ is $\mathrm W_n=\U(n)_{k,n+k}$, with $k=N-n$. Using this value of the level in~\eqref{eq:Z_U_IR} indeed reproduces the Witten index computed in the ultraviolet, cf.~\eqref{eq:I_SU_UV}.

We now proceed to computing the trace twisted by the charge conjugation symmetry $\mathsf c$ of $\U(n)_{k,n+k}$. Consider first the case of odd $k$, where the theory is naturally bosonic. In this case, computing the trace amounts to counting the real representations of $\U(n)_{k,n+k}$. A representation of $\U(n)_{k,n+k}$ can be labeled by the pair $(R,q)$, where $R$ is an $\SU(n)_k$ representation, and $q\in[0,n(n+k))$, subject to $|R|=q\mod n$, where $|R|$ denotes the number of boxes in the Young diagram of $R$. 
Representations $(R,q)$ and $(\sigma^\ell\cdot R,q+\ell(n+k))$, with $\sigma^\ell\cdot R$ the $\SU(n)$ representation with Dynkin labels $(\sigma^\ell\cdot\lambda)_i=\lambda_{i-\ell\mod n}$, are identified by $\mathbb Z_n$ spectral flow.

% The pairs $(R,q)$ and $(\sigma\cdot R,q+n+k)$, with $\sigma\cdot R$ the representation with Dynkin labels $(\sigma\cdot\lambda)_i=\lambda_{i-1\mod n}$, are identified by spectral flow. More generally, the $\mathbb Z_n$ orbits are $(\sigma^\ell\cdot R,q+\ell(n+k))$ for $\ell=0,1,\dots,n-1$ (recall that all orbits are long). These representations are identified by $\mathbb Z_n$ spectral flow.

%A pair $(R,q)$ is real if
%\begin{equation}
%\begin{aligned}
%[\lambda_0,\lambda_1,\lambda_2,\dots,\lambda_{n-2},\lambda_{n-1}]&=\sigma^\ell\cdot[\lambda_0,\lambda_{n-1},\lambda_{n-2},\dots,\lambda_2,\lambda_1]\\
%q&=-q+\ell(n+k)\mod n(n+k)
%\end{aligned}
%\end{equation}
%for some $\ell$. In other words, if $\lambda_i=\lambda_{\ell-i\mod n}$ and $2q=\ell(n+k)\mod n(n+k)$.
%

The abelian charge $q$ is correlated with the $\SU(n)$ representation. Indeed, if $n$ is even and $R$ is real modulo $\sigma^\ell$, there is a single charge $q\in[0,n(n+k))$ that makes $(R,q)$ real; if $n$ is odd, there are two such charges.\footnote{Namely, we are looking for solutions to $2q=\ell(n+k)\mod n(n+k)$. These are $q=\frac{\ell}{2}(n+k)$ and $q=\frac{\ell+n}{2}(n+k)$ (except for $(n,\ell)=\text{(even, odd)}$, where there is no solution; see also~\eqref{eq:real_up_to_l}). For $n$ even, only one of these two solutions is valid, depending on the parity of $|R|$ (recall that we require $q=|R|\mod n$).} Therefore, the number of real representations in $\U(n)_{k,n+k}$ is the number of representations of $\SU(n)_k$ that are real up to the action of $\sigma$, divided by $n$ (the length of the orbits), and multiplied by $2$ if $n$ is odd. Let us now count the $\SU(n)_k$ representations.

An $\SU(n)_k$ representation $[\lambda_0,\lambda_1,\dots,\lambda_{n-1}]$ is real up to the action of $\sigma$ if $\lambda_i=\lambda_{\ell-i\mod n}$ for some $\ell$. The number of such representations is the number of integer solutions to
\begin{equation}\label{eq:C_up_to_SF}
\begin{aligned}
&2\lambda_0+2\lambda_1+\dots+2\lambda_{\lfloor(\ell-1)/2\rfloor}+\left\{\begin{array}{cc}
0 & \ell\text{ odd}\\
\lambda_{\ell/2} & \ell\text{ even}
\end{array}\right\}+\\
&\qquad+2\lambda_{\ell+1}+\dots+2\lambda_{\lfloor (n+\ell-1)/2\rfloor}+\left\{\begin{array}{cc}
0 & n+\ell\text{ odd}\\
\lambda_{(n+\ell)/2} & n+\ell\text{ even}
\end{array}\right\}=k\,.
\end{aligned}
\end{equation}
The number of solutions to this equation is
\begin{equation}\label{eq:real_up_to_l}
N_\ell=\begin{cases}
\displaystyle\binom{(n+k)/2-1}{(k-1)/2} & \text{$n$ odd, $k$ odd,}\\[+15pt] 
\displaystyle0 &\text{$n$ even, $\ell$ odd, $k$ odd,}\\[+15pt] 
\displaystyle2\binom{(n+k-1)/2}{(k-1)/2} & \text{$n$ even, $\ell$ even, $k$ odd,}\\[+15pt] 
\displaystyle\binom{(n+k)/2}{k/2}+\binom{(n+k)/2-1}{k/2-1} & \text{$n$ even, $\ell$ even, $k$ even,}\\[+15pt] 
\displaystyle\binom{(n+k)/2-1}{k/2} & \text{$n$ even, $\ell$ odd, $k$ even,}
\end{cases}
\end{equation}
where, for future reference, we have also included the case of $k$ even.

%If $N$ is odd, this is $(N-1)/2$ nodes with comark $2$, and one with comark $1$, and thus $\binom{(N+k)/2-1}{(k-1)/2}$ solutions. If $N$ is even and $\ell$ is odd, this is $N/2$ nodes with comark $2$, and thus there are no solutions; and, finally, if $N$ is even and $\ell$ is also even, this is $(N-2)/2$ nodes with comark $2$, and two with comark $1$, and thus $2\binom{(N+k-1)/2}{(k-1)/2}$ solutions.

We now sum over all $\ell=0,1,\dots,n-1$. For $n$ odd, this just multiplies $\binom{(n+k)/2-1}{(k-1)/2}$ by $n$. If $n$ is even, it multiplies $2\binom{(n+k-1)/2}{(k-1)/2}$ by $n/2$, because half the cases yield no solutions. Next, we divide by $n$ (due to the quotient), and multiply by $2$ if $n$ is odd. This yields the number of real representations of $\U(n)_{k,n+k}$ with $k$ odd as\footnote{Note that the expression for $n$ odd is invariant under $n\leftrightarrow k$, as required by level-rank duality.}
\begin{equation}\label{eq:U_IR_C}
\tr_{\U(n)_{k,n+k}}(\mathsf c)=\begin{cases}
\displaystyle2\binom{(n+k)/2-1}{(k-1)/2}&\text{$n$ odd, $k$ odd,}\\ \\
\displaystyle\binom{(n+k-1)/2}{(k-1)/2}&\text{$n$ even, $k$ odd.}
\end{cases}
\end{equation}
By plugging $k=N-n$ in~\eqref{eq:U_IR_C}, the partition function reproduces the Witten index twisted by charge conjugation computed in the ultraviolet, cf.~\eqref{eq:I_UV_C}.

The case of $k$ even is slightly more complicated because the theory is naturally spin. For $n$ odd we can obtain the twisted trace from the $k$ odd case by using level-rank duality $\U(n)_{k,n+k}\leftrightarrow \U(k)_{-n,-(n+k)}$. But for $n,k$ both even, the theory is spin, and cannot be written as a bosonic theory times a trivial spin theory -- at least not using the standard level-rank duality. Thus, we have to explicitly compute the trace of $\mathsf c$ in the R-R sector. This is non-trivial because, among other things, $\mathsf c$ may act as $-1$ on some states (see the discussion around~\eqref{eq:c_eq_-1}), and thus it is not enough to just count real representations.

A shortcut to compute the trace of $\mathsf c$ over the odd spin structure, for $k,n$ both even, is to sum over all spin structures:
\begin{equation}
\frac12\bigg(\sum_\sigma\tr_\sigma\mathsf c+\tr_\sigma(-1)^F\mathsf c\bigg)=\tr_{B}\mathsf c\,,
\end{equation}
where $\tr_B$ denotes the trace over the bosonic parent. From this expression, and noting that $(-1)^F$ is trivial in $\U(n)_{k,n+k}$ theories (due to the lack of Majorana lines), we can solve for the trace we are after:
\begin{equation}\label{eq:odd_trace_C_u}
\tr_\text{R-R}\mathsf c=\tr_B\mathsf c-3\tr_\text{NS-NS}\mathsf c\,.
\end{equation}

Let us begin with the first term. As this is a trace over a bosonic Hilbert space, we are just to count real representations of $\SU(n)_k\times U(1)_{n(n+k)}$. The first factor corresponds to $\ell=0$ in~\eqref{eq:real_up_to_l}, while the second factor has two real representations (namely, $q=0$ and $q=n(n+k)/2$). The end result is
\begin{equation}
\tr_B\mathsf c=2\bigg[\binom{(n+k)/2}{k/2}+\binom{(n+k)/2-1}{k/2-1}\bigg]\,.
\end{equation}

Let us now compute the second term in~\eqref{eq:odd_trace_C_u}. This is a trace over a fermionic Hilbert space, but over the NS-NS sector, and so we only have to count fixed-points, as they all contribute with $\mathsf c=+1$. In other words, the trace is just the number of real representations of $\U(n)_{k,n+k}$, that is, the number of solutions to~\eqref{eq:C_up_to_SF}, summed over $\ell=0,1,\dots,n-1$, and divided by $n$ due to the quotient. Using~\eqref{eq:real_up_to_l}, we get
\begin{equation}
\begin{aligned}
\tr_\text{NS-NS}\mathsf c&\equiv \frac1n\bigg(\frac n2\bigg[\binom{(n+k)/2}{k/2}+\binom{(n+k)/2-1}{k/2-1}\bigg]+\frac n2\binom{(n+k)/2-1}{k/2}\bigg)\\[+4pt]
&=\frac 12\binom{(n+k)/2}{k/2}+\frac12\binom{(n+k)/2-1}{k/2-1}+\frac 12\binom{(n+k)/2-1}{k/2}\,.
\end{aligned}
\end{equation}

Plugging these two traces into~\eqref{eq:odd_trace_C_u}, the twisted index, for $n,k$ even, becomes
\begin{equation}
\tr_{\U(n)_{k,n+k}}(\mathsf c)=\binom{(n+k)/2-1}{n/2}-\binom{(n+k)/2-1}{k/2}\,.
\end{equation}
Note that this is invariant under $n\leftrightarrow k$, as required by level-rank duality.\footnote{That is, invariant up to a sign. This is due to the fact that $\mathsf c=-1$ in the R-R sector of $\U(1)_1$ (cf.~\eqref{eq:c_eq_-1}), and the level-rank pair has a difference in their framing anomaly equal to $nk+1\equiv 1\mod2$, cf.~\cite{Hsin:2016blu}.} This expression for the twisted partition function of $\U(n)_{k,n+k}$ with $k=N-n$ matches the twisted Witten index computed in the ultraviolet, cf.~\eqref{eq:I_UV_C}.

Finally, we briefly sketch the computation of the trace twisted by the one-form symmetry $\mathsf g^t\in\mathbb Z_{n+k}$ of $\U(n)_{k,n+k}$, where $\mathsf g$ denotes a primitive root of unity, and $t\in[0,n+k)$. The states of $\U(n)_{k,n+k}$ are orbits of the form
\begin{equation}
%(R,q),\ (\sigma\cdot R,q+N+k),\ (\sigma^2\cdot R,q+2(N+k)),\dots,\ (\sigma^{N-1}\cdot R,q+(N-1)(N+k))
\{(\sigma^\ell \cdot R,q+\ell(n+k))\}\,,
\end{equation}
where $\ell$ ranges from $0$ to $n-1$. All the orbits are of length-$n$. The theory has a $\mathbb Z_{n+k}$ one-form symmetry that acts as $q\mapsto q+t n$, where $t\in[0,n+k)$. A state is invariant if and only if this transformation cyclically permutes the elements of the orbit, i.e., if a representative $(R,q)$ is mapped into itself up to spectral flow,
\begin{equation}
(R,q+tn)\equiv (\sigma^\ell \cdot R,q+\ell(n+k))\,.
\end{equation}

It is clear that if $tn$ is not of the form $\ell(n+k)$ for some $\ell\in\mathbb Z$, then no state is invariant, and the twisted trace vanishes. So let us assume that such an $\ell$ exists; it is clear that it is unique, so counting invariant orbits reduces to counting appropriate $\SU(n)_k$ representations. More specifically, the number of invariant states is
\begin{equation}
\tr_{\U(n)_{k,n+k}}(\mathsf g^t)=\frac{n(n+k)}{n^2}\hat N_\ell\,,
\end{equation}
where $n(n+k)$ denotes the number of states in $\U(1)_{n(n+k)}$, and the factor of $n^2$ is due to the $\mathbb Z_n$ quotient. $\hat N_\ell$ denotes the number of $\SU(n)_k$ representations that satisfy $R=\sigma^\ell\cdot R$, with $\ell:=t n/(n+k)\in\mathbb Z$.

Counting such $\SU(n)_k$ representations is easy, because this is a simply-connected group, so the states are labelled by the Dynkin labels, $\lambda_0,\lambda_1,\dots,\lambda_{n-1}$, which can be thought of a collection of independent bosons (cf.~\eqref{eq:Z_bos_twist}). The most efficient way to count the representations that are invariant under $\sigma^\ell$ is to recall that the associated diagonal phase is just the charge under the center~\eqref{eq:twist_SU}, which is a multiplicative phase, so the partition function factorizes:
\begin{equation}
\prod_{j=0}^{n-1}(1-e^{2\pi ij\ell/n}q)^{-1}\equiv\sum_{k\ge0} \hat N_\ell q^k\,,
\end{equation}
and therefore
\begin{equation}
%\hat N_\ell=\frac{n}{n+k}\binom{n+k}{n}_{\!\!e^{2 \pi i\ell/n}}\,.
\hat N_\ell=\binom{n+k-1}{k}_{\!\!e^{2 \pi i\ell/n}}\,.
%\\&=\frac{N}{N+k}\binom{\gcd(N+k,\mathsf z)}{N \gcd (N+k,\mathsf z)/(N+k)}
\end{equation}
Recall that the $q$-binomial coefficient at a root of unity can be expressed as a regular binomial coefficient, cf.~\eqref{eq:q_binom_UV} and~\eqref{eq:q_gcd_binom_UV}.

Putting everything together, the one-form symmetry twisted trace reads
\begin{equation}\label{eq:Z_IR_t_tw}
\begin{aligned}
\tr_{\U(n)_{k,n+k}}(\mathsf g^t)&=\binom{n+k}{n}_{\!\!\mathsf g^t}\\[+5pt]
&\equiv\begin{cases}
\displaystyle\binom{\gcd(n+k,t)}{n \gcd (n+k,t)/(n+k)} & tn\equiv0\mod n+k,\\[+15pt]
\displaystyle0 & \text{otherwise.}
\end{cases}
\end{aligned}
\end{equation}
It is easily checked that, if we plug $k=N-n$ in~\eqref{eq:Z_IR_t_tw}, the twisted trace for $\U(n)_{k,n+k}$ exactly reproduces the twisted Witten index of the $n$-domain wall of $\SU(N)$ computed in the ultraviolet, cf.~\eqref{eq:q_gcd_binom_UV}.

\subsection[$G=\O(n)$]{$\boldsymbol{G=\O(n)}$}\label{sec:orthogonal_chern_simons}

The $n$-domain wall theory for $4d$ ${\mathcal N}=1$ SYM with $G=\Spin(N)$ is proposed to be $\O(n)^1_{N-2-n,N-n+1}$ Chern-Simons theory. Let us proceed to study the partition functions of $\O(n)^1_{k,L}$.

The $\O(n)^1_{k,L}$ Chern-Simons theory is defined as~\cite{Cordova:2017vab}
\begin{equation}
\O(n)^1_{k,L}:=\frac{\O(n)^1_{k,0}\times(\mathbb Z_2)_L}{\mathbb Z_2}\,,
\end{equation}
where $(\mathbb Z_2)_L\leftrightarrow \Spin(L)_{-1}$ denotes a $\mathbb Z_2$ gauge theory with twist $L$, and the quotient denotes the gauging of a diagonal $\mathbb Z_2$ one-form symmetry. The value of the level we shall be interested in is $L=k+3$. On the other hand, the first factor is given by the following:
\begin{itemize}
\item If $n$ is even, the theory $\O(n)^1_{k,0}$ is defined as the $\mathsf{CM}$-orbifold of $\SO(n)_k$. Here $\mathsf C$ denotes the charge-conjugation $\mathbb Z_2$ zero-form symmetry that acts by permuting the last two Dynkin labels in $\SO(n)$, and $\mathsf M$ is the magnetic $\mathbb Z_2$ zero-form symmetry that is dual to the gauged one-form $\mathbb Z_2$ symmetry in the denominator of $\SO(n)_k\equiv\Spin(n)_k/\mathbb Z_2$. As such, it permutes the lines that split in the quotient, i.e., the lines of $\Spin(n)_k$ that are fixed by fusion with the extending simple current.

\item If $n$ is odd, the group $\O(n)$ is a direct product of $\SO(n)$ and $\mathbb Z_2$. The Chern-Simons theory $\O(n)^1_{k,0}$ itself does not necessarily factorize, because of the convention of which $\mathbb Z_2$ subgroup the reflection represents. The choice in~\cite{Cordova:2017vab} was
\begin{equation}
\O(n)^1_{k,0}:=\begin{cases}
\displaystyle \frac{\Spin(n)_k\times(\mathbb Z_2)_{(k-2)(n-1)}}{\mathbb Z_2}&\text{$n$ odd, $k$ even,}\\ \\
\displaystyle \SO(n)_k\times(\mathbb Z_2)_{(k-2)(n-1)}&\text{$n$ odd, $k$ odd.}
\end{cases}
\end{equation}

\end{itemize}

Let us compute the different traces in this theory. As above, the details depend sensitively on the parity of $n$ and $k$, so we consider each case separately.

%\note{The one-form symmetry is as follows:
%\begin{itemize}
%\item In $\O(4n)_{4k}$, the symmetry is $\mathbb Z_4$.
%\item In $\O(4n)_{4k+2}$, the symmetry is $\mathbb Z_2^2$.
%\item In $\O(4n+2)_{4k}$, the symmetry is $\mathbb Z_2^2$.
%\item In $\O(4n+2)_{4k+2}$, the symmetry is $\mathbb Z_4$.
%\item In $\O(4n+1)_{4k+1}$, the symmetry is $\mathbb Z_2^2$.
%\item In $\O(4n+3)_{4k+3}$, the symmetry is $\mathbb Z_2^2$.
%\item In $\O(4n+1)_{4k+3}$, the symmetry is $\mathbb Z_4$.
%\item In $\O(4n+3)_{4k+1}$, the symmetry is $\mathbb Z_4$.
%\item In $\O(2n+1)_{2k}$, the symmetry is $\mathbb Z_2$.
%
%\end{itemize}
%}

\paragraph{Even/Even.} We begin with the theory $\Spin(2n)_{2k}$. Its integrable representations satisfy
\begin{equation}
\lambda_0+\lambda_1+2(\lambda_2+\cdots+\lambda_{n-2})+\lambda_{n-1}+\lambda_n=2k\,,
\end{equation}
which has $\tr_{\Spin(2n)_{2k}}(\boldsymbol1)$ solutions (cf.~\eqref{eq:Z_Spin_unt}).

We now construct $\SO(2n)_{2k}$, i.e., we gauge a $\mathbb Z_2$ one-form symmetry, which acts as $\lambda_0\leftrightarrow\lambda_1$ and $\lambda_{n-1}\leftrightarrow\lambda_n$. This is a bosonic quotient. The neutral representations satisfy $\lambda_{n-1}+\lambda_n=\text{even}$, which has
\begin{equation}
\mathbb N:=\binom{n+k}{k}+2\binom{n+k-1}{k-1}+\binom{n+k-2}{k-2}\,,
\end{equation}
solutions. These are divided into length-2 orbits, and fixed points. The former satisfy $\lambda_0\neq\lambda_1\lor\lambda_{n-1}\neq \lambda_n$, and the latter $\lambda_0=\lambda_1\land\lambda_{n-1}= \lambda_n$. The number of fixed points is
\begin{equation}
\mathbb F:=\binom{n+k-2}{k}\,,
\end{equation}
and the number of length-2 orbits is $\frac12(\mathbb N-\mathbb F)$. Finally, the number of representations of $\SO(2n)_{2k}$ is
\begin{equation}
\tr_{\SO(2n)_{2k}}(\boldsymbol1)=2\mathbb F+\frac12(\mathbb N-\mathbb F)\equiv 2\binom{n+k-2}{k}+\binom{n+k-2}{k-1}+(n\leftrightarrow k)\,,
\end{equation}
which is invariant under $n\leftrightarrow k$, as expected by level-rank duality. This also agrees with expression~\eqref{eq:Z_SO_unt}.

We now orbifold by $\mathsf{CM}$, which acts by swapping the lines in $2\mathbb F$ pairwise, and as $\lambda_{n-1}\leftrightarrow \lambda_n$. The representations that are fixed under $\mathsf{CM}$ are the subset of the length-2 orbits that satisfy either $\lambda_0\neq \lambda_1\land\lambda_{n-1}=\lambda_n$ or $\lambda_0=\lambda_1\land \lambda_{n-1}\neq \lambda_n$. In other words, the lines that satisfy either of
\begin{equation}\label{eq:orb_fix}
\begin{aligned}
\lambda_0+\lambda_1+2(\lambda_2+\cdots+\lambda_{n-2}+\lambda_{n-1})&=2k,\qquad \ \ \lambda_0\ \neq \lambda_1\,,\\
2(\lambda_1+\lambda_2+\cdots+\lambda_{n-2})+\lambda_{n-1}+\lambda_n&=2k,\qquad \lambda_{n-1}\neq \lambda_n\,.\\
\end{aligned}
\end{equation}
By symmetry, both conditions have the same number of solutions. In total,
\begin{equation}
\mathbb A:=2\bigg[\binom{n+k-1}{k}+\binom{n+k-2}{k-1}-\binom{n+k-2}{k}\bigg]
\end{equation}
solutions. Note that these are length-2 orbits of $\Spin(2n)_{2k}$, so the number of lines is $\mathbb A/2$.

The representations that are interchanged under $\mathsf{CM}$ are all of $\mathbb F$, plus the subset of the length-2 orbits that satisfy $\lambda_{n-1}+\lambda_n=\text{even}$ and $\lambda_{n-1}\neq \lambda_n$, minus the solutions to the second line in~\eqref{eq:orb_fix}. The latter are
\begin{equation}
\begin{aligned}
\mathbb B&:=\binom{n+k}{k}+2\binom{n+k-1}{k-1}+\binom{n+k-2}{k-2}\\
&\quad-\binom{n+k-1}{k}-\binom{n+k-2}{k-1}-\frac12\mathbb A\\
&=4 \binom{n+k-2}{k-2}\,.
\end{aligned}
\end{equation}
Note that $\mathbb F$ are fixed points of $\mathbb Z_2$, while $\mathbb B$ are length-2 orbits, so the number of lines is $2\mathbb F+\mathbb B/2$. Adding the lines in $\mathbb A$ we get $\frac12(\mathbb A+\mathbb B)+2\mathbb F\equiv \tr_{\SO(2n)_{2k}}(\boldsymbol1)$, as one would expect.

Putting all these results together, we see that the number of twisted and untwisted lines in the orbifold is~\cite{Dijkgraaf:1989hb}
\begin{equation}
\begin{aligned}
N_\text{twisted}&=\mathbb A\,,\\
N_\text{untwisted}&=\mathbb A+\frac14\mathbb B+\mathbb F\,,
\end{aligned}
\end{equation}
and so the theory has
\begin{equation}
\begin{aligned}
\tr_{\O(2n)^1_{2k,0}}(\boldsymbol1)&=N_\text{twisted}+N_\text{untwisted}\\
%&=\binom{k+n-2}{k}+\frac{9}{4} \binom{k+n-2}{k-1}+(n\leftrightarrow k)
%&=\frac{1}{8}\binom{n+k}{k}+2 \binom{n+k-1}{k-1}-\frac{5}{4} \binom{n+k-2}{k}\\
%&+\frac{7}{4} \binom{n+k-2}{k-1}+(n\leftrightarrow k)
&=-\frac{9}{8} \binom{n+k-2}{k}+4 \binom{n+k-2}{k-1}+\frac{17}{8} \binom{n+k-2}{k-2}+(n\leftrightarrow k)
\end{aligned}
\end{equation}
lines. %Note that this is invariant under $n\leftrightarrow k$, as expected by the level-rank duality.
 This expression agrees with~\cite{Cordova:2017vab}.

We now move on to $\O(2n)^1_{2k,2k+3}$. This is obtained by taking the theory we just constructed, $\O(2n)^1_{2k,0}$, tensoring with $\Spin(2k+3)_{-1}$, and gauging a diagonal $\mathbb Z_2$ one-form symmetry:
\begin{equation}
\O(2n)^1_{2k,2k+3}=\frac{\O(2n)^1_{2k,0}\times \Spin(2k+3)_{-1}}{\mathbb Z_2}\,,
\end{equation}
where the quotient is fermionic. %First, recall that the number of states in $\O(2n)^1_{2k,0}$ is $N_\text{twisted}+N_\text{untwisted}$, where
%\begin{equation}
%\begin{aligned}
%N_\text{twisted}&=\mathbb A\\
%N_\text{untwisted}&=\mathbb A+\frac14\mathbb B+\mathbb F
%\end{aligned}
%\end{equation}
%where $\mathbb A/2$ is the number of $\SO(2n)_{2k}$ lines that are fixed by $\mathsf{CM}$, and $\frac14\mathbb B+\mathbb F$ is the number of (pairs of) lines that are interchanged under $\mathsf{CM}$.

Take the states of $\O(2n)^1_{2k,0}$ as above, i.e., $N_\text{twisted}$ and $N_\text{untwisted}$, and tensor by $\Spin(2k+3)_{-1}=\{\boldsymbol1,\sigma,\chi\}$. The NS and R lines are as follows:
\begin{equation}
\begin{aligned}
\text{NS}&:\qquad N_\text{untwisted}\otimes \boldsymbol1,\quad N_\text{twisted}\otimes \sigma,\quad N_\text{untwisted}\otimes \chi\\
\text{R}&:\qquad\ N_\text{twisted}\otimes \boldsymbol1,\quad N_\text{untwisted}\otimes \sigma,\quad N_\text{twisted}\otimes \chi\,.
\end{aligned}
\end{equation}
We now quotient by the $\mathbb Z_2$ one-form symmetry. This symmetry maps $\boldsymbol 1\leftrightarrow\chi$, and it fixes $\sigma$; and, also, it permutes lines in $\mathbb A$ pairwise, $a\leftrightarrow a'$, and it fixes those in $\frac14\mathbb B+\mathbb F$. Therefore, in the NS sector it acts as
\begin{equation}\label{eq:R_states_O_even}
\begin{aligned}
\mathbb A\otimes \boldsymbol1&\leftrightarrow \mathbb A'\otimes \chi\\
(\tfrac14\mathbb B+\mathbb F)\otimes \boldsymbol1&\leftrightarrow (\tfrac14\mathbb B+\mathbb F)\otimes \chi\\
\mathbb A\otimes \sigma&\leftrightarrow \mathbb A'\otimes \sigma
\end{aligned}
\end{equation}
which are all length-two orbits (recall that there are never fixed-points in the NS sector). Thus, the dimension of the Hilbert space is
\begin{equation}\label{eq:dim_O_NS}
\begin{aligned}
\dim(\O(2n)^1_{2k,2k+3})&=\mathbb A+(\tfrac14\mathbb B+\mathbb F)+\tfrac12\mathbb A\\
&\equiv \tfrac12N_\text{twisted}+N_\text{untwisted}\,.
\end{aligned}
\end{equation}
This corresponds to the trace of $\boldsymbol1$ over the Hilbert space on any of the spatial spin structures.

Consider now the R sector. The one-form symmetry acts as
\begin{equation}
\begin{aligned}
\mathbb A\otimes \boldsymbol1&\leftrightarrow \mathbb A'\otimes \chi\\
(\tfrac14\mathbb B+\mathbb F)\otimes \sigma&\leftrightarrow (\tfrac14\mathbb B+\mathbb F)\otimes \sigma\\
\mathbb A\otimes \sigma&\leftrightarrow \mathbb A'\otimes \sigma\,,
\end{aligned}
\end{equation}
and so all of $\mathbb A$ are in length-two orbits, while all of $\tfrac14\mathbb B+\mathbb F$ are fixed-points. Thus, the number of fermions and bosons is
\begin{equation}
\begin{aligned}
N_\text{boson}&=\mathbb A+\tfrac12\mathbb A\equiv \tfrac32N_\text{twisted}\,,\\
N_\text{fermion}&=\tfrac14\mathbb B+\mathbb F\equiv N_\text{untwisted}-N_\text{twisted}\,.
\end{aligned}
\end{equation}

Note that $N_\text{boson}+N_\text{fermion}$ agrees with the dimension of the Hilbert space as computed in the NS sector (cf.~\eqref{eq:dim_O_NS}). On the other hand, the trace in the odd spin structure, weighted by fermion parity, is $N_\text{boson}-N_\text{fermion}$:
\begin{equation}\label{eq:Z_O_IR}
\tr_{\O(2n)^1_{2k,2k+3}}(-1)^F\equiv \tfrac52N_\text{twisted}-N_\text{untwisted}\,.
\end{equation}

As a consistency check, recall that one can also express the fermionic trace as $\tr_{{\mathcal H}_\text{R-R}}(-1)^F=2\dim(\mathcal H_B)-7\dim(\mathcal H_F)$ (cf.~\eqref{eq:index_from_dims}). The dimension of the bosonic Hilbert space is
\begin{equation}
\dim(\O(2n)^1_{2k,0}\times \Spin(2k+3)_{-1})\equiv 3(N_\text{untwisted}+N_\text{twisted})\,,
\end{equation}
while the dimension of the fermionic Hilbert space is half the number of lines, i.e., $\frac12(2N_\text{untwisted}+N_\text{twisted})$. Thus,
\begin{equation}
\tr_{\O(2n)^1_{2k,2k+3}}(-1)^F=6N_\text{twisted}+6N_\text{untwisted}-\tfrac72N_\text{twisted}-7N_\text{untwisted}\,,
\end{equation}
which indeed matches the expression above.

Recall that the conjectured infrared theory corresponding to the $n$-domain wall of $\Spin(N)$ was $\mathrm W_n=\O(n)^1_{k,k+3}$, with $k=N-2-n$. Using this value of the level in~\eqref{eq:Z_O_IR} indeed reproduces the Witten indices computed in the ultraviolet, cf.~\eqref{eq:I_Spin_even_UV}.

\paragraph{Odd/Odd.} We consider 
\begin{equation}
\O(2n+1)^1_{2k+1,2k+4}=\SO(2n+1)_{2k+1}\times (\mathbb Z_2)_{2(n+k)}\,.
\end{equation}
As the theory is a tensor product, the traces factorize:
\begin{equation}
\tr_{\O(2n+1)^1_{2k+1,2k+4}}(\mathcal O_1\otimes\mathcal O_2)\equiv \tr_{\SO(2n+1)_{2k+1}}(\mathcal O_1)\cdot \tr_{(\mathbb Z_2)_{2(n+k)}}(\mathcal O_2)\,.
\end{equation}

For example, the $\mathbb Z_2$ gauge theory has four states, all bosonic, $\tr_{(\mathbb Z_2)_{2(n+k)}}(-1)^F\equiv 4$, which means that the untwisted index is
\begin{equation}\label{eq:Z_O_IR_odd}
\tr_{\O(2n+1)^1_{2k+1,2k+4}}(-1)^F=4\bigg[\binom{n+k}{k}-2\binom{n+k-1}{k}\bigg]\,,
\end{equation}
where we have used the trace of $\SO(2n+1)_{2k+1}$ as given in~\eqref{eq:index_SO_odd_odd}.

Similarly, the index twisted by the zero-form symmetry $\mathsf c$ has $\tr_{(\mathbb Z_2)_{2(n+k)}}(\mathsf c)\equiv 2$, where $\mathsf c$ acts by permuting the two spinors (this is the only zero-form symmetry of this $\mathbb Z_2$ gauge theory, cf.~\cite{KITAEV20032,Delmastro:2019vnj}; it fixes both the identity and the vector). On the other hand, the only zero-form symmetry of $\SO(2n+1)_{2k+1}$ is fermion parity,\footnote{The Dynkin diagram of $\SO(N)$ for $N$ odd has no reflection symmetries, i.e., its outer automorphism group is trivial. Thus, the zero-form symmetries of $\SO(N)$, if any, must be due to the global structure of the group, as its algebra has no symmetries. Indeed, the zero-form symmetry comes from $\pi_1(\SO(N))=\mathbb Z_2$, but this is just the magnetic dual to the gauged $\mathbb Z_2$ one-form symmetry, which means that the magnetic symmetry is formally just $(-1)^F$. If we were to gauge this symmetry, we would recover $\Spin(N)$.} and there is in fact a natural identification $\mathsf c=(-1)^F$ (cf.~\cite{Cordova:2017vab}). Thus, the $\mathsf c$-twisted trace weighted by fermion parity actually computes the untwisted trace, with antiperiodic (NS) boundary conditions on the time circle:
\begin{equation}
\tr_{\SO(2n+1)_{2k+1}}((-1)^F\mathsf c)\equiv \tr_{\SO(2n+1)_{2k+1}}(\boldsymbol 1)\equiv \binom{n+k}{k}\,,
\end{equation}
where we have used~\eqref{eq:Z_SO_odd_NS}. All in all, the twisted trace of $\O(2n+1)^1_{2k+1,2k+4}$ is
\begin{equation}\label{eq:Z_O_IR_odd_c}
\tr_{\O(2n+1)^1_{2k+1,2k+4}}((-1)^F\mathsf c)=2\binom{n+k}{k}\,.
\end{equation}

The index twisted by the one-form symmetry is also straightforward. This symmetry is $\mathbb Z_2^2$ for $\O(4n+1)_{4k+1}$ and $\O(4n+3)_{4k+3}$, and $\mathbb Z_4$ for $\O(4n+1)_{4k+3}$ and $\O(4n+3)_{4k+1}$. These correspond to fusion with the abelian anyons of $\Spin(L)_{-1}$, with $L=0\mod4$ and $L=2\mod4$ respectively, which indeed have a $\mathbb Z_2^2/\mathbb Z_4$ fusion algebra. As abelian fusion has no fixed-points, all the twisted traces vanish:
\begin{equation}\label{eq:Z_O_IR_odd_g}
\begin{aligned}
\tr_{\O(2n+1)^1_{2k+1,2k+4}}((-1)^F\mathsf g_1\mathsf g_2)&\equiv 0\,,\\
\tr_{\O(2n+1)^1_{2k+1,2k+4}}((-1)^F\mathsf g)&\equiv 0\,,
\end{aligned}
\end{equation}
where $(\mathsf g_1,\mathsf g_2)\in\mathbb Z_2^2$ and $\mathsf g\in\mathbb Z_4$, respectively.

Recall that the conjectured infrared theory corresponding to the $n$-domain wall of $\Spin(N)$ was $\mathrm W_n=\O(n)^1_{k,k+3}$, with $k=N-2-n$. Using this value of the level in~\eqref{eq:Z_O_IR_odd}, \eqref{eq:Z_O_IR_odd_c}, \eqref{eq:Z_O_IR_odd_g} indeed reproduces the Witten indices computed in the ultraviolet, cf.~\eqref{eq:I_Spin_even_UV}, \eqref{eq:I_Spin_even_c_UV}, \eqref{eq:I_Spin_even_g_UV_even}, \eqref{eq:I_Spin_even_g_UV_odd}.

\paragraph{Odd/Even \& Even/Odd.} We only need to consider one; the other follows by the level-rank duality. Take
\begin{equation}
\O(2n+1)^1_{2k,0}=\frac{\Spin(2n+1)_{2k}\times (\mathbb Z_2)_{4n(k-1)}}{\mathbb Z_2}\,,
\end{equation}
where the gauged one-form symmetry is generated by $a\otimes\mathsf e$, where $a=[0,2k,0,\dots,0]$ and %$\mathsf e=(0,1)$. This is a bosonic quotient.
$\mathsf e$ is the electric line of the toric code. This is a bosonic quotient.

The one-form symmetry acts as $\lambda_0\leftrightarrow \lambda_1$ and $\mathsf e\colon \mathsf m\leftrightarrow \mathsf{em}$. The neutral lines are of the form
\begin{equation}
\begin{aligned}
\lambda&\otimes\boldsymbol 1,\hspace{14pt} \lambda\otimes\mathsf e,\hspace{33pt} \lambda_n=\text{even}\\
\lambda&\otimes\mathsf m, \quad \lambda\otimes\mathsf{em},\qquad \lambda_n=\text{odd}\,.
\end{aligned}
\end{equation}
Note that there are no fixed points, and all orbits are of length $2$:
\begin{equation}
\{\lambda\otimes\boldsymbol 1,\quad(a\times\lambda)\otimes\mathsf e\},\qquad \{\lambda\otimes\mathsf m, \quad (a\times\lambda)\otimes\mathsf{em}\}\,.
\end{equation}

Therefore, a set of representatives can be taken as $\lambda_\text{tensor}\otimes\boldsymbol 1$ and $\lambda_\text{spinor}\otimes\mathsf m$. In what follows we drop the second label, as it is correlated with $\lambda$ in a unique way. The number of tensors and spinors is (cf.~\eqref{eq:tensor_spinor_B})
\begin{equation}
N_\text{tensor}=\binom{n+k-1}{k-1}+\binom{n+k}{k},\qquad N_\text{spinor}=2\binom{n+k-1}{k-1}\,.
\end{equation}

We now tensor the theory by a factor of $\Spin(2k+3)_{-1}=\{\boldsymbol1,\sigma,\chi\}$, and gauge the fermionic one-form symmetry generated by $f=a\otimes\chi$. The Ramond sector requires $h_{\alpha\times f}=h_\alpha\mod 1$, which means that the lines are
\begin{equation}
(\lambda_{\text{tensor}},\sigma),\qquad (\lambda_{\text{spinor}},\boldsymbol 1\text{ or }\chi)\,.
\end{equation}
Note that only the former can be a fixed-point under the fermionic quotient, inasmuch as $\chi\times \sigma=\sigma$ while $\chi\colon \boldsymbol1\leftrightarrow\chi$. In particular, the fixed-points are
\begin{equation}
\lambda_{\text{tensor}},\qquad \lambda_0=\lambda_1\,,
\end{equation}
while the rest of lines are all in length-2 orbits. The fixed-points satisfy $\lambda_1+\lambda_2+\cdots+\lambda_{n-1}+\lambda_n/2=k$, which has
\begin{equation}
\mathbb F:=\binom{n+k-1}{k}\,,
\end{equation}
solutions. Thus, finally
\begin{equation}\label{eq:Z_O_IR_mix}
\begin{aligned}
\tr_{\O(2n+1)^1_{2k,2k+3}}(-1)^F&=N_\text{spinor}+\tfrac12(N_\text{tensor}-\mathbb F)-\mathbb F\\
&=\tfrac52\binom{n+k-1}{k-1}+\tfrac12\binom{n+k}{k}-\tfrac32\binom{n+k-1}{k}\,.
\end{aligned}
\end{equation}

The trace over $\O(2n)^1_{2k+1,2k+4}$ can be obtained by using the orthogonal level-rank duality $\O(2n)^1_{2k+1,2k+4}\leftrightarrow \O(2k+1)^1_{-2n,-(2n+3)}$.

One can similarly compute the index twisted by the $\mathbb Z_2$ one-form symmetry, which acts via fusion with the electric line $\mathsf e$. The charged states are those that include the magnetic line $\mathsf m$, to wit, the spinors. In other words, the one-form symmetry correlates (gauge) spin and (spacetime) statistics, so that the states with $(-1)^F\mathsf e=+1$ are the tensor bosons and spinor fermions, and states with $(-1)^F\mathsf e=-1$ are the spinor bosons and the tensor fermions. With this,
\begin{equation}\label{eq:Z_O_IR_mix_g}
\begin{aligned}
\tr_{\O(2n+1)^1_{2k,2k+3}}((-1)^F\mathsf e)&=-N_\text{spinor}+\tfrac12(N_\text{tensor}-\mathbb F)-\mathbb F\\
&=-\binom{n+k}{k}\,.
\end{aligned}
\end{equation}
As above, the trace for $\O(2n)^1_{2k+1,2k+4}$ is obtained by level-rank duality.

Recall that the conjectured infrared theory corresponding to the $n$-domain wall of $\Spin(N)$ was $\mathrm W_n=\O(n)^1_{k,k+3}$, with $k=N-2-n$. Using this value of the level in~\eqref{eq:Z_O_IR_mix}, \eqref{eq:Z_O_IR_mix_g} indeed reproduces the Witten indices computed in the ultraviolet, cf.~\eqref{eq:I_UV_Spin_odd}, \eqref{eq:I_UV_Spin_odd_g}.

\subsection[$G=G_2$]{$\boldsymbol{G=G_2}$}\label{sec:G2_TQFT}

The $2$-domain wall theory for $4d$ ${\mathcal N}=1$ SYM with $G=G_2$ is $\SO(3)_3\times S^1$, where $S^1$ denotes then nonlinear sigma model with $S^1$ target space. We already proved in section~\ref{sec:Z_spin_TQFT} that the theory $\SO(3)_3$ has vanishing Witten index, and since there is a unique vacuum of the $S^1$ sigma model on the torus, the infrared index vanishes. This matches the Witten index computed in the ultraviolet, which is given by the coefficient of $q^2$ in~\eqref{eq:g_2_UV_Z}. Indeed, expanding this polynomial one finds that the index vanishes.

The domain wall with $n=1$ (and $n=3$, which is the anti-wall of $n=1$) is addressed below.

\subsection{Minimal Wall for Arbitrary Gauge Group}

The $n=1$ domain wall theory for $4d$ ${\mathcal N}=1$ SYM with arbitrary $G$ is proposed to be $G_{-1}$ Chern-Simons theory. 

As $G$ is simply-connected, the theory is naturally bosonic, and the trace $\tr_{G_{-1}}(-1)^F$ computes the dimension of the Hilbert space, that is, the number of integrable representations at level $1$. In other words, the trace is the number of solutions to~\eqref{eq:boson_int_reps} with $k=1$, namely
\begin{equation}
\sum_{i=0}^r \lambda_i a_i^\vee=1\,,
\end{equation}
which, as in~\eqref{eq:minimal_wall_G}, requires $\lambda_i=1$ for some $i$ with $a^\vee_i=1$, and $\lambda_j=0$ for all $j\neq i$. Therefore, the trace is
\begin{equation}
\tr_{G_{-1}}(-1)^F=m_1
\end{equation}
where $m_1$ denotes the number of nodes in the Dynkin diagram of $G$ that have comark equal to $1$. This clearly reproduces the ultraviolet index~\eqref{eq:minimal_wall_G_index}, as required. 

For simply-laced $G$, $G_{-1}$ Chern-Simons theory is in fact an abelian TQFT, and all the lines generate one-form symmetries. The number of lines is the number of one-form symmetries, that is, the order of $\Gamma$, which indeed agrees with $m_1$. Equivalently, it is known that simply-laced theories at level $1$ admit a $K$-matrix representation, where one can take $K$ as the Cartan matrix of $\mathfrak g$. The number of states is indeed $\det(K)\equiv |\Gamma|$.\footnote{Note that abelian systems typically have a very large number of zero-form symmetries~\cite{Delmastro:2019vnj}, most of which are emergent in our picture, inasmuch as the ultraviolet theory only has $\mathsf C$ as its zero-form symmetry group.}

One can define a zero-form twisted index for $(E_6)_{-1}$. The only node with comark $1$ preserved by the charge conjugation symmetry of this theory is the extended node, and thus
\begin{equation}
\tr_{(E_6)_{-1}}(-1)^F=1\,.
\end{equation}
Since $(E_6)_{-1}$ and $(E_7)_{-1}$ are abelian, twisting by a one-form symmetry has no fixed points and 
 \begin{equation}
\begin{array}{rll}
\tr_{(E_6)_{-1}}((-1)^F\mathsf g)=0&\quad\mathsf g\in \Gamma=\mathbb Z_3\\
\tr_{(E_7)_{-1}}((-1)^F\mathsf g)=0&\quad\mathsf g\in \Gamma=\mathbb Z_2\,.
\end{array}
\end{equation}
These reproduce the twisted Witten indices on the walls~\eqref{twistEE}.

\section{Concluding Remarks and Open Questions}
\label{sec:fin}

In this paper we have proposed the explicit infrared theories on the domain walls of $4d$ ${\mathcal N}=1$ SYM with gauge group $G$. We have found precise agreement between computations carried out in the ultraviolet of the domain walls and the TQFTs we propose emerge in the infrared. We have highlighted the importance in identifying the infrared of the domain wall theories of studying the Hilbert space of spin TQFTs, in particular the partition function in the R-R sector and identifying the fermionic states in the Hilbert space, and not merely counting states. The nontrivial matching of the twisted Witten indices provides strong support for our proposal.

A heuristic argument can be made in favor of our proposal that the $n$-domain wall in $4d$ ${\mathcal N}=1$ SYM with gauge group $G$ is the infrared of $3d$ ${\mathcal N}=1$ $G_{h/2-n}$ SYM (see equation~\eqref{proposal}).\footnote{We would like to thank D.~Gaiotto     for an interesting discussion regarding this point.} Consider $4d$ SYM on $\mathbb R^3\times S^1$ with the YM $\theta$-angle linear in the $S^1$ coordinate and winding number $n$ around the circle. This theory can be defined while preserving half of the supersymmetry.\footnote{The Lagrangian of this theory can be written as ${\mathcal L}=\int d^2\vartheta\ X W_\alpha W^\alpha$, where $X$ is a background chiral multiplet and $W_\alpha$ the chiral gauge field strength. $\operatorname{Re}(X)$ determines the gauge coupling and $\operatorname{Im}(X)$ the $\theta$-angle. The background $\operatorname{Im}(X)\propto nx_3$ with $F_X\propto i n$ preserves half of the supersymmetries. The background for $F_X$ induces a mass term for the gaugino $\propto in \bar \lambda \gamma_{(5)} \gamma^3\lambda$, where $\lambda$ is Majorana.} When the radius of the circle is large one can expect the theory to be gapped everywhere except at the location of the wall $\mathrm W_n$. For small radius, the theory reduces to $3d$ ${\mathcal N}=1$ $G_{-n}$ SYM with an adjoint real multiplet (the scalar is compact, as it arises from reducing the gauge field along a circle). It was argued in~\cite{Bashmakov:2018wts} that with a suitable superpotential for the real multiplet, the multiplet gaps out and flows to $3d$ ${\mathcal N}=1$ $G_{h/2-n}$ SYM, where the shift is induced by integrating out the massive fermion in the real multiplet. Assuming that there is no phase transition as the size of the circle is reduced leads to the proposal. However, the lack of control over the superpotential upon reduction makes the argument suggestive but heuristic.

The $n>1$ domain wall theories for the groups $G=F_4, E_6, E_7$ and $E_8$ remain to be discovered. Equivalently, the phase diagram of the corresponding $3d$ ${\mathcal N}=1$ $G_k$ SYM with $k<h/2-1$ remains elusive. We collect in appendix~\ref{sec:exceptional_indices} the twisted partition functions computed in the ultraviolet for future reference. One strategy towards the identification of the infrared domain wall theory is to search for novel level-rank dualities in $G_k$ Chern-Simons theories that go beyond the ones that follow from conformal embeddings. In general, level-rank dualities follow from embeddings into holomorphic theories (theories with only one state), and this approach could lead to suitable level-rank dualities and in turn to explicit proposals for the remaining $3d$ ${\mathcal N}=1$ $G_k$ SYM phase diagrams (and associated $4d$ domain walls). 

In this paper we have made an intriguing connection between the Hilbert space of Chern-Simons theories on the torus and the Hilbert space of fermions in $0+1$ dimensions labeled by the extended Dynkin diagram $\mathfrak g^{(1)}$ corresponding to a Lie group $G$. That is, the fermionic Hilbert space ${\mathcal H}^n_\text{F}$ with energy $n$ is isomorphic as super-vector spaces to the R-R Hilbert space of a suitable spin TQFT, which we denote by TQFT$_{n}$
\begin{equation}
{\mathcal H}^n_\text{F}\simeq {\mathcal H}^{\text{TQFT}_{n}}_\text{R-R}\,.
\end{equation} 
Consequently, the partition functions with periodic and antiperiodic boundary conditions on the time circle also match. Specifically we have established the correspondence (the $A^{(1)}_{N-1}$ case was studied by Douglas in \cite{Douglas:1994ex})\footnote{In writing this we use the duality $(G_2)_{1}\leftrightarrow \U(2)_{3,1}$ and the notation $\U(2)_{6,0}\equiv \SO(3)_3\times S^1$.}
 \begin{equation}
\begin{aligned}
A^{(1)}_{N-1}&\ \longleftrightarrow\ \U(n)_{N-n,N}\\
B^{(1)}_{N}&\ \longleftrightarrow\ \O(n)^1_{2N-1-n,2N-n+2}\\
C^{(1)}_{N}&\ \longleftrightarrow\ \Sp(n)_{N+1-n}\\ 
 D^{(1)}_{N}&\ \longleftrightarrow\ \O(n)^1_{2N-2-n,2N-n+1}\\
 G_2^{(1)}&\ \longleftrightarrow\ \U(2)_{3n,2-n}\,.
\end{aligned}
\end{equation}
Another route to constructing the domain walls for $G=F_4, E_6, E_7$ and $E_8$ is to identify the TQFT whose R-R Hilbert space on the torus is that of the collection of free fermions based on the corresponding affine Dynkin diagram $\mathfrak{g}^{(1)}$.

\section*{Acknowledgments}

We would like to thank Davide Gaiotto, Zohar Komargodski, Bruno Le Floch and Nathan Seiberg for useful discussions. Research at Perimeter Institute is supported in part by the Government of Canada through the Department of Innovation, Science and Economic Development Canada and by the Province of Ontario through the Ministry of Colleges and Universities. Any opinions, findings, and conclusions or recommendations expressed in this material are those of the authors and do not necessarily reflect the views of the funding agencies. 
 \vfill\eject

\appendix
\section{Chern-Simons with Unitary and Orthogonal groups}\label{sec:app_su_spin}

In this appendix we compute several traces on the torus Hilbert space of Chern-Simons theories over simply-connected Lie groups. These traces are useful when studying more complicated theories over non-simply-connected groups.

\subsection[$G=\SU(n)$]{$\boldsymbol{G=\SU(n)}$}

Consider the algebra $A_{n-1}=\mathfrak{su}_n$. The comarks are all $a_i^\vee=1$. Plugging this into~\eqref{eq:Z_bos_twist} we get the generating function as
\begin{equation}
Z(\SU(n),q)=(1-q)^{-n}\,,
\end{equation}
and, by expanding, the untwisted trace
\begin{equation}\label{eq:Z_SU_TQFT}
\tr_{\SU(n)_k}(\boldsymbol1)=\binom{n+k-1}{k}\,.
\end{equation}

This is the number of integrable representations of $\SU(n)_k$, that is, the dimension of the torus Hilbert space of this Chern-Simons theory. This result will be useful when we discuss the Chern-Simons theory over the unitary group $\U(n)$, see section~\ref{sec:unitary_chern_simons}.

\subsection[$G=\Spin(2n+1)$]{$\boldsymbol{G=\Spin(2n+1)}$}

Consider the algebra $B_n=\mathfrak{so}_{2n+1}$. The comarks are $a_i^\vee=1$ for $i=0,1,n$, and $a_i^\vee=2$ for $i=2,\dots,n-1$. Plugging this into~\eqref{eq:Z_bos_twist} we get the generating function as
\begin{equation}
Z(\Spin(2n+1),q)=(1-q)^{-3}(1-q^2)^{-(n-2)}\,,
\end{equation}
and, by expanding, the untwisted trace
\begin{equation}\label{eq:Z_Spin_odd}
\begin{aligned}
\tr_{\Spin(2n+1)_{2k}}(\boldsymbol1)&=\binom{n+k}{k}+3 \binom{n+k-1}{k-1}\,,\\
\tr_{\Spin(2n+1)_{2k+1}}(\boldsymbol1)&=\binom{n+k-1}{k-1}+3 \binom{n+k}{k}\,.
\end{aligned}
\end{equation}

This is the number of integrable representations of $\Spin(2n+1)_k$, that is, the dimension of the torus Hilbert space of this Chern-Simons theory. For future reference, it is also useful to break up the states into the tensors and spinors. In other words, we shall be interested in knowing how many of the states of $\Spin(2n+1)$ are tensorial representations, and how many are spinorial representations. These are defined by $\lambda_n=\text{even}$ and $\lambda_n=\text{odd}$, respectively, which yields the following:
\begin{equation}\label{eq:tensor_spinor_B}
\begin{aligned}
N_\text{tensor}^{\Spin(2n+1)_{2k}}&=\binom{n+k}{k}+\binom{n+k-1}{k-1}\,,\\
N_\text{spinor}^{\Spin(2n+1)_{2k}}&=2 \binom{n+k-1}{k-1}\,,\\
N_\text{tensor}^{\Spin(2n+1)_{2k+1}}&=2 \binom{n+k}{k}\,,\\
N_\text{spinor}^{\Spin(2n+1)_{2k+1}}&=\binom{n+k}{k}+\binom{n+k-1}{k-1}\,,
\end{aligned}
\end{equation}
so that
\begin{equation}
\tr_{\Spin(2n+1)_k}(\boldsymbol1)\equiv N_\text{tensor}^{\Spin(2n+1)_k}+N_\text{spinor}^{\Spin(2n+1)_k}\,.
\end{equation}

For a more interesting example, let us now compute the partition function of $\SO(2n+1)_k=\Spin(2n+1)_k/\mathbb Z_2$, which corresponds to the algebra $\mathfrak{so}_{2n+1}$ extended by the simple current $\chi=[0,k,0,\dots,0]$. This current has spin $h_\chi=k/2$, and so the extension is fermionic for odd $k$. The current acts on a given representation $[\lambda_0,\lambda_1,\dots,\lambda_n]$ as $\lambda_0\leftrightarrow \lambda_1$.

Consider first the case of even $k$, so that $\SO(2n+1)_k$ makes sense as a bosonic theory. The extension has two effects: first, it projects out all the spinors, and second, it organizes the tensors into $\mathbb Z_2$-orbits. Such an orbit may have length two or one; the latter corresponds to a fixed-point under spectral flow, i.e., to a tensor with $\lambda_0=\lambda_1$, which splits into two primaries in the quotient. The number of fixed-points corresponds to the number of solutions to $\lambda_0+\lambda_1+2(\lambda_2+\cdots+\lambda_{n-1})+\lambda_n=k$ with $\lambda_0=\lambda_1$ and $\lambda_n$ even, i.e., $\binom{n+k/2-1}{k/2}$. Therefore, the number of conformal blocks is
\begin{equation}
\tr_{\SO(2n+1)_{2k}}(\boldsymbol1)=\frac12\bigg(N_\text{tensor}^{\Spin(2n+1)_{2k}}-\binom{n+k-1}{k}\bigg)+2\binom{n+k-1}{k}\,.
\end{equation}

Let now $k$ be odd, which makes $\SO(2n+1)_k$ a spin theory. The total number of states is the same on every spin structure, so we shall count the bosons and fermions in the Ramond sector (which is the richest case, as only this sector may contain fermions). The total number of states is the sum, while the Witten index is the difference. In the Ramond sector, the quotient projects out the tensors, and it organizes the spinors into $\mathbb Z_2$-orbits. The bosons are the length-two orbits, and the fermions are the fixed-points. The latter are the representations with $\lambda_0+\lambda_1+2(\lambda_2+\cdots+\lambda_{n-1})+\lambda_n=k$ with $\lambda_0=\lambda_1$ and $\lambda_n$ odd, which has $\binom{n+(k-1)/2-1}{(k-1)/2}$ solutions. Thus, the number of bosons and fermions is
\begin{equation}
\begin{aligned}
N_\text{boson}^{\SO(2n+1)_{2k+1}}&=\frac12\bigg(N_\text{spinor}^{\Spin(2n+1)_{2k+1}}-\binom{n+k-1}{k}\bigg)\,,\\
N_\text{fermion}^{\SO(2n+1)_{2k+1}}&=\binom{n+k-1}{k}\,,
\end{aligned}
\end{equation}
from where it follows that
\begin{equation}\label{eq:Z_SO_odd_NS}
\begin{aligned}
\tr_{\SO(2n+1)_{2k+1}}(\boldsymbol1)&=\tr_{\SO(2n+1)_{2k+1}}(-1)^F=N_\text{boson}^{\SO(2n+1)_{2k+1}}+N_\text{fermion}^{\SO(2n+1)_{2k+1}}\\
&=\binom{n+k}{k}\,,
\end{aligned}
\end{equation}
for all spatial spin structures, except for the odd structure for which
\begin{equation}\label{eq:index_SO_odd_odd}
\begin{aligned}
\tr_{\SO(2n+1)_{2k+1}}(-1)^F&=N_\text{boson}^{\SO(2n+1)_{2k+1}}-N_\text{fermion}^{\SO(2n+1)_{2k+1}}\\
&=\binom{n+k}{k}-2 \binom{n+k-1}{k}\,.
\end{aligned}
\end{equation}

We see that $\tr(\boldsymbol1)$ is invariant under $n\leftrightarrow k$, as required by level-rank duality. Similarly, $\tr(-1)^F$ is invariant up to a sign, which is due to the difference in the framing anomalies (i.e., the precise level-rank duality~\cite{Aharony:2016jvv} is $\SO(2n+1)_{2k+1}\leftrightarrow\SO(2k+1)_{-2n-1}\times \SO((2n+1)(2k+1))_1$, with the invertible factor contributing with a global factor of $(-1)^{(2n+1)(2k+1)}\equiv-1$ to the trace, cf.~\eqref{eq:inv_phase_sign}).

%\note{For fun, let us compute the partition function twisted by the magnetic symmetry $\mathsf M$ (in the bosonic case only for now, to keep things simple). This symmetry permutes the lines that split in the one-form quotient. Thus, the partition function with $\mathsf M$ inserted along any of the homology cycles (or both at the same time) computes $\tr_{\SO(2N+1)_{2k}}(\mathsf M)\equiv\binom{N+k-1}{k}$. Summing over the $2^8$ possible insertions, we get
%\begin{equation}
%\frac{1}{|\mathbb Z_2|}\sum_{\mathsf M_{\mathsf a},\mathsf M_{\mathsf b}}\tr_{\mathsf M_{\mathsf a},\mathsf M_{\mathsf b}}(\boldsymbol1)\equiv\frac12(\tr\boldsymbol1+7\tr\mathsf M)
%\end{equation}
%which equals $Z^\text{tensor}(\Spin(2N+1)_{2k})+Z^\text{spinor}(\Spin(2N+1)_{2k})\equiv Z(\Spin(2N+1)_{2k})$, as expected. It would be nice to have a similar formula for spin theories, so as to have a simpler expression for the twisted indices that doesn't require explicitly counting Majorana lines and understanding the action of $\mathsf M$ on them. Perhaps, the index in $G$ twisted by $\mathsf M$ is given by
%\begin{equation}
%I^{\mathsf M}_G\sim2I_{G/\mathsf M}-I_G
%\end{equation}
%where $I_G$ denotes the untwisted index, and $I_{G/\mathsf M}$ the untwisted index of the theory after the $\mathsf M$-orbifold.
%}

\subsection[$G=\Spin(2n)$]{$\boldsymbol{G=\Spin(2n)}$}

Consider the algebra $D_n=\mathfrak{so}_{2n}$. The comarks are $a_i^\vee=1$ for $i=0,1,n-1,n$, and $a_i^\vee=2$ for $i=2,\dots,n-2$. Plugging this into~\eqref{eq:Z_bos_twist} we get the generating function as
\begin{equation}
Z(\Spin(2n),q)=(1-q)^{-4}(1-q^2)^{-(n-3)}\,,\\
\end{equation}
and, by expanding, the untwisted trace
\begin{equation}\label{eq:Z_Spin_unt}
\begin{aligned}
\tr_{\Spin(2n)_{2k}}(\boldsymbol1)&=\binom{n+k}{k}+6 \binom{n+k-1}{k-1}+\binom{n+k-2}{k-2}\,,\\
\tr_{\Spin(2n)_{2k+1}}(\boldsymbol1)&=4 \binom{n+k}{k}+4 \binom{n+k-1}{k-1}\,.
\end{aligned}
\end{equation}

This is the number of integrable representations of $\Spin(2n)_k$, that is, the dimension of the torus Hilbert space of this Chern-Simons theory. For future reference, it is also useful to break up the states into the tensors and spinors. In other words, we shall be interested in knowing how many of the states of $\Spin(2n)$ are tensorial representations, and how many are spinorial representations. These are defined by $\lambda_{n-1}+\lambda_n=\text{even}$ and $\lambda_{n-1}+\lambda_n=\text{odd}$, respectively, which yields the following:
\begin{equation}
\begin{aligned}
N_\text{tensor}^{\Spin(2n)_{2k}}&=\binom{n+k}{k}+2 \binom{n+k-1}{k-1}+\binom{n+k-2}{k-2}\,,\\
N_\text{spinor}^{\Spin(2n)_{2k}}&=4 \binom{n+k-1}{k-1}\,,\\
N_\text{tensor}^{\Spin(2n)_{2k+1}}&=N_\text{spinor}^{\Spin(2n)_{2k+1}}=2 \binom{n+k}{k}+2 \binom{n+k-1}{k-1}\,,
\end{aligned}
\end{equation}
so that
\begin{equation}
\tr_{\Spin(2n)_k}(\boldsymbol1)\equiv N_\text{tensor}^{\Spin(2n)_k}+N_\text{spinor}^{\Spin(2n)_k}\,.
\end{equation}

For a more interesting example, let us now compute the partition function $\SO(2n)_k=\Spin(2n)_k/\mathbb Z_2$, which corresponds to the algebra $\mathfrak{so}_{2n}$ extended by the simple current $\chi=[0,k,0,\dots,0]$. This current has spin $h_\chi=k/2$, and so the extension is fermionic for odd $k$. The current acts on a given representation $[\lambda_0,\lambda_1,\dots,\lambda_n]$ as $\lambda_0\leftrightarrow \lambda_1$ and $\lambda_{n-1}\leftrightarrow \lambda_n$.

Consider first the case of even $k$, so that $\SO(2n)_k$ makes sense as a bosonic theory. The extension has two effects: first, it projects out all the spinors, and second, it organizes the tensors into $\mathbb Z_2$-orbits. Such an orbit may have length two or one; the latter corresponds to a fixed-point under spectral flow, i.e., to a tensor with $\lambda_0=\lambda_1$ and $\lambda_{n-1}=\lambda_n$, which splits into two primaries in the quotient. The number of fixed-points corresponds to the number of solutions to $\lambda_0+\lambda_1+2(\lambda_2+\cdots+\lambda_{n-2})+\lambda_{n-1}+\lambda_n=k$ with $\lambda_0=\lambda_1$ and $\lambda_{n-1}=\lambda_n$, i.e., $\binom{n+k/2-2}{k/2}$. Therefore, the number of conformal blocks is
\begin{equation}\label{eq:Z_SO_unt}
\tr_{\SO(2n)_{2k}}(\boldsymbol1)=\frac12\bigg(N_\text{tensor}^{\Spin(2n)_{2k}}-\binom{n+k-2}{k}\bigg)+2\binom{n+k-2}{k}\,.
\end{equation}

Let now $k$ be odd, which makes $\SO(2n)_k$ a spin theory. The number of states is the same on every spin structure, so we shall count the bosons and fermions in the Ramond sector (which is the richest case, as only this sector may contain fermions). The total number of states is the sum, while the Witten index is the difference. In the Ramond sector, the quotient projects out the tensors, and it organizes the spinors into $\mathbb Z_2$-orbits. The bosons are the length-two orbits, and the fermions are the fixed-points. Note that the spinors have $\lambda_{n-1}+\lambda_n=\text{odd}$, which is incompatible with the fixed-point condition $\lambda_{n-1}=\lambda_n$, and so there are no fixed-points. Thus, the number of bosons and fermions is
\begin{equation}
\begin{aligned}
N_\text{boson}^{\SO(2n)_{2k+1}}&=\frac12N_\text{spinor}^{\Spin(2n)_{2k+1}}\,,\\
N_\text{fermion}^{\SO(2n)_{2k+1}}&=0\,,
\end{aligned}
\end{equation}
from where it follows that
\begin{equation}\label{eq:dim_SO_even_NS}
\tr_{\SO(2n)_{2k+1}}(\boldsymbol1)=\tr_{\SO(2n)_{2k+1}}(-1)^F=N_\text{boson}^{\SO(2n)_{2k+1}}\,,
\end{equation}
for all spatial spin structures. Note that the equality of $\tr(\boldsymbol1),\tr(-1)^F$ on all spin structures was in fact expected from the level-rank duality $\SO(2n)_{2k+1}\leftrightarrow \SO(2k+1)_{-2n}$, the \emph{r.h.s.}~being fermionic only due to a trivial $\SO(2n(2k+1))_1=\{\boldsymbol1,\psi\}$ factor (which contains an even number of fermions, so not even the sign of $\tr(-1)^F$ may depend on the spin structure).

\section{The Exceptional Groups}\label{sec:exceptional_indices}

In this appendix we gather the different indices for the exceptional groups, whose domain wall theory is yet to be identified. Any given proposal for the dynamics of such walls ought to be consistent with the indices below. By particle-hole symmetry, the indices satisfy $I^{\mathsf s}_n=\pm I^{\mathsf s}_{h-n}$, and therefore we only show the first $\lceil h/2\rceil$ indices, so as to avoid repetition.

We compute the untwisted indices, and the indices twisted by the zero-form and one-form symmetries (see table~\ref{tab:comarksone}). The symmetries $\mathsf c\in\mathsf C=\mathbb Z_2$ and $\mathsf g\in\Gamma=\mathbb Z_3$ act on the Dynkin diagram of $E_6$ as follows:
\begin{equation}
\begin{tikzpicture}[baseline=(current bounding box.center)]

\node at (-1.6,-2) {$E_6^{{\color{blue}(1)}}\colon\ \begin{cases}
\\[+180pt]
\end{cases}$};

\node at (4.5,.1) {$\mapsto\quad E_6^{{\color{blue}(2)}}\colon$};
\node at (4.5,-3.9) {$\mapsto\quad G_2^{{\color{blue}(1)}}\colon$};

\draw[thick,blue] (0,0) -- (.6,0);
\draw[thick] (.6,0) -- (1.2,0) -- ({1.2+1.2*cos(60)},{1.2*sin(60)});
\draw[thick] (1.2,0) -- ({1.2+1.2*cos(60)},{-1.2*sin(60)});
\draw[thick,fill=white,draw=blue] (0,0) circle (.1cm) node[anchor=south,inner ysep=10pt,scale=.6] {${\color{blue}1}$};
\draw[thick,fill=white] (.6,0) circle (.1cm) node[anchor=south,inner ysep=10pt,scale=.6] {$2$};
\draw[thick,fill=white] (1.2,0) circle (.1cm) node[anchor=south,inner ysep=10pt,scale=.6] {$3$};
\draw[thick,fill=white] ({1.2+.6*cos(60)},{.6*sin(60)}) circle (.1cm) node[anchor=south east,inner ysep=10pt,scale=.6] {$2$};
\draw[thick,fill=white] ({1.2+1.2*cos(60)},{1.2*sin(60)}) circle (.1cm) node[anchor=south east,inner ysep=10pt,scale=.6] {$1$};
\draw[thick,fill=white] ({1.2+.6*cos(60)},{-.6*sin(60)}) circle (.1cm) node[anchor=north east,inner ysep=7pt,inner xsep=7pt,scale=.6] {$2$};
\draw[thick,fill=white] ({1.2+1.2*cos(60)},{-1.2*sin(60)}) circle (.1cm) node[anchor=north east,inner ysep=7pt,inner xsep=7pt,scale=.6] {$1$};

\draw[<->,thick,red,>=stealth] (1.8,.4) arc[radius=.8, start angle=30, end angle=-30];
\draw[<->,thick,red,>=stealth] (2.2,.9) arc[radius=1.6, start angle=35, end angle=-35];

\begin{scope}[shift={(6,0)}]
\draw[thick,blue] (0,0) -- (.6,0);
\draw[thick] (.6,0) -- (1.2,0);
\draw[thick] (1.2,.1) -- (1.8,.1);
\draw[thick] (1.2,-.1) -- (1.8,-.1);
\draw[thick] (1.8,0) -- (2.4,0);
\draw[thick] (2-.6,-.2) -- (2.2-.6,0) -- (2-.6,.2);

\draw[thick,fill=white,draw=blue] (0,0) circle (.1cm) node[anchor=south,inner ysep=10pt,scale=.6] {${\color{blue}1}$};
\draw[thick,fill=white] (.6,0) circle (.1cm) node[anchor=south,inner ysep=10pt,scale=.6] {$2$};
\draw[thick,fill=white] (1.2,0) circle (.1cm) node[anchor=south,inner ysep=10pt,scale=.6] {$3$};
\draw[thick,fill=white] (1.8,0) circle (.1cm) node[anchor=south,inner ysep=10pt,scale=.6] {$4$};
\draw[thick,fill=white] (2.4,0) circle (.1cm) node[anchor=south,inner ysep=10pt,scale=.6] {$2$};
\end{scope}

\begin{scope}[shift={(0,-4)}]
\draw[thick,blue] (0,0) -- (.6,0);
\draw[thick] (.6,0) -- (1.2,0) -- ({1.2+1.2*cos(60)},{1.2*sin(60)});
\draw[thick] (1.2,0) -- ({1.2+1.2*cos(60)},{-1.2*sin(60)});
\draw[thick,fill=white,draw=blue] (0,0) circle (.1cm) node[anchor=south,inner ysep=10pt,scale=.6] {${\color{blue}1}$};
\draw[thick,fill=white] (.6,0) circle (.1cm) node[anchor=south,inner ysep=10pt,scale=.6] {$2$};
\draw[thick,fill=white] (1.2,0) circle (.1cm) node[anchor=south,inner ysep=10pt,scale=.6] {$3$};
\draw[thick,fill=white] ({1.2+.6*cos(60)},{.6*sin(60)}) circle (.1cm) node[anchor=south east,inner ysep=10pt,scale=.6] {$2$};
\draw[thick,fill=white] ({1.2+1.2*cos(60)},{1.2*sin(60)}) circle (.1cm) node[anchor=south east,inner ysep=10pt,scale=.6] {$1$};
\draw[thick,fill=white] ({1.2+.6*cos(60)},{-.6*sin(60)}) circle (.1cm) node[anchor=north east,inner ysep=7pt,inner xsep=7pt,scale=.6] {$2$};
\draw[thick,fill=white] ({1.2+1.2*cos(60)},{-1.2*sin(60)}) circle (.1cm) node[anchor=north east,inner ysep=7pt,inner xsep=7pt,scale=.6] {$1$};

\draw[->,thick,red,>=stealth] (1.8,.4) arc[radius=.8, start angle=30, end angle=-30];
\draw[->,thick,red,>=stealth] (2.2,.9) arc[radius=1.6, start angle=35, end angle=-35];
\draw[->,thick,red,>=stealth] (1.8-1.2,.4) arc[radius=.8, start angle=30+120-8, end angle=-30+120-8];
\draw[->,thick,red,>=stealth] (1.8-.6,.4-1.1) arc[radius=.8, start angle=30-120-3, end angle=-30-120-3];
\draw[->,thick,red,>=stealth] (2.2-2.3,.9-.5) arc[radius=1.6, start angle=35+120, end angle=-35+120];
\draw[->,thick,red,>=stealth] (2.2-.8,.9-2.25) arc[radius=1.6, start angle=35-120-4, end angle=-35-120-4];

\begin{scope}[shift={(6,0)}]
\draw[thick,blue] (.6,0) -- (1.2,0);
\draw[thick] (1.2,.1) -- (1.8,.1);
\draw[thick] (1.2,-.1) -- (1.8,-.1);
\draw[thick] (2.2-.6,-.2) -- (2-.6,0) -- (2.2-.6,.2);
\draw[thick] (1.2,0) -- (1.8,0);

\draw[thick,fill=white,draw=blue] (.6,0) circle (.1cm) node[anchor=south,inner ysep=10pt,scale=.6] {${\color{blue}3}$};
\draw[thick,fill=white] (1.2,0) circle (.1cm) node[anchor=south,inner ysep=10pt,scale=.6] {$6$};
\draw[thick,fill=white] (1.8,0) circle (.1cm) node[anchor=south,inner ysep=10pt,scale=.6] {$3$};
\end{scope}

\end{scope}

\end{tikzpicture}
\end{equation}

The symmetry $\mathsf g\in\Gamma$ acts on $E_7$ as follows:
\begin{equation}
\begin{tikzpicture}[baseline=(current bounding box.center)]

\node at (-1.6,.1) {$E_7^{{\color{blue}(1)}}\colon$};
\node at (5.5,.1) {$\mapsto\quad F_4^{{\color{blue}(1)}}\colon$};

\draw[thick,blue] (0,0) -- (.6,0);
\draw[thick] (.6,0) -- (3.6,0);
\draw[thick] (1.8,0) -- (1.8,.6);
\draw[thick,fill=white,draw=blue] (0,0) circle (.1cm) node[anchor=south,inner ysep=10pt,scale=.6] {${\color{blue}1}$};
\draw[thick,fill=white] (.6,0) circle (.1cm) node[anchor=south,inner ysep=10pt,scale=.6] {$2$};
\draw[thick,fill=white] (1.2,0) circle (.1cm) node[anchor=south,inner ysep=10pt,scale=.6] {$3$};
\draw[thick,fill=white] (1.8,0) circle (.1cm) node[anchor=north,inner ysep=10pt,scale=.6] {$4$};
\draw[thick,fill=white] (2.4,0) circle (.1cm) node[anchor=south,inner ysep=10pt,scale=.6] {$3$};
\draw[thick,fill=white] (3,0) circle (.1cm) node[anchor=south,inner ysep=10pt,scale=.6] {$2$};
\draw[thick,fill=white] (3.6,0) circle (.1cm) node[anchor=south,inner ysep=10pt,scale=.6] {$1$};
\draw[thick,fill=white] (1.8,.6) circle (.1cm) node[anchor=south,inner ysep=10pt,scale=.6] {$2$};

\begin{scope}[yscale=-1,shift={(-1.12,-3.1)}]
\draw[<->,thick,red,>=stealth] (4.75,3.5) arc[radius=1.95, start angle=20, end angle=160];
\draw[<->,thick,red,>=stealth] (4.2,3.5) arc[radius=1.35, start angle=20, end angle=160];
\draw[<->,thick,red,>=stealth] (3.6,3.5) arc[radius=.7, start angle=20, end angle=160];
\end{scope}

\begin{scope}[shift={(7.5,0)}]
\draw[thick,blue] (0,0) -- (.6,0);
\draw[thick] (.6,0) -- (1.2,0);
\draw[thick] (1.2,.1) -- (1.8,.1);
\draw[thick] (1.2,-.1) -- (1.8,-.1);
\draw[thick] (1.8,0) -- (2.4,0);
\draw[thick] (2.2-.6,-.2) -- (2-.6,0) -- (2.2-.6,.2);

\draw[thick,fill=white,draw=blue] (0,0) circle (.1cm) node[anchor=south,inner ysep=10pt,scale=.6] {${\color{blue}2}$};
\draw[thick,fill=white] (.6,0) circle (.1cm) node[anchor=south,inner ysep=10pt,scale=.6] {$4$};
\draw[thick,fill=white] (1.2,0) circle (.1cm) node[anchor=south,inner ysep=10pt,scale=.6] {$6$};
\draw[thick,fill=white] (1.8,0) circle (.1cm) node[anchor=south,inner ysep=10pt,scale=.6] {$4$};
\draw[thick,fill=white] (2.4,0) circle (.1cm) node[anchor=south,inner ysep=10pt,scale=.6] {$2$};
\end{scope}

\end{tikzpicture}
\end{equation}

Using these diagrams we find:

$\bullet$ $E_6$:
\begin{equation}
\begin{aligned}
Z(q)&=1 - 3 q + 7 q^3 - 3 q^4 - 6 q^5 +\cdots\\
Z^{\mathsf c}(q)&=1 - q - 2 q^2 + q^3 + q^4 + 2 q^5+\cdots\\
Z^{\mathsf g}(q)&=1 - 2 q^3 +\cdots
\end{aligned}
\end{equation}

$\bullet$ $E_7$:
\begin{equation}
\begin{aligned}
Z(q)&=1 - 2 q - 2 q^2 + 4 q^3 + 3 q^4 - 7 q^6 - 4 q^7 + 5 q^8 + 4 q^9 +\cdots\\
Z^{\mathsf g}(q)&=1 - 2 q^2 - q^4 + 3 q^6 + q^8+\cdots
\end{aligned}
\end{equation}

$\bullet$ $E_8$:
\begin{equation}
\begin{aligned}
Z(q)&=1 - q - 2 q^2 + q^4 + 4 q^5 + q^6 - 3 q^8\\
&\quad - 6 q^9 - q^{10} + 4 q^{12} + 5 q^{13} + 5 q^{14}+\cdots
\end{aligned}
\end{equation}

$\bullet$ $F_4$:
\begin{equation}
Z(q)=1 - 2 q - q^2 + 3 q^3 + q^4+\cdots
\end{equation}

%\bibliography{references}
\clearpage
\printbibliography
\end{document}